\documentclass[apjl]{emulateapj}
\usepackage{color}

\newbox\grsign \setbox\grsign=\hbox{$>$} \newdimen\grdimen \grdimen=\ht\grsign
\newbox\simlessbox \newbox\simgreatbox
\setbox\simgreatbox=\hbox{\raise.5ex\hbox{$>$}\llap
     {\lower.5ex\hbox{$\sim$}}}\ht1=\grdimen\dp1=0pt
\setbox\simlessbox=\hbox{\raise.5ex\hbox{$<$}\llap
     {\lower.5ex\hbox{$\sim$}}}\ht2=\grdimen\dp2=0pt

\def\simless{\mathrel{\copy\simlessbox}}
\newbox\simppropto
\setbox\simppropto=\hbox{\raise.5ex\hbox{$\sim$}\llap
     {\lower.5ex\hbox{$\propto$}}}\ht2=\grdimen\dp2=0pt

\def\alwaysmath#1{\ifmmode{#1}\else{$#1$}\fi}

\lefthead{Majewski et al.} 
\righthead{APOGEE}

\begin{document}

\title{The Apache Point Observatory Galactic Evolution Experiment (APOGEE)} 
\author{Steven R. Majewski\altaffilmark{1}, 
Ricardo P. Schiavon\altaffilmark{2,3},  
Peter M. Frinchaboy\altaffilmark{4}, 
Carlos Allende Prieto\altaffilmark{5,6},  
Robert Barkhouser\altaffilmark{7},  
Dmitry Bizyaev\altaffilmark{8,9},
Basil Blank\altaffilmark{10},  
Sophia Brunner\altaffilmark{1}, 
Adam Burton\altaffilmark{1},
Ricardo Carrera\altaffilmark{5,6},  
S. Drew Chojnowski\altaffilmark{1,11}, 
K\'atia Cunha\altaffilmark{12,13}, 
Courtney Epstein\altaffilmark{14}, 
Greg Fitzgerald\altaffilmark{15},  
Ana E. Garc\'ia P\'erez\altaffilmark{1,5},  
Fred R. Hearty\altaffilmark{1,16}, 
Chuck Henderson\altaffilmark{10},  
Jon A. Holtzman\altaffilmark{11},  
Jennifer A. Johnson\altaffilmark{14},  
Charles R. Lam\altaffilmark{1}, 
James E. Lawler\altaffilmark{17}, 
Paul Maseman\altaffilmark{18},  
Szabolcs M\'esz\'aros\altaffilmark{5,6,19},  
Matthew Nelson\altaffilmark{1},   
Duy Coung Nguyen\altaffilmark{20},  
David L. Nidever\altaffilmark{1,21}, 
Marc Pinsonneault\altaffilmark{14}, 
Matthew Shetrone\altaffilmark{22},  
Stephen Smee\altaffilmark{7}, 
Verne V. Smith\altaffilmark{13,23}, 
Todd Stolberg\altaffilmark{15}, 
Michael F. Skrutskie\altaffilmark{1}, 
Eric Walker\altaffilmark{1},
John C. Wilson\altaffilmark{1}, 
Gail Zasowski\altaffilmark{1,7}, 
Friedrich Anders\altaffilmark{24},
Sarbani Basu\altaffilmark{25},
Stephane Beland\altaffilmark{26,27},
Michael R. Blanton\altaffilmark{28},
Jo Bovy\altaffilmark{29,30}, 
Joel R. Brownstein\altaffilmark{31},
Joleen Carlberg\altaffilmark{1,32},
William Chaplin\altaffilmark{33,34},
Cristina Chiappini\altaffilmark{24},
Daniel J. Eisenstein\altaffilmark{35},
Yvonne Elsworth\altaffilmark{33},
Diane Feuillet\altaffilmark{11},  
Scott W. Fleming\altaffilmark{36,37},
Jessica Galbraith-Frew\altaffilmark{31},
Rafael A. Garc\'\i a\altaffilmark{38},
D. An\'\i bal Garc\'\i a-Hern\'andez\altaffilmark{5,6},
Bruce A. Gillespie\altaffilmark{7},
L\'eo Girardi\altaffilmark{39,40},
James E. Gunn\altaffilmark{41},
Sten Hasselquist\altaffilmark{1,11},  
Michael R. Hayden\altaffilmark{11}, 
Saskia Hekker\altaffilmark{34,42},
Inese Ivans\altaffilmark{31},
Karen Kinemuchi\altaffilmark{8},
Mark Klaene\altaffilmark{8},
Suvrath Mahadevan\altaffilmark{16},
Savita Mathur\altaffilmark{43},
Beno\^\i t Mosser\altaffilmark{44},
Demitri Muna\altaffilmark{14},
Jeffrey A. Munn\altaffilmark{45}
Robert C. Nichol\altaffilmark{46},
Robert W. O'Connell\altaffilmark{1},
A.C. Robin\altaffilmark{47},
Helio Rocha-Pinto\altaffilmark{40,48},
Matthias Schultheis\altaffilmark{49},
Aldo M. Serenelli\altaffilmark{50},
Neville Shane\altaffilmark{1},
Victor Silva Aguirre\altaffilmark{34},
Jennifer S. Sobeck\altaffilmark{1},
Benjamin Thompson\altaffilmark{4},
Nicholas W. Troup\altaffilmark{1},
David H. Weinberg\altaffilmark{14},  
Olga Zamora\altaffilmark{5,6} }

\affil{$^1$ Dept. of Astronomy, University of Virginia, Charlottesville, VA 22904-4325, USA} 
\affil{$^2$ Gemini Observatory, 670 N. A'Ohoku Place, Hilo, HI 96720, USA}
\affil{$^3$ Astrophysics Research Institute, Liverpool John Moores University, 146 Brownlow Hill, Liverpool, L3 5RF, UK}
\affil{$^4$ Department of Physics and Astronomy, Texas Christian University, Fort Worth, TX 76129, USA}
\affil{$^5$ Instituto de Astrof\'isica de Canarias, E-38200 La Laguna,Tenerife, Spain}
\affil{$^6$ 16 Departamento de Astrof\'isica, Universidad de La Laguna, E-38206 La Laguna, Tenerife, Spain}
\affil{$^7$ Department of Physics and Astronomy, Johns Hopkins University, Baltimore, MD 21218, USA}
\affil{$^8$ Apache Point Observatory and New Mexico State University, P.O. Box 59, Sunspot, NM, 88349-0059, USA}
\affil{$^9$ Sternberg Astronomical Institute, Moscow State University, Universitetsky prosp. 13, Moscow, Russia}
\affil{$^{10}$ Pulse Ray Machining \& Design, 4583 State Route 414, Beaver Dams, NY 14812 USA}
\affil{$^{11}$ New Mexico State University, Las Cruces, NM 88003, USA}
\affil{$^{12}$ Observat\'orio Nacional, Rio de Janeiro, RJ 20921-400, Brazil}
\affil{$^{13}$ Steward Observatory, University of Arizona, Tucson, AZ 85721, USA}
\affil{$^{14}$ The Ohio State University, Columbus, OH 43210, USA}
\affil{$^{15}$ New England Optical Systems, 237 Cedar Hill Street, Marlborough, MA 01752 USA}
\affil{$^{16}$ Department of Astronomy \& Astrophysics, The Pennsylvania State University, 525 Davey Laboratory, University Park PA 16802, USA}
\affil{$^{17}$ Department of Physics, University of Wisconsin-Madison, 1150 University Avenue, Madison, WI 53706, USA}
\affil{$^{18}$ Steward Observatory, University of Arizona, Tucson, AZ 85721, USA}
\affil{$^{19}$ ELTE Gothard Astrophysical Observatory, H-9704 Szombathely, Szent Imre Herceg St. 112, Hungary}
\affil{$^{20}$ Dunlap Institute for Astronomy and Astrophysics, University of Toronto, Toronto, Ontario, Canada}
\affil{$^{21}$ Department of Astronomy, University of Michigan, Ann Arbor, MI 48109, USA}
\affil{$^{22}$ University of Texas at Austin, McDonald Observatory, Fort Davis, TX 79734, USA}
\affil{$^{23}$ National Optical Astronomy Observatories, PO Box 26732, Tucson, AZ 85719, USA}
\affil{$^{24}$ Leibniz-Institut f\"ur Astrophysik Potsdam (AIP), An der Sternwarte 16, 14482 Potsdam, Germany}
\affil{$^{25}$ Department of Astronomy, Yale University, PO Box 208101, New Haven, CT 06520-8101 USA}
\affil{$^{26}$ Laboratory for Atmospheric and Space Physics, University of Colorado, Boulder, CO 80303, USA}
\affil{$^{27}$ Center for Astrophysics and Space Astronomy, University of Colorado Boulder, Boulder, CO 80303, USA}
\affil{$^{28}$ Center for Cosmology and Particle Physics, Department of Physics, New York University, 4 Washington Place, New York, NY 10003, USA}
\affil{$^{29}$ Institute for Advanced Study, Einstein Drive, Princeton, NJ 08540, USA}
\affil{$^{30}$ John Bahcall Fellow}
\affil{$^{31}$ Department of Physics and Astronomy, University of Utah, 115 S 1400 E \#201 Salt Lake City, UT 84112 USA}
\affil{$^{32}$ NASA Goddard Space Flight Center, Code 667, Greenbelt, MD 20771, USA}
\affil{$^{33}$ School of Physics and Astronomy, University of Birmingham, Birmingham B15 2TT, UK}
\affil{$^{34}$ Stellar Astrophysics Centre (SAC), Department of Physics and Astronomy, Aarhus University, Ny Munkegade 120, DK-8000 Aarhus C, Denmark}
\affil{$^{35}$ Harvard-Smithsonian Center for Astrophysics, 60 Garden St., MS \#20, Cambridge, MA 02138, USA}
\affil{$^{36}$ Computer Sciences Corporation, 3700 San Martin Dr, Baltimore, MD 21218, USA}
\affil{$^{37}$ Space Telescope Science Institute, 3700 San Martin Dr, Baltimore, MD 21218, USA}
\affil{$^{38}$ Laboratoire AIM, CEA/DSM -- CNRS - Univ. Paris Diderot -- IRFU/SAp, Centre de Saclay, 91191 Gif-sur-Yvette Cedex, France}
\affil{$^{39}$ Osservatorio Astronomico di Padova -- INAF, Vicolo dell'Osservatorio 5, I-35122 Padova, Italy}
\affil{$^{40}$ Laborat\'orio Interinstitucional de e-Astronomia - LIneA, Rua Gal. Jos\'e Cristino 77, Rio de Janeiro, RJ - 20921-400, Brazil}
\affil{$^{41}$ Department of Astrophysical Sciences, Peyton Hall, Princeton University 08544, USA}
\affil{$^{42}$ Max-Planck-Institut f\"ur Sonnensystemforschung, Justus-von-Liebig-Weg 3, 37077 G\"ottingen, Germany}
\affil{$^{43}$ Space Science Institute, 4750 Walnut street, Suite 205, Boulder, CO 80301 USA}
\affil{$^{44}$ LESIA, CNRS, Université Pierre et Marie Curie, Université Denis Diderot, Observatoire de Paris, 92195 Meudon Cedex, France}
\affil{$^{45}$ US Naval Observatory, Flagstaff Station, 10391 West Naval Observatory Road, Flagstaff, AZ 86005-8521, USA}
\affil{$^{46}$ Institute of Cosmology and Gravitation, University of Portsmouth, Portsmouth, UK}
\affil{$^{47}$ Institut Utinam, CNRS UMR6213, Universit\'e de Franche-Comt\'e, OSU THETA Franche-Comt\'e-Bourgogne, Observatoire de Besan\c con, BP 1615, 25010 Besan\c con Cedex, France}
\affil{$^{48}$ Universidade Federal do Rio de Janeiro, Observat\'orio do Valongo, Ladeira do Pedro Ant\^onio 43, 20080-090 Rio de Janeiro, Brazil}
\affil{$^{49}$ Universit\'e de Nice Sophia-Antipolis, CNRS, Observatoire de C\^ote d'Azur, Laboratoire Lagrange, 06304 Nice Cedex 4, France}
\affil{$^{50}$ Institute of Space Sciences (CSIC-IEEC) Campus UAB, Torre C5 parell 2 Bellaterra, 08193 Spain}

\begin{abstract} The Apache Point Observatory Galactic Evolution
Experiment (APOGEE), one of the programs in the Sloan Digital Sky
Survey III (SDSS-III), has now completed its systematic, homogeneous
spectroscopic survey sampling all major populations of the Milky
Way.  After a three year observing campaign on the Sloan 2.5-m
Telescope,
APOGEE has collected a half million high resolution ($R\sim 22,500$),
high $S/N$ ($>$100), infrared ($1.51$-$1.70$~$\mu$m) spectra for
146,000 stars, with time series information via repeat visits to
most of these stars.  This paper describes the motivations for the
survey and its overall design --- hardware, field placement, target
selection, operations --- and gives an overview of these aspects
as well as the data reduction, analysis and products.  An index
is also given to the complement of technical papers that describe
various critical survey components in detail.  Finally, we discuss
the achieved survey performance and illustrate the variety of
potential uses of the data products by way of a number of science
demonstrations, which span from time series analysis of stellar
spectral variations and radial velocity variations from stellar
companions, to spatial maps of kinematics, metallicity and abundance
patterns across the Galaxy and as a function of age, to new views
of the interstellar medium, the chemistry of star clusters, and
the discovery of rare stellar species.  As part of SDSS-III Data
Release 12, all of the APOGEE data products are now publicly
available.  \end{abstract}


\section{Introduction}\label{sec:intro}

\subsection{Galactic Archaeology Surveys}\label{sec:surveys}

Modern astrophysics has taken two general observational approaches to
understand the evolution of galaxies.
On the one hand, increasingly larger aperture telescopes, on the ground
and in space, give access to the high redshift universe and offer ``low resolution"
snapshots of ever earlier phases of galaxy evolution.   
On the other hand, increasingly efficient,
multiplexing photometric and spectroscopic instrumentation, often on 
smaller, workhorse telescopes, has made
possible enormous, definitive surveys of nearby galaxies yielding a
``high resolution'' view of the present state of these systems.  
These data can be tested against ``end state" predictions for the growth
of large structures in the universe to provide critical constraints on
cosmological models --- so-called ``near-field cosmology".  
These two observational approaches 
 --- overviews of global properties at high
redshift versus more detailed information at low redshift ---
provide complementary information that must
be accommodated by evolutionary theories.

The highest-granularity information about galaxy evolution is provided
by stars in our own Milky Way, whose present spatial distributions, ages, chemistry
and kinematics contain fossilized clues to its formation.
Guided by detailed models for the chemical and dynamical evolution
of stellar populations,
critical telltale signatures and correlations within the above
observables provide constraints on the model predictions for physical
quantities that cannot be observed directly, such as the history
of star formation, the early stellar initial mass function, and the
merger history of Galactic subsystems.  This ``Galactic archaeology''
remains the principal basis by which models for the formation and
chemodynamical evolution of the Milky Way and analogous systems are
formulated and refined.  The vast literature on Milky Way stellar populations as
tools for understanding Galactic evolution has been reviewed in the
past by, e.g., \citet{Gilmore89}, \citet{Majewski93}, \citet{Freeman02}, 
and more recently by \citet{Ivezic12} and \citet{Rix13}.

These efforts are of course greatly aided by access to expansive,
carefully designed, homogeneous, and precise databases of properties for
stellar samples that span large regions of the Galaxy and include
all of the principal stellar populations.  Modern archetypes of
such databases are large photometric surveys like the Two Micron
All-Sky Survey (2MASS; \citealt{Skrutskie06}) and the Sloan Digital Sky Survey 
(SDSS; \citealt{York00}).
Over the past decade, these photometric catalogs have been
widely exploited for insights into the nature of the Milky Way and
probing the complexities of Galactic structure 
--- e.g., 
halo substructure \citep[e.g.,][]{Majewski03,Rocha-Pinto04, Belokurov06,Grillmair09}, 
satellite galaxies \citep[e.g.,][]{Willman05,Belokurov07}, 
the warp of the disk \citep[e.g.,][]{LopezCorredoira02,Reyle09}, 
and the still unresolved, composite anatomy of the bulge  \citep[e.g.,][]{Robin12}, which includes 
the recently found X-shaped feature \citep[e.g.,][]{McWilliam10}, 
and one or more central bars \citep[e.g.,][]{Alard01,Hammersley00,Cabrera-Lavers07}.  
Follow-on, low and medium resolution spectroscopic
programs provide  additional dynamical discrimination of, and context for,
these structures as well as general
information on their chemical make-up (e.g., mean metallicities
and, in some cases, an additional dimension of chemistry, such as [$\alpha$/Fe]);
these broad brushstrokes represent an important  step in characterizing
stellar populations and constraining galactic evolution models.

Meanwhile, high-resolution stellar spectroscopy has become an increasingly
indispensable tool for providing the necessary detail to discriminate
galaxy evolution models.  Accurate multi-element chemical abundances
provide insight into the stellar initial mass functions, and
histories of star formation and  chemical enrichment of stellar
populations, which, in turn, fuel ever more sophisticated galactic
dynamical and chemodynamical models 
\citep[e.g.,][]{Chiappini01,Chiappini03,Sellwood02,Abadi03,Bournaud09,
Schonrich09,Minchev10,Bird13,Minchev13,Minchev14,Kubryk14}.
Coupled with orbital information derived from precise radial
velocities, these data probe the role of dynamical phenomena such
as large-scale dissipative collapses, mergers, gas flows,
bars, spiral arms, dynamical heating and radial migration.

Conventional echelle spectroscopy programs to deliver high resolution
spectroscopic data useful for Galactic archaeology demand substantial
resources, often on the world's largest telescopes. Consequently,
while heroic efforts have been devoted to surveying stars in a wide
variety of environments --- including, e.g., dwarf spheroidals,
globular clusters, the Magellanic Clouds, tidal streams, and the
Galactic bulge --- until very recently the solar neighborhood was
the only region for which multiple hundreds or thousands of
observations had been assembled for ``Galactic field stars" (e.g.,
\citealt{Edvardsson93},
\citealt{Bensby03},
\citealt{Fuhrmann04}, 
\citealt{Venn04}, 
\citealt{Nissen10},
\citealt{Soubiran10},
\citealt{Adibekyan12,Adibekyan13},
\citealt{Bensby14}) --- 
and with these samples 
traditionally relying on kinematically-selected samples to harvest from the nearby
stars of accessible apparent brightnesses a broad spread
of stellar ages and population classes.
For stellar populations not represented in the solar
neighborhood, like the Galactic bulge, and for {\it in situ} studies of field stars outside of the
solar neighborhood, 
high resolution observations are only now generating samples with hundreds of stars.
In the inner Galaxy where foreground dust obscuration is a formidable challenge, 
many previous samples were concentrated to a 
handful of low extinction sightlines, such as Baade's
Window.  Unfortunately,
the aggregate of these piecemeal collections of spectroscopic data, heterogeneously 
assembled, can give a biased and incomplete view of the Milky Way.

\begin{figure}[h]
\includegraphics[width=0.50\textwidth]{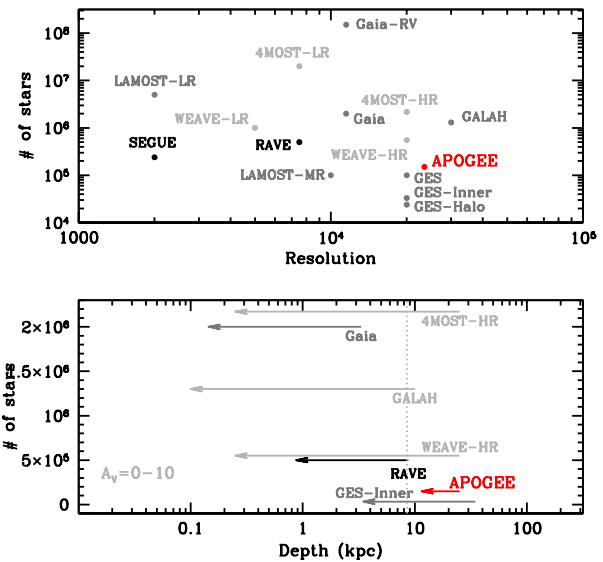}
\vspace{-25mm}
\caption{\footnotesize 
APOGEE in the context of other Galactic archaeology surveys, past,
present and future.  
The top panel shows the
number of Milky Way stars, observed or anticipated, as a function
of survey resolution.  For those surveys with at least a resolution
of $R = 10,000$, the bottom panel shows the expected nominal depth
of the survey for a star with $M_{\rm V}=-1$ in the case of no
extinction ({\it right end of arrows}) and in the case of $A_V =
10$ ({\it left end of arrows}). In both panels, already
completed surveys are shown in black,
ongoing surveys in dark gray, and planned surveys in light gray.
For surveys with multiple resolution modes, data in the top panel are plotted
separately for high resolution (HR), medium resolution (MR) and/or
low resolution (LR).  For the Gaia/ESO survey, data for ``Inner
Galaxy'' and ``Halo'' subsamples are shown separately as well.
``Gaia-RV'' includes Gaia high resolution spectra of enough $S/N$
to deliver radial velocities, whereas ``Gaia'' indicates only those
with $S/N$ high enough for abundance work.   For Gaia we adopted $A_G/A_V$ from
\cite{Jordi2010}, assuming $(V-I_C)_0=1.7$; sample numbers were taken from
{\tt http://www.cosmos.esa.int/web/gaia/science-performance.}
}
\label{fig:surveys}
\end{figure}

Truly comprehensive evolutionary models for the Milky Way must be
informed and constrained by statistically reliable, complete or at
least unbiased Galactic archaeology studies, which requires the
construction of large, truly systematic and homogeneous chemokinematical
surveys covering expansive volumes of the Milky Way and sampling
all stellar populations, including, in particular, those dust
obscured inner regions where the bulk of the Galactic stellar mass
is concentrated.
A number of ambitious
``Galactic archaeology" spectroscopic surveys that aim to fill this need
have been previously undertaken --- such as RAVE \citep{Steinmetz06},
SEGUE-1 \citep{Yanny09}, 
SEGUE-2 \citep{Rockosi09}, and ARGOS \citep{Freeman13} --- are currently
underway --- such as LAMOST \citep{Cui12}, Gaia/ESO \citep{Gilmore12},
GALAH \citep{Zucker12}, and Gaia \citep{Perryman01} --- 
or are envisaged --- e.g., those associated with the WEAVE \citep{Dalton14},
4MOST \citep{deJong14}, and MOONS \citep{Cirasuolo14} instruments.
While each of these surveys focuses on large samples of $\gtrsim100,000$
stars, all of those surveys past and ongoing are based on optical
observations and are therefore strongly hampered by
interstellar obscuration in the Galactic plane (Fig.~\ref{fig:surveys}, {\it bottom}); 
this makes it challenging to sample significant numbers of stars
within the very dusty regions of the Milky Way that are
both central to constraining formation models and encompass most
of the Galactic stellar mass (and some projects, like the RAVE, SEGUE
and GALAH surveys,
specifically avoid low Galactic latitudes).
Therefore, with optical wavelength surveys 
it is challenging to assemble a systematic census having
comparable or sufficient representation
of all Galactic stellar populations and across wide expanses of
the Galactic disk and bulge.

While other surveys, such as BRAVA \citep{Rich07}, ARGOS \citep{Freeman13},
and \cite{Gonzalez11}
aim to fill at least part of this void by specifically focusing on the Galactic
bulge, they utilize target selection criteria that differ from those
of surveys of other parts of the Milky Way, 
which makes it difficult to generate a holistic
picture of stellar populations and their potential
connections.
Moreover, apart from GALAH and the Gaia/ESO survey,
these other studies are limited to medium resolution spectroscopy ($R<10,000$;
Fig.~\ref{fig:surveys}),
so are unable to provide reliably the kind of detailed elemental
abundance information that is now a key input to the models, while
at the same time the moderate velocity precisions can limit their
sensitivity to more subtle, second order dynamical effects (e.g.,
perturbations by spiral arms and the bar, dynamical resonances,
velocity coherent moving groups and streams).

\vspace{-10mm}

\subsection{APOGEE: Basic Architecture and
Motivations}\label{sec:motivations}

In contrast to previous and ongoing surveys,
the Apache Point Observatory Galactic Evolution
Experiment (APOGEE) in Sloan Digital Sky Survey III (SDSS-III) 
was designed to tackle the fundamental problem of galaxy formation
through the first large-scale, systematic, precision chemical and
kinematical study specifically optimized to include exploration of
the ``dust-hidden" populations in the Milky Way.  As planned,
APOGEE aimed 
to build a database of high-resolution ($R$$\sim$22,500),
near-infrared (1.6~$\mu$m $H$-band) spectra for over $10^5$ stars --- 
predominantly red giant branch (RGB) and other luminous post-main-sequence stars 
--- across the Milky Way, but with particular
emphasis on obtaining significant representation from heavily
dust-obscured parts of the Galactic disk and bulge.  Operationally,
this plan, now successfully executed, exploits several key advantages:
\begin{itemize}
\vskip-0cm
\item Near-infrared observations profit from a selective extinction
  many times lower (for $H$-band, a factor of 6)
  in magnitudes (i.e., 250 times in flux) than that at visual wavelengths.
\vskip-0cm
\item High resolution spectra provide the
  chemical abundance and radial
  velocity precision needed for constraining Galactic evolutionary models
  and, in the $H$ band, sample 
  lines of numerous elements up to and including the iron peak and for which
  non-LTE departures are typically small.
\vskip-0cm
\item Collectively, RGB stars, asymptotic giant branch 
(AGB) and red
supergiant (RSG) stars are good tracers of the disk and bulge, together sample
populations of all ages and metallicities, and are luminous in the
NIR.  Meanwhile the high sky density of these stellar types --- combined 
with the large, 3 deg diameter Sloan 2.5-m telescope field-of-view (FOV) and
high throughput, multifiber plugplate handling system --- allows
simultaneous observation of several hundred targets at a time, and
thousands per night.

\vskip-0cm
\end{itemize}
Together these advantages translate into a Milky Way survey trade-space
``sweet spot" that permits efficient, high resolution, near-infrared spectroscopic
study of large numbers of stars that are not easily accessible to
optical programs, and enables a consistent database of stellar spectra
to be assembled across {\it all} Galactic stellar populations, from
the inner bulge to the more distant Galactic halo.  
Thus, with APOGEE, it is possible for the first time to explore and compare
with great statistical significance the chemokinematical character of all Milky Way 
stellar subsystems using the same set of chemical elements and line transitions
represented in data of uniform high quality that has been gathered, 
processed and analyzed identically.

\subsection{High Level Science Goals}
\label{sec:sciencegoals}

The principal scientific goals of APOGEE, which together provide a broad but integrated 
approach to furthering our understanding of galaxy evolution, are:

\begin{enumerate}

\item To measure high precision abundances for multiple elements in 
$\sim 10^5$ stars across the Galaxy, and derive
distributions
of these chemical properties to
constrain Galactic chemical evolution
(GCE) models.  Among the target elements are the preferred GCE tracers
and most common metals --- i.e., carbon, nitrogen and oxygen --- as well
as other metals with particular sensitivity to the star formation
history (SFH) and the initial mass function (IMF) of stellar populations.

\item To derive high precision kinematical data useful for 
constraining dynamical models for the disk, bulge, bar and  
halo, and for discriminating substructures within these components.

\item To access the often ignored, dust-obscured inner regions of the 
Galaxy, and for the observed stars in these regions
derive the same data as is available for other, more accessible 
stellar populations, which will also be included in the survey; furthermore, 
by collecting large survey samples, provide statistically reliable 
measures of chemistry and kinematics in dozens of Galactic zones 
($R$, $\theta$, $Z$) at the level currently available in the solar 
neighborhood. 

\item To contribute to explorations of the early Galaxy 
by inferring the properties of the earliest stars, thought to reside or to 
have resided in the Galactic bulge \citep{Tumlinson10}.  This can be
achieved either by detecting them 
directly if they survive to the present day, or (more likely) by measuring their 
nucleosynthetic products in the most metal-poor stars that do survive.

\item To achieve a dramatic (more than two orders of magnitude) leap
in the total number of available high resolution, high $S/N$ infrared stellar 
spectra, which will enable substantial advances in stellar astrophysics,
GCE modeling, and dynamical modeling of the Milky Way.

\end{enumerate}

Among the more specific issues that APOGEE
 addresses are:
\begin{itemize}
\vskip-0cm
  \item Completing the first systematic determination of the 3-D chemical
    abundance distribution --- for numerous elements ---
    across the Galactic disk,
    determining the Galactic rotation curve and examining
    correlations between abundances and stellar kinematics at all disk radii.
\vskip-0cm
  \item Determining distribution functions of chemical abundances for
  a variety of elements in the bulge, bar(s) and inner
  disk, and probing correlations
    between abundances and kinematics there, with the goal of
    investigating the physical mechanisms that connect these structures and
    	determining the origin(s) of the bulge.
\vskip-0cm
  \item Establishing the nature of the Galactic bar and spiral arms and their influence on the disk
  through detailed assessment of
    abundances and velocities of stars in and around them.

\vskip-0cm
  \item Assessing the properties that discriminate the thin and thick disks to clarify the nature and 
  origin of the latter.
  
\vskip-0cm
  \item Drawing a comprehensive picture of the chemical
	  evolution of the Galaxy via the placement of strong constraints
	  on the initial mass function and star formation rates of the bulge, 
  disk and halo as a function of position and time.

\vskip-0cm
  \item Searching for and probing the chemistry and dynamics of low-latitude 
  substructures in both the disk and halo, whether from dynamical resonances 
  or the accretion of satellites.
\vskip-0cm  \item Investigating the kinematics and chemistry of the
Galactic halo and its substructure, and using them to assess the
relative contribution of accreted versus stars formed {\it in situ}. 
\vskip-0cm
  \item By reference to other available optical, near-IR, mid-IR and radio
	data, exploring the interstellar medium, mapping the Galactic dust
    distribution and constraining variations in the interstellar extinction law.
\vskip-0cm
  \item By combining spectroscopic data with the detailed information on stellar structure provided by asteroseismology surveys, deriving accurate ages for Galactic field stars, which provide key timestamps for
the exploration of all manner of Galactic evolutionary phenomena.

 \item Through the marriage of accurate stellar parameters and detailed
chemical compositions from APOGEE with accurate asteroseismological data,
providing fundamental constraints on models of the structure of
stellar interiors, opening doors to progress in important areas of stellar
physics.

\vskip-0cm
\end{itemize}

\subsection{Goals of this Paper}\label{sec:presentgoals}

The goal of the present paper is to give a broad overview of the APOGEE survey, with particular focus
on the scientific motivations and technical rationale that led to the instrument and survey design choices
(\S \ref{sec:topleveltechnical}).
The ``birds-eye'' descriptions of the APOGEE project given here are at a level intended to give the potential 
user of APOGEE data sufficient background to understand the basic structure of the instrument
(\S \ref{sec:instrumentoverview}) and
survey (\S \ref{sec:surveydesign}), 
how the data were collected (\S \ref{sec:operations}) and 
processed (\S \ref{sec:handling}), 
and what the data look like 
and how they may be accessed (\S \ref{sec:products_docs}).
We also summarize how the survey met its intended goals 
(\S \ref{sec:performance}), further  illustrated via several science 
demonstrations (\S \ref{sec:science_demo}).  
 The latter also introduce some of the variety of applications to which APOGEE 
 data may be directed.
Based on the success of the APOGEE project, a new collaboration has been
formed to expand upon this initial survey via the APOGEE-2 project; these and related future
efforts are discussed briefly in \S \ref{sec:future}. 

This paper is a primer to those interested in a general understanding 
of the overall structure of the APOGEE survey.  
For more details on particular
aspects of the survey, users are encouraged
to consult a series of focused technical papers that address various
specific elements of the survey, the software and algorithms used
to produce the publicly released databases, and post-survey assessments
of the calibration and accuracy of the data (Table \ref{tab:techpapers}).
These papers and other survey documentation are described further in \S
\ref{sec:documentation}.
On-line information describing the data
release file formats and available on-line tools for data visualization and
download may be found at {\tt http://www.sdss.org}.

\begin{table}[htdp]
\caption{APOGEE Survey Technical Papers}
\begin{center}
\begin{tabular}{l l }
\hline 
Topic  & Reference \\
\hline 
Spectrograph  					& \citet{Wilson15} \\
Target Selection 				& \citet{Zasowski13} \\
Data Reduction Pipeline			& \citet{Nidever15} \\
Stellar Atmosphere Models 		& \citet{Meszaros12} \\
Stellar Spectral Libraries			& \citet{Zamora15} \\
APOGEE Line List   & \citet{Shetrone15}\\
Tests of the APOGEE Line List 	& \citet{Smith13} \\
Stellar Parameters and Chemical \\
\ \ \ \ \ Abundances Pipeline (ASPCAP) & \citet{GarciaPerez15} \\
ASPCAP Calibration for DR10  		&  \citet{Meszaros13} \\
Tests of Individual Element  \\
\ \ \ \ \ Abundances from ASPCAP & \citet{Cunha15} \\
Overview of DR12 APOGEE data                   & \citet{Holtzman15}\\
Kepler Asteroseismology Collaboration 	& \citet{Pinsonneault14}\\
CoRoT Asteroseismology Collaboration	& \citet{Anders15} \\
\hline 
\end{tabular}
\end{center}
\label{tab:techpapers}
\end{table}

\section{Top Level Technical Requirements}
\label{sec:topleveltechnical}

The requirement for accurate abundances of a large number of
chemical elements 
necessitates
an intricate interplay between $S/N$,
spectral coverage and spectral resolution, which are the most
fundamental factors that drove the APOGEE instrumental design. On
one hand, the desire to obtain abundances for a large number of
chemical elements calls for a wide wavelength baseline, so that
numerous absorption lines from many chemical species are represented
in the observed spectra. On the other hand, the accuracy achievable
in abundance analysis work is strongly dependent on spectral
resolution, which, for a fixed detector format in the limit of
Nyquist sampling, is inversely proportional to spectral bandwidth.
Additionally, the lower the resolution, the higher is the $S/N$
required to achieve a given abundance accuracy goal. Finally, the
higher the $S/N$, the fewer the stars that can be observed in a
given time period, for a given multiplexing power.
We discuss
here the scientific considerations that led to the final instrument
technical requirements for APOGEE.

\subsection{Wavelength Window of Operation}
\label{sec:wavelengths}

Recent technology development has made high resolution NIR spectroscopy
a new and vigorous area of astrophysical investigation, particularly
in the area of stellar atmospheres analysis.  The value and promise
of high resolution NIR spectroscopy for exploring stellar abundances
is attested by the growing number of papers on the subject over the
past decade using instruments suitable 
for the purpose on the world's
largest telescopes --- e.g.,  CRIRES on the VLT, NIRSPEC at
Keck, IRCS at Subaru, and, formerly, Phoenix at Gemini-South
\citep[e.g.,][]{Rich05,Cunha06,Cunha07,Ryde10,Tsuji14}.  While the
flow of high resolution NIR data has recently seen a dramatic upturn,
the study of stellar photospheres on the basis of NIR spectroscopy
has a long tradition \citep[e.g., see the early review by][]{Merrill79}.
The current state of the art in interpreting these data is proving
highly successful, competitive with, and complementary to, traditional
analyses in the optical (see references below).  

To probe the largest distances in the Galaxy most easily one should
focus on the intrinsically brightest population tracers.  
A particular advantage realized by working in the NIR is that the intrinsically
brightest common stars found in different aged populations --- 
RGB, AGB and RSG stars (collectively referred to as ``giants''  
throughout this paper) --- all have cool atmospheres, and are 
even brighter in the infrared than at optical wavelengths.
Moreover, selecting for red stars in
dereddened color-magnitude diagrams made from a magnitude-limited
survey like 2MASS 
guarantees a virtually giant-dominated sample.
Fortunately, the analysis of 
giant star atmospheres is an area that
has received particular attention in high resolution NIR spectroscopy,
given that these stars are the most accessible in star clusters,
resolved galaxies (like the Magellanic Clouds), and fields towards
the Galactic Center, like Baade's Window.  The earlier papers by
\citet{Smith85,Smith86,Smith90}
 focusing on
the CNO abundances in red giant stars were among the first efforts
to explore chemical abundances from high-resolution spectra in the
infrared.  More recently, the analysis of high-resolution spectra
in the $H$ band 
for stars in the Magellanic Clouds as well as the Galactic
bulge and center \citep{Smith02,Rich05,Cunha06,Cunha07,Ryde10}
have helped to demonstrate
the feasibility of determining precise chemical abundances in the
$H$-band and have helped to lay the foundation for the APOGEE Survey.

Choice of the {\it specific} NIR wavelength range to be used for APOGEE
involved optimizing a trade-off between competing desires:

\begin{itemize}
\item {\bf Penetration of Interstellar Dust:}  
The longer the infrared wavelength observed, the smaller is the
sensitivity of the light to the extinguishing effects of interstellar
dust, and the greater is the ability of the survey to penetrate
highly obscured regions of the inner Galaxy.

\item {\bf Thermal Background:}  At longer
wavelengths the contribution of the thermal background increases,
and becomes significant in the $K$-band and beyond.

\item {\bf Airglow:} The intensity of airglow emission (particularly
from OH) varies across the near infrared, with the weakest lines
in the $J$-band, and the strongest in the $H$-band.

\item {\bf Telluric Absorption:} The ranges of the ground-based
NIR bands are defined by major telluric absorption bands,
most especially from CO$_2$ and H$_2$O; however, bands of
various strengths from these molecules, as well as from CH$_4$,
O$_2$ and
O$_3$, are found all across the near-infrared.

\item {\bf Available Line Transitions:} Some key atomic elements,
like Fe, C, N and O (the latter expressed in molecular line absorption
from diatomics like CO, OH and CN) are represented by spectral
features all over the NIR,
whereas other interesting elements, like K, F, Al and Sc, 
have only a few lines.
\end{itemize}

Weighing the various aspects of this trade-space led to the selection of
the $H$-band for APOGEE, with relatively strong weighting given to
the first two considerations above:  While the $K$-band is less
sensitive to dust extinction than is the $H$-band \citep[$A_K/A_V
\sim 1/9$ compared to $A_H / A_V \sim 1/6$; e.g., ][]{Cardelli89},
the $H$-band still confers a powerful degree of insensitivity to dust,
whereas, in the meantime, $S/N$ considerations motivate avoiding
the large $K$-band backgrounds. Moreover, a $K$-band instrument
requires much greater consideration to mitigating contamination from
local sources of thermal background
than does an instrument working in the $H$-band.\footnote{Indeed,
initial designs for the APOGEE spectrograph considered the notion
of a highly accessible bench spectrograph operating in a commercial-grade
food storage freezer,
but eventually converged toward the conventional
liquid-nitrogren-cooled cryostat design described in \S \ref{sec:spectrograph}
(not least because of
problems with the significant heat dumping into the telescope
environment that the freezer would contribute).}

\begin{figure}[h]
\includegraphics[width=0.50\textwidth]{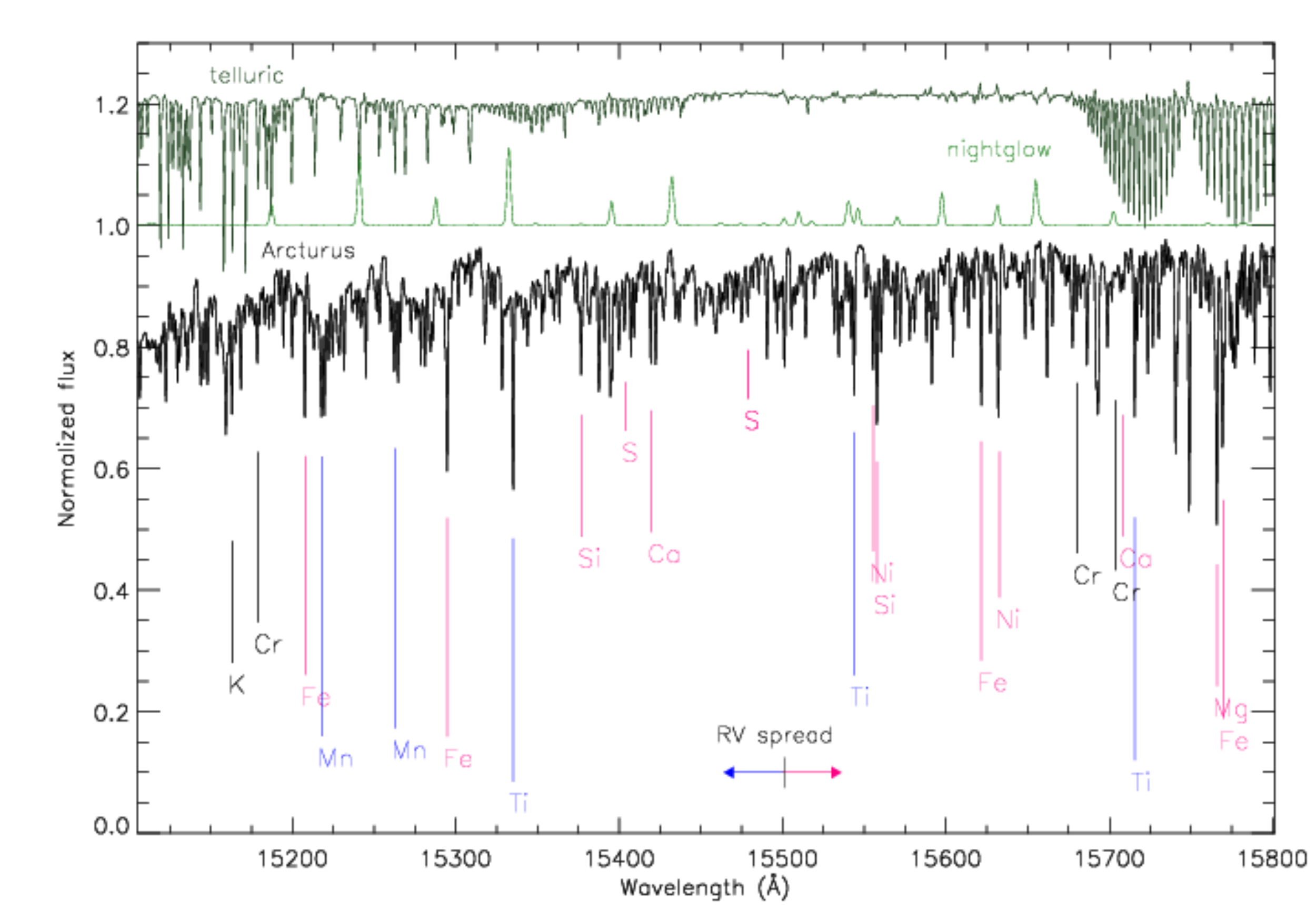}
\includegraphics[width=0.50\textwidth]{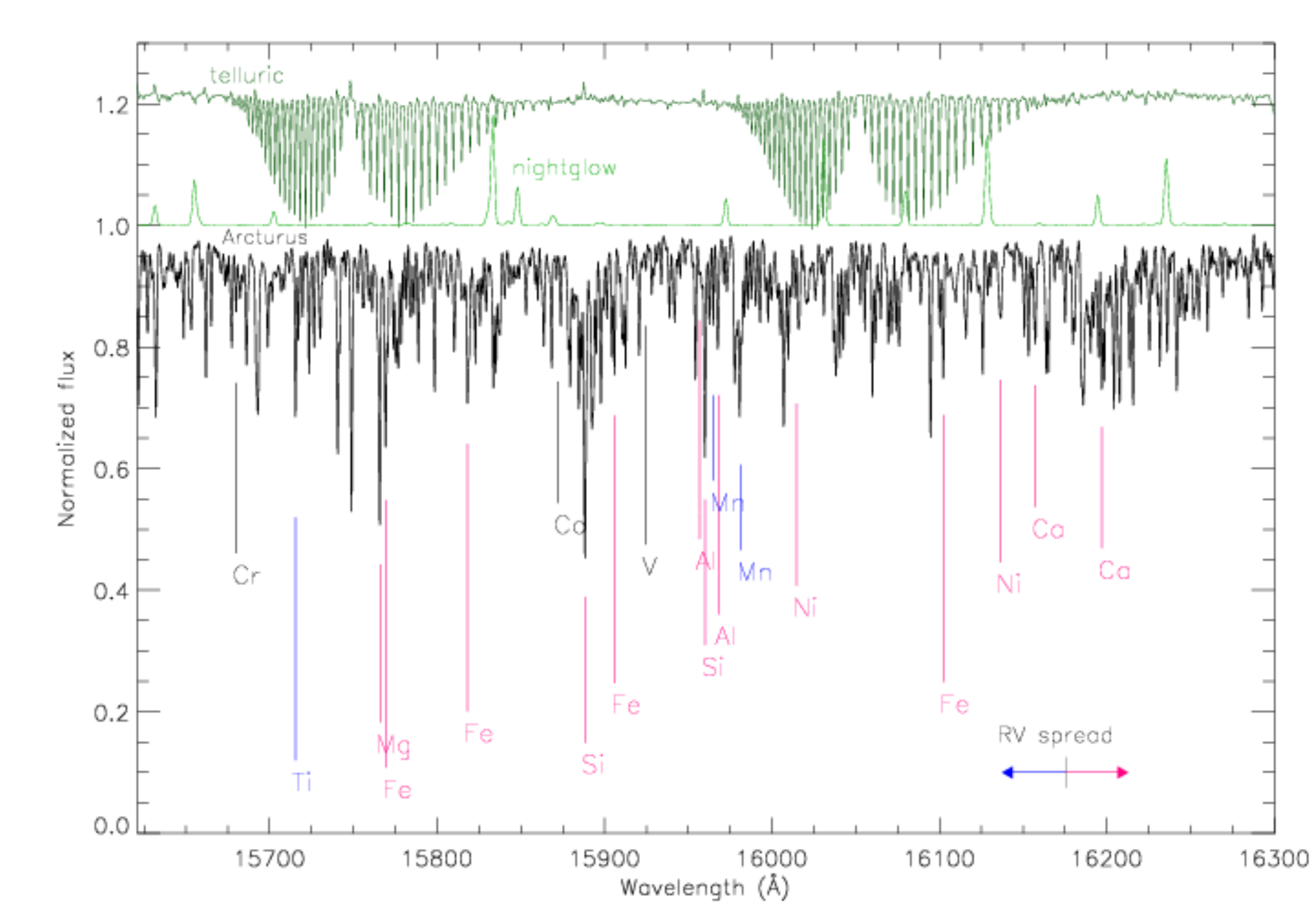}
\includegraphics[width=0.50\textwidth]{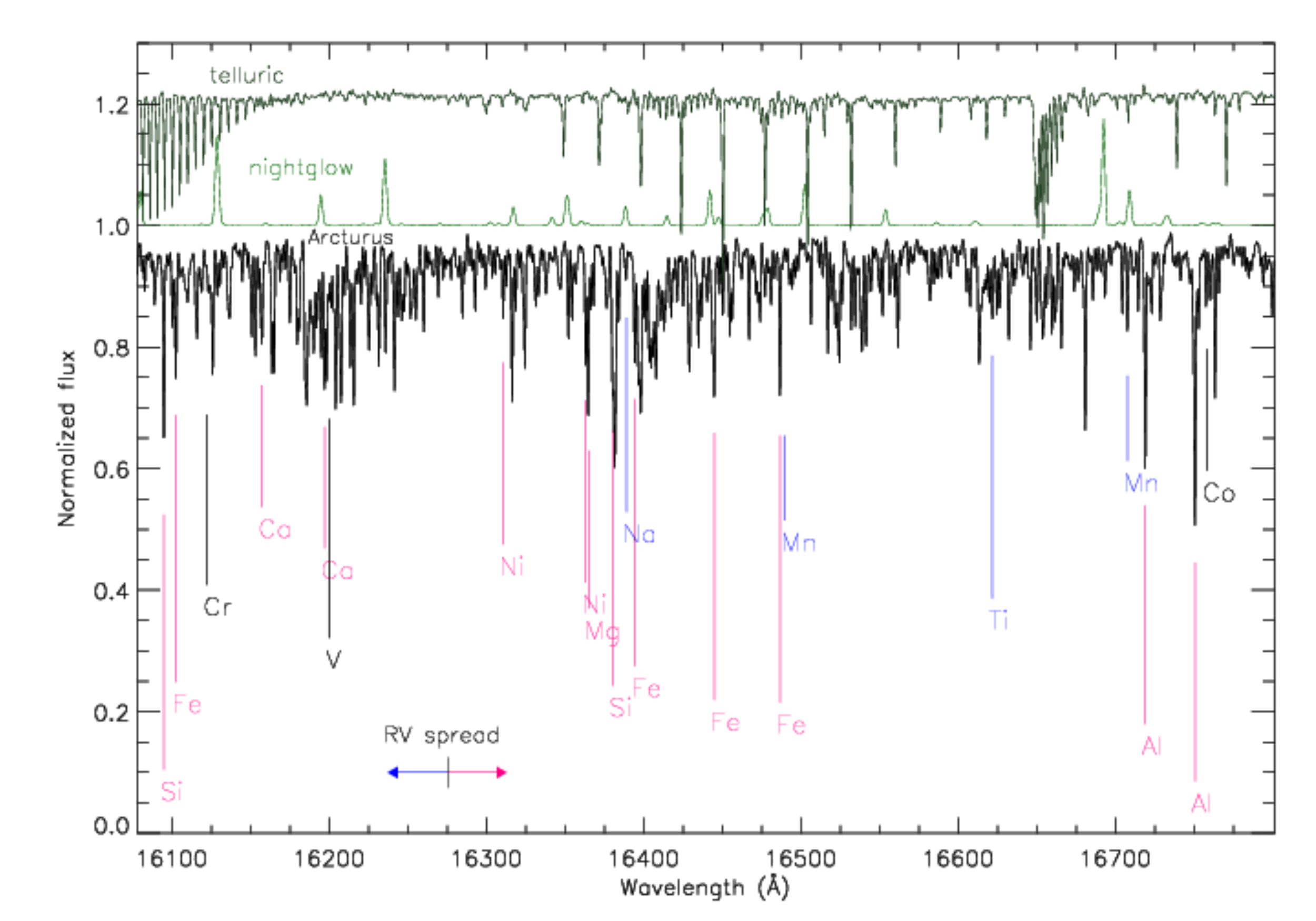}
\caption{\footnotesize 
In three overlapping wavelength regions, the distribution of telluric
absorption ({\it top spectra in each panel}), airglow ({\it middle
spectra}), and atomic lines in the spectrum of the star Arcturus
({\it bottom spectra}).  Some prominent atomic lines in the Arcturus
spectrum that guided the ultimate selection of the APOGEE wavelength
region are identified and color-coded as high priority ({\it red}),
medium priority ({\it blue}) and lower priority ({\it black}).  Also
indicated are the extremes in the potential shift in the lines from
extremes in radial velocity variation for potential (e.g., halo)
Milky Way stars (adopted as $\pm700$ km s$^{-1}$ in the lines). }
\label{fig:wavelengths} \end{figure}

Unfortunately, while the above thermal background issues favor it,
the $H$-band does include by far the strongest lines of the
OH airglow spectrum.  
On the other hand, in principle,
with high enough resolution the impact of those airglow lines could be
confined to a small fraction of the total spectrum, whereas in the
$K$-band the thermal
background would affect all pixels.  In the ultimately selected
APOGEE spectral range, the airglow spectrum includes about a dozen
strong lines and a few dozen weaker lines (e.g., Fig.
\ref{fig:wavelengths}); coincidentally, these lines span the entire 
APOGEE spectral region, which 
makes them potentially useful for wavelength calibration.

The shape of the telluric absorption spectrum strongly drove the primary part of the $H$-band worth 
considering for APOGEE.  The $H$-band itself was defined as
the atmospheric
transmission window between the strong and broad water absorption bands at
$\sim$1.4$~\mu$m 
and $\sim$1.9~$\mu$m.
By far, the lowest absorption in this region is in the range of approximately 
1.5-1.75~$\mu$m, although this region is punctuated by the 30013
$\leftarrow$ 00001 and 30012 $\leftarrow$ 00001\footnote{The
notation for the vibrational states follows the convention established by
HITRAN \citep{Rothman13}.} bands of the CO$_2$ 
molecule \citep{Miller04},
which cover roughly the $\lambda\lambda1.568$-$1.586$ and $\lambda\lambda
1.598$-$1.617~\mu$m spectral intervals, respectively
(Fig. \ref{fig:wavelengths}).  An initial, two-detector
design of APOGEE sought to avoid most of these bands, but eventually these bands
were almost fully included in the near-contiguous wavelength 
coverage of the final, three-detector
APOGEE instrument (\S \ref{sec:resolution}).

\subsection{Chemical Elements}
\label{sec:chemicalelements}

In principle, different near-infrared windows offer some variance
in available elements, but for many important elements (C, N, O ---
the most abundant metals in the universe --- and the fiducial element
Fe) there is ample representation in all three of the NIR bands
($J$, $H$ and $K$).  Inspection of the \citet[][]{Hinkle95} infrared
atlas reveals the $J$-band to have lines for almost the same set
of elements as the $H$-band, but the $H$-band lines tend to be
stronger in the spectra of giant stars than their $J$-band counterparts,
as attested by inspection of medium resolution NIR spectra
from the IRTF library \citep[see, e.g.,][in particular their Figures
10 and 11]{Rayner09}.
And while a number of $\alpha$-elements are represented
in either the $H$ or $K$ bands, other atoms with few transitions
are represented in only one or the other (e.g., the $H$-band offers
the important odd-$Z$ elements Al and K).
While these trade-offs --- typically between elements tracking similar
nucleosynthetic families --- were not strong drivers in the decision process
leading to the choice of the broadband NIR bandpass in which to
operate (i.e., $J$ versus $H$ versus $K$), they did play a larger 
role in fine tuning the precise limits of the wavelength coverage
(see below).
Fortunately, the $H$-band, preferred for other reasons given above,
was determined to offer an appealingly wide range of chemical
elements that could be sampled, covering a range of nucleosynthetic
pathways.

A detailed visual inspection of the infrared spectrum of Arcturus by
\citet[][Fig.~\ref{fig:wavelengths}]{Hinkle95} was used to define
the specific limits of the APOGEE spectral range.  Initially, a
survey of potentially accessible elements (atomic and in molecular
combinations) in the $H$-band was made, and showed potentially
useful representation from the following elements: C, N, O, Na, Mg,
Al, Si, S, K, Ca, Ti, V, Cr, Mn, Fe, Co and Ni (element by element
maps are shown in Fig. \ref{fig:element_maps} in Appendix A).
This is a useful subset of atomic species with which to probe most
types of nucleosynthesis.  Moreover, many of these elements 
are now accessible to integrated spectroscopy of extragalactic
systems, which makes it possible to place the Milky Way 
in context with other
galaxies having a range of masses and morphological types.
Unfortunately, conspicuously absent from this 
initial assessment are any significant
lines from neutron-capture elements, a general problem across the
NIR.\footnote{Subsequent work (e.g., Appendix E) 
has
resulted in the identification of weak lines from several neutron-capture
elements --- e.g., associated with Nd II and Ce III --- in the
APOGEE spectra of some s-process enhanced stars
\citep[e.g.,][]{Majewski15,Shetrone15}.}

The above panoply of $H$-band-accessible
elements offers a number of potentially interesting
insights into various aspects of Galactic chemical evolution  (see,
e.g., \citealt{Matteucci01} and the recent review of nucleosynthesis
and chemical evolution by \citealt{Nomoto13}):

\begin{itemize}

\item {\bf C, N}: Important elements produced in significant amounts
in intermediate-mass stars \citep{Ventura13}, and thus sensitive
to $\sim$100 Myr timescales of star formation and chemical
evolution.  Carbon is synthesized in both massive stars
($M\ge10M_{\odot}$) and lower-mass AGB stars ($M\sim1-4M_{\odot}$),
in roughly equal amounts \citep{Nomoto13}.  Because AGB stars produce
no Fe, [C/Fe] can present an interesting behavior as a
function of time in systems with ongoing star formation and chemical
enrichment: initially increasing due to the contribution by
core-collapse type II supernovae (SN II) and AGB stars, then declining
as a result of the onset of enrichment by Type Ia supernovae (SN
Ia).  Moreover, because oxygen is produced in large amounts by SN II,
the C/O ratio tracks
the relative contributions of low to intermediate-mass stars versus
massive stars in a given stellar population.  Nitrogen is produced
efficiently in intermediate-mass AGB stars \citep{Karakas10}, and
there are suggestions in the literature \citep[][and references
therein]{Chiappini13} for an important contribution by massive stars
as well.  Analysis of integrated spectra of M31 globular clusters
\citep{Schiavon13} and early-type galaxies \citep{Schiavon07,Conroy14}
suggests that secondary enrichment was important in these systems.
Although N can exhibit complicated behavior as a result of chemical
evolution, it provides information on the relative importance of
intermediate-mass stars
to chemical evolution.  Finally, because the [C/N] ratio
is affected by internal mixing, it is a function of stellar mass,
metallicity, and evolutionary stage, which suggests that it might be
useful for relative age determinations of stellar populations
\citep[e.g.,][]{Masseron15}.

\item {\bf O}: The quintessential SN II yield from hydrostatic
He-burning in massive stars and the most abundant element in the
universe, after hydrogen and helium. The timescale for the
release of oxygen by SN II is much shorter than that of iron by SN
Ia \citep[e.g.,][]{Tinsley79}.  Therefore, one can be argue that
[O/H] is a more suitable and sensible chronometer and independent
variable than [Fe/H] as a surrogate for ``metallicity'' in
investigations of temporal abundance ratio variations benchmarked by
overall enrichment level.
That iron is more commonly used to indicate
stellar metallicity is at least partly historically-rooted in
the relative ease with which [Fe/H] can be estimated from
analysis of high resolution blue/optical spectra of solar type
stars.  
However, because the $H$-band includes many OH and CO lines that can be
easily measured (and modeled) in the spectra of cool giants, APOGEE
can provide reliable and precise [O/H] abundances for large stellar samples
to lend better insights into 
crucial observables such as, e.g., the age-metallicity relation
in different Galactic subcomponents.  Moreover, stellar oxygen
abundances can be more directly compared with gas-phase metallicities,
which are predominantly based on measurements of oxygen lines
\citep[e.g.,][]{Kewley08}.  The [O/Fe] ratio has been extensively
used as an indicator of the relative contribution of SN II and SN
Ia to chemical enrichment, which makes it sensitive to the timescale
and/or efficiency for star formation as well as the shape of the
high-mass end of the IMF 
\citep[e.g.,][]{Tinsley79,Tinsley80,Wheeler89,McWilliam97}.

\item {\bf Mg}: Another important $\alpha$-element, Mg is an excellent
complement to O.  Its main isotope, $^{24}$Mg is produced in massive
stars during carbon burning.  Therefore, magnesium can also
constrain enrichment by SN II, having become commonly
used in part because it is relatively easier to measure than oxygen
in optical spectra, with early abundances being based on medium
resolution spectra \citep{Wallerstein62,Tomkin85,Laird86}.  When
combined with oxygen, magnesium can both probe the importance of Wolf-Rayet
winds in chemical evolution and provide insights on the slope of
the stellar initial mass function (IMF) \citep[e.g.,][and references
therein]{Fulbright07,Stasinska12,Nomoto13}.  Magnesium is also
important as the main element constraining the [$\alpha$/Fe] ratio
from integrated-light studies of extragalactic stellar systems
\citep[e.g.,][]{Worthey92,Schiavon07}. Thus, Mg measurements may
provide a key bridge between Galactic and extragalactic chemical
composition studies and facilitate the placement of the Milky Way
within the broader context of galaxy evolution.
In early-type galaxies \citep{Worthey92} and, to a lesser extent, in the bulges
of spirals \citep{Proctor02} magnesium is found to be enhanced
relative to iron, which is commonly interpreted as due to a short
timescale for star formation in those systems.

\item {\bf Na, Al}: Odd-$Z$ elements.  Sodium is produced during
carbon burning and returned to the ISM via SN II.  Aluminum, in
turn, is expected to be produced mostly during neon burning, with
only a small contribution from carbon burning.  The SN II yields
for these elements are moderately dependent on metallicity
\citep{Nomoto13}. Both Na and Al also particiapte in H-burning in
intermediate-mass stars \citep[e.g.,][]{Karakas10}, so these elements
can also monitor the impact of intermediate-mass stars on chemical
evolution.  Interestingly, studies of chemical evolution in the
Galactic thin and thick disk and halo reveal different trends for
the abundances of these elements as a function of [Fe/H]
\citep[e.g.,][]{Bensby14}.

\item {\bf Si, S}: These $\alpha$-elements are produced
mostly in SN II (with small amounts in
SN Ia).  Silicon, as $^{28}$Si, is the
most abundant product of oxygen burning, with the dominant sulfur
isotope, $^{32}$S, also synthesized in oxygen burning
\citep[e.g.,][]{Francois04,Nomoto13}.  
The abundances of these elements, in principle,
provide constraints on the stellar IMF
by comparison to the abundances of
lighter $\alpha$-elements O and Mg \citep[e.g.,][]{McWilliam97}.

\item {\bf K}: Another odd-$Z$ element whose chemical evolution is poorly
understood. \citet{Shimansky03} suggest that the evolution
of K comes from hydrostatic oxygen burning and we expect an increase
in [K/Fe] with [Fe/H].

\item {\bf Ca, Ti}: Two more elements with strong ties to SN II
yields, but which may also have some fraction produced in SN Ia
\citep[e.g.,][]{Francois04,Nomoto13}.  In Galactic populations,
these elements display similar trends to those of O, Mg, Si, and
S, but there has been debate in the literature as to whether they
behave like SN Ia products in early-type galaxies
\citep[e.g.,][]{Milone00,Saglia02,Cenarro04,Schiavon10,Conroy14}.

\item {\bf V}: Produced in both explosive oxygen-burning
and silicon burning, $^{51}$V is synthesized through radioactive
parents, $^{51}$Cr and $^{51}$Mn, and is made in both SN II
and SN Ia \citep{Nomoto13}.  \cite{Reddy06} find
[V/Fe] to be approximately solar
in the thin disk and slightly enhanced in
the thick disk (by about 0.1 dex).

\item {\bf Mn}: While most iron-peak elements follow iron, Mn does
not, with [Mn/Fe] decreasing with decreasing [Fe/H].  Manganese is
produced mainly from radioactive decay of $^{55}$Co in both
core-collapse and Type Ia supernovae \citep{Nomoto13};
the dominant source of Mn has not been definitively identified.

\item {\bf Cr, Fe, Co, Ni}:  These elements represent
the Fe-peak in APOGEE spectra and are produced in varying
amounts in both SN Ia and SN II.

\end{itemize}

The mere presence of a line transition, of course, is not sufficient
for it to provide scientifically useful abundance measurements.
As a means to assess the identified lines, extensive tests were
made of model RGB spectra of different metallicities ([Fe/H]
= $-2$, $-1$, 0) at a number of potential spectrograph resolutions
to determine their suitability for 0.1 dex precision measurements
(see \S \ref{sec:resolution}).  Given the results of these tests,
and to inform the final selection of the specific spectral
coverage, these elements were ranked in a prioritization scheme
that considered not only the nucleosynthetic family to which the
element belonged
and their value to mapping Galactic chemical evolution, but the
strength and number of the available transitions: \begin{itemize}

\item Top priority (i.e., ``must have'' elements): C, N, O, Mg, Al,
Si, Ca, Fe, Ni.

\item Medium priority (i.e., valuable elements
worth trying to include in APOGEE, but that should not necessarily drive
requirements for the 
survey): Na, S, Ti, Mn, K.

\item Lower priority (i.e., ``if at all possible'' elements ---
interesting elements but
not deemed essential for success):
V, Cr, Co.
\end{itemize}

A census of the $H$-band shows that the reddest third (approximately
1.7-1.8~$\mu$m) is the
most deficient in interesting spectral lines whereas the middle third 
(approximately 1.6-1.7~$\mu$m) has the highest density. 
Moreover, the 1.7-1.8~$\mu$m
subwindow has significantly worse telluric absorption (Fig.
\ref{fig:element_maps}).  This ultimately drove the primary APOGEE
wavelength of interest to roughly 1.5-1.7~$\mu$m.  The precise
wavelength limits were set by the specific line transitions desired,
after detailed assessment of resolution and $S/N$ considerations.

The ultimately adopted wavelength
setting includes sufficient lines for abundance work on all of the top and
medium priority elements listed above.  However, a subsequent assessment of 
the available lines for the low priority elements suggested that
abundances for Cr and Co would be very difficult to obtain reliably,
given the excitation potential, $\log gf$ and strength in
the Arcturus spectrum of these lines.  Therefore, abundances of Co and Cr were not
attempted in the first round of elemental abundance determinations
leading up to DR12.  The additional element Cu, on the other hand
was not considered as a viable APOGEE product 
when the survey was initially conceived, but later Cu was successfully
explored in FTS spectra of standard stars in the APOGEE region by
\cite{Smith13}.  The situation of these elements will be reevaluated
in the near future as a better understanding of available line
transitions in the APOGEE spectral range is
achieved and as ASPCAP's performance is improved.

\subsection{Resolution, S/N and Specific Wavelength Limits}
\label{sec:resolution}

As with most spectrographs, the precise specifications of the APOGEE spectrograph were the product of 
balancing the competing benefits of high resolution, high $S/N$ and a broad wavelength range. 
To model these factors we calculated a series of synthetic $H$-band spectra for RGB stars 
($T_{eff} = 4000$ K, $\log{g} = 1$) with [Fe/H] = $-2$, $-1$, 0, at a number of values for resolving 
power between $R=15,000$ and 30,000.  For each case we computed two spectra, one with
solar-scaled composition, and a second in which the abundance of a particular element, X, was modified by
$\Delta$[X/Fe] = 0.1.  These calculations were used to derive an estimate of the $S/N$ required to measure
abundance variations of the order of 0.1 dex at each resolution, as described in Appendix B.
The results are summarized in Figure \ref{fig:SNR}.

\vspace{-25mm}
\begin{figure}[h]
\includegraphics[width=0.50\textwidth]{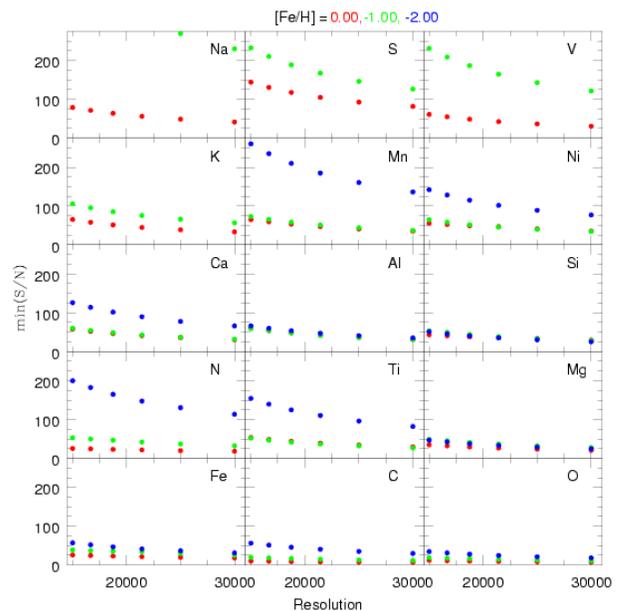}
\caption{\footnotesize Summary of the $S/N$ experiments described in Appendix B for each of 15 chemical elements.  
For each, the minimum required $S/N$ to measure 0.1 dex precision abundances is plotted for a variety of
resolutions from $R=15,000$ to 30,000, and for three metallicities, [Fe/H] $=-2$, $-1$, and 0. 
For Al, Si, and Mg the data points for all three modeled metallicities fall on top of one another. }
\label{fig:SNR}
\end{figure}

These calculations give rise to a number of general considerations:

\begin{itemize} 

\item Clearly the highest $S/N$ are required at the
lowest metallicities and resolutions, with metallicity being the
stronger driver.  For instance, measuring the Mg abundance to 0.1
dex at [Fe/H] $= -2.0$ would require $S/N>50$ at $R = 15,000$ and
$S/N>25$ at $R = 30,000$. At the other extreme, measuring K to 0.1
dex requires $S/N>700$ at $R = 15,000$ and $S/N>400$ at $R = 30,000$
for the same metal-poor star (outside the range shown for this
element in Fig.\ \ref{fig:SNR}).  

\item The Galactic thin disk is dominated by stars with
[Fe/H] $> -1$, for which the number of elemental abundances
that can be determined with 0.1 dex precision is maximum for a given $S/N$. For
example, at $R = 21,000$ and $S/N = 100$ we are able to measure all
of the listed elements except Na, S and V for thin disk stars.

\item For more metal-poor
stars, the challenging elements (at the top of Fig.\ \ref{fig:SNR}
and Tables \ref{tab:sntab} and  \ref{tab:snjon}) are measurable
with less demanding precision.
It might also be possible to do at least a statistical 
analysis of abundance patterns in metal-poor stars with the minimum nominal $S/N$ 
by combining spectra for multiple stars of similar chemistry or
position in phase space.

\item Obviously, for a constant exposure time, 
we can achieve higher $S/N$ by probing stars of brighter magnitudes and thereby
recover more of the challenging lines.

\end{itemize}

Even more specifically, this analysis led to the following considerations:
\begin{itemize}

\item Na is challenging for all but the most metal-rich
stars (even ignoring that the available Na lines are affected by non-negligible 
blending by molecular lines),
but we have Al as a substitute. Therefore Na
was not used as a requirement driver.

\item V is similar in chemical character to Al, and behaves similarly
to the $\alpha$-elements Ca and Ti
\citep{Reddy06}.
Therefore, loss of this element for some stars was not considered
a substantial setback.  

\item S is perhaps the most valuable element
with weak lines in the potential APOGEE line list.  The S I lines at
15422\AA\ and 15469\AA\  are the cleanest two lines, whereas the
strongest line at 15478\AA\  is blended with a strong Fe I feature.
In some ways Si can play the same role in terms of constraining the
high mass end of the IMF, though the combination of S and Si
is better.  While it was expected that S could be measured for
bright stars, it was accepted that S should not be a requirement
driver at the nominal survey magnitude limit.  

\item Given the above
logic that we would not use Na, V or S to drive the survey
specifications, it seemed reasonable to adopt the measurement of
the stellar K abundance for [Fe/H] $> -1$ stars as a requirements
driver.  

\item For metal-poor stars ([Fe/H] $\simless$ --1) it was
considered desirable to have, at minimum, O, C, Fe, Mg, Si, Al, Ca
and Ni, making the measurement of Ni in all stars a requirements
driver.

\item Overall improved resolution lowers the $S/N$ requirements,
but the gains from $R = 15,000$ to $R = 21,000$ are modest, according
to the calculations.  However, the  above estimates were assumed
to be somewhat optimistic, given that telluric lines, sky emission,
and blends of stellar lines were not considered.  Telluric and sky
lines will be better removed at higher resolution.  All elements
studied have at least some lines that are free of telluric or sky
interference for most stellar RVs, and fairly isolated at
solar metallicity and intermediate temperatures ($T_{\rm eff} \simeq
4000$ K). However, at cooler temperatures and similar metallicities,
molecular lines due to CN, CO, and/or OH affect virtually all
wavelengths in the $H$ band.

\end{itemize}

Taking into consideration these calculations and the wavelengths
of the transitions of the target elements (all those listed above,
except Na, V, and S), we obtained the following constraints on
wavelength coverage:
The blue limit of the APOGEE range was set to capture the single
available K I line at 15160 \AA\  as well as the best Mn I lines at
15157-15263\AA\ , for reasons discussed
above.  Meanwhile, the
red limit was set by the goal to make sure to include at least one
of the three Al I lines at 16720-16770\AA\ .\footnote{In addition,
there is a weak Co line at ~16764\AA\ and
an atomic C I line at 16895\AA\ .  Although CO should be fine
as a carbon abundance indicator, the atomic carbon line
provides a check on C abundances derived from molecules. 
While
not put as a requirement, the C I line lies within
the wavelength range recorded by the spectrograph (see \S
\ref{sec:inst_performance} and Fig. \ref{fig:array_schematic}).} The
specified wavelength range also needed to account for potential heliocentric
velocity variations in Galactic stars, and a contingency of $\pm700$
km s$^{-1}$ was adopted.

Initially it was thought that the goals for the APOGEE science might
be met with a baseline, single grating instrument sampling
two disjoint $H$-band windows, 
but a desire to sample multiple lines for each element for redundancy, 
as well as the greater than linear gains of increased spectral
resolution drove to a three-detector design
with nearly continuous coverage 
from the K I to Al I lines.
Nevertheless, even with three detectors, the desired minimal spectral resolution leaves
the short wavelength end slightly undersampled.  To address this problem, 
it was decided 
that the three detector spectrograph would include a mechanism by which the focal plane arrays can be 
dithered precisely by half pixel steps.  By taking exposures in dithered pairs, the spectral resolution can be 
recovered as properly (Nyquist) sampled through interpolation of the paired exposures during post-processing.

A final issue that had no bearing on the instrument design
but did bear on the allocation of survey resources is that
of unidentified lines.  At the start of the survey, approximately
6\% of all lines deeper than 5\% of the continuum within the APOGEE
wavelength interval were still not identified with a
transition from a given excitation and ionization state of a known
chemical element.  This number went up to 20\% when all lines deeper than
1\% of the continuum were considered.
To improve this situation, 
the  APOGEE team initiated a collaboration with a team of
laboratory astrophysicists.  For details, we refer the reader to Appendix E.

\subsection{Kinematical Precision}
\label{sec:kinematicalprecision}

For many problems in large-scale Galactic dynamics --- e.g.,
measuring the disk rotation curve or the velocity dispersions of
stellar populations, sorting stars into populations, looking for 
kinematical substructures --- velocity precision at the level of 1
km s$^{-1}$ per star is not only suitable, but substantially better
than has been available in these kinds of investigations heretofore.  However,
the combination of high resolution and a very stable instrument platform
made possible achieving kinematical precision beyond these initial
survey specifications.  In fact, the APOGEE instrument and the
existing radial velocity software routinely deliver radial velocities
at a precision of $\sim0.07$ km s$^{-1}$ for $S/N$ $>$ 20, while
the survey provides external calibration sufficient to ensure
accuracies at the level of $\sim0.35$ km s$^{-1}$ 
(\citealt{Nidever15}; \S \ref{sec:perf_stelpar}),
which allows more subtle dynamical effects to be measured.  For example,
the detection of pattern speeds of --- or
kinematical substructure in the disk due to perturbations and resonances from --- 
spiral arms, the bar, or other (e.g., dark matter) substructure 
\citep[e.g.,][]{Dehnen98,Famaey05,Junqueira15}, 
the detection of stellar binary companions \citep[e.g., ][]{Terrien14}, 
the assessment of stellar membership in star clusters \citep[e.g., ][]{Terrien14,Carlberg15}
or extended stellar kinematic groups (i.e., ``moving groups" or ``superclusters'') 
in the disk \citep[e.g.,][]{Eggen58,Eggen98,Montes01,Malo13},
and the 
accurate measurement of stellar velocity dispersions in star clusters
or satellite galaxies \citep{Majewski13}
are all made possible with radial velocity measurements
of the RMS precision and external accuracies routinely achieved
by APOGEE for main survey program stars.  But it has been 
shown that even greater precision and accuracy may be obtained
from APOGEE spectra, which greatly benefits 
sensitivity
to low mass stellar companions \citep{Deshpande13} and the exploration
of the intricate dynamics of young star clusters \citep{Cottaar14,Foster15}
greatly benefits from even greater precision and accuracy.

\subsection{Sample Size and Field Coverage}
\label{sec:samplesize}

The largest detailed chemical abundance studies are typically focused
on stars in the solar neighborhood, and include samples of order
$10^3$ stars \citep{Venn04,Bensby03}.  A primary goal of APOGEE
is to obtain similar-sized samples of several thousand stars in
many dozens of Galactic zones across the Galaxy, and this led to a
basic technical requirement to obtain data on 100,000 stars distributed
across all major Galactic populations.
For example, a typical prediction from GCE models that we aim to
test are gradients in mean abundance for critical elements (Fe, C,
N, O, Al), with differences in the models seen at the level of a
few 0.01 dex at each radial or vertical point in the Milky Way.  Discriminating the
present models demands an accuracy in mean abundances of $\sim 0.01$
dex per Galactic zone, or more than 100 stars with 0.1 dex accurate
abundances in that zone assuming $\sqrt{N}$ statistics.  Similar
precisions are needed to determine, within each zone, the variation
of [X/Fe] with [Fe/H] or [O/H] (which are important discriminants
of the IMF and SFH), and therefore require 100 stars with 0.1 dex
accurate abundances in each metallicity bin.  Thus, deriving not
only mean abundances but accurate and useful multi-dimensional
abundance distribution functions (such as [$\alpha$/Fe] and [Fe/H])
in each zone requires orders of magnitude more stars per zone. Such
accounting (e.g., [dozens of Galactic zones][$\sim 20$ metallicity
bins][100 stars/bin]) leads to samples of $\sim 10^5$ stars.  Fortunately,
such numbers were estimated to be achievable {\it if} a three year
observing campaign were feasible within the duration of SDSS-III
(which had a well-defined end of mountain operations
in the summer of 2014; \S
\ref{sec:throughput}).

While a $\sim 10^5$ sample of stars with $R \approx 22,500$ spectra
is orders of magnitude larger than had been previously available
for Galactic archaeology, implicit to making this a true
milestone is that the stars be distributed systematically and widely
across the Galaxy, to include: (a) fields that cover a substantial part
of the Galactic bulge including the Galactic Center, (b) fields
that span a substantial fraction of the Galactic disk from the
Galactic Center to and beyond the longitude of the Galactic Anticenter,
(c) high latitude fields to map the halo, and (d) fields that probe
a variety of specific targets of interest, such as star clusters
(valuable as both science targets and calibration standards) and
known Galactic substructures (e.g., the bar, disk warp/flare, tidal
streams).  In addition, a small fraction of the survey time/fibers
would be available for potential ancillary science projects
\S \ref{sec:ancillary}), though
these would drive neither the science requirements nor instrument
design, nor significantly impact the net throughput of
the main survey.

\subsection{Survey Depth and MARVELS Co-Observing} 
\label{sec:depth}

For APOGEE's primary target --- evolved stars --- the survey
seeks to
reach across the Galactic disk in moderate extinction, to the
Galactic Center in fairly heavy extinction, and to the outer halo
in low extinction.  With some variation due to metallicity, the tip
of the red giant branch (TRGB) lies at $M_H \sim -5.5$.   AGB 
stars extend still brighter, whereas red clump
stars have 
$M_H \sim -1.5$.  To achieve the goal of readily
and abundantly sampling all Galactic populations {\it in situ}, it
was required that for ``typical'' survey fields and exposure times
that APOGEE routinely be able to reach to a depth of $H = 12.2$,
which translates to probing the TRGB to 35 kpc for no extinction
and to $>8.5$ kpc (i.e., to at least the distance of the Galactic
center) through $\sim 3$ magnitudes of $H$-band extinction ($A_V \sim
18$).  Thus $H=12.2$ was adopted as the ``baseline'' magnitude limit
for ``normal" APOGEE fields.

Special consideration was required for bulge fields, for which the
considerable zenith distance even at culmination translates to short
observing windows and more extreme differential refraction at APO.
To enable greater bulge spatial coverage, a baseline magnitude
limit of $H=11.1$ was implemented for these fields to reduce the
integration time by a factor of three.  Nevertheless, Galactic
center distances are reachable for TRGB stars when $A_H \lesssim
2$.

However, in other fields longer integrations were anticipated 
to enable APOGEE to probe red clump stars in low extinction fields
to $>$8.5 kpc or TRGB stars to $>$50 kpc, or TRGB stars to the
Galactic Center in fields with $A_H \sim 4$ ($A_V \sim 25$).  Such
longer fields were not only desired for APOGEE, but they were a
necessary part of the observing plan because, initially, APOGEE shared
bright time observations with the Multi-Object APO Radial Velocity
Exoplanet Large-area Survey (MARVELS) project \citep{Ge08,Zhao09}.
The baseline MARVELS program observed fields for $\ge24$
epochs at about 1 hour per visit; thus, at least some APOGEE fibers
on these same cartridges could sample fainter stars by accumulating
integrations of up to 24 hours.  
At first, MARVELS ``controlled" 
half of the bright time,\footnote{This control included some
choice in field location, but primarily the cadence of observations.}
 so that about half of the APOGEE time was
in these ``long fields".  Subsequent
termination of the MARVELS observing campaign in the second year
of APOGEE observations enabled some reconfiguration of our observing
plan (\S \ref{sec:fieldselectionevolution}).

\subsection{Throughput} 
\label{sec:throughput}

For throughput and target selection requirements, the APOGEE team
assumed that it would be able to observe during 95\% of the available
bright time (i.e., after accounting for a $\sim 50\%$ loss for weather
plus one SDSS-III-wide engineering night per month) for the final three
years of SDSS-III.  This high fraction would be achieved by carrying
out simultaneous MARVELS/APOGEE observing with the two instruments
sharing the focal plane.  The ability to carry out such simultaneous
observations was thus a technical requirement for achieving the
desired survey size and depth.

It was determined from the onset that APOGEE would feature a 300-fiber
spectrometer, which is the number of spectra that can be
maximally packed along the spatial direction on a 2048
pixel-wide detector (allowing $\sim$7 pixels per spectrum, assumed
to be sufficient to span both the PSF of each spectrum and leave
dark gaps between).  Initially it was assumed that 50 fibers
would be needed for simultaneous observations of sky and
telluric calibration stars.\footnote{In the end, this number was
increased to 35 fibers for sky plus 35 for telluric absorption
stars (\S \ref{sec:calfibers}).}

As discussed above, the requirement of detailed and precise chemical
composition determinations drives requirements of $S/N\approx
100$ per pixel at resolution $R \approx 22,500$.  The requirement
to observe $\sim 10^5$ stars then implied that, after adopting
conservative estimates for all variables, the instrument would have
to achieve this $S/N$ at the typical observation depth $H=12.2$ in
3 hours of total integration time:   $N_{stars} \approx $ \\
\begin{itemize} 
\item (3 year observing campaign) $\times$ 
\item (11 months per year\footnote{One month is lost to summer shutdown during
monsoon season.}) $\times$
\item (30 nights per month) $\times$
\item (11 hours per night) $\times$ 
\item (40\% bright time) $\times$
\item (95\% of bright time to APOGEE) $\times$ 
\item(50\% clear weather)
$\times$ \item (250 targets per field) $/$ 
\item (3 + 1.5 hours per
field\footnote{This is assuming 1.5 hours of overhead per 3 hours
of exposure (30 minutes for each of three one hour visits; see \S
\ref{sec:binaries}) --- a generous overhead but one that includes
some allowance for longer exposures in sub-optimal conditions.})
$= 1.15 \times 10^5.$ 
\end{itemize} 
More detailed analyses that,
for example, included lost nights for engineering time, various
weather models, and more sophisticated observing plans all yielded
estimates within 15-20\% of this conservative estimate.

\subsection{Binary Stars, Field Visit Duration and Field Visit Cadence}
\label{sec:binaries}

Because the majority of APOGEE targets are RGB stars, a substantial fraction 
are expected to be single-lined binaries.  The amplitudes of radial velocity variations in binary
stars can reach as much as 10-20 km s$^{-1}$; thus it is useful 
for such binary 
systems to be identified and flagged so that they can, when necessary, be removed from APOGEE
kinematical samples to minimize 
deterioration of the precision of derived bulk dynamical
quantities for stellar populations --- e.g., the inflation of measured velocity dispersions.

Identification of the radial velocity variability associated with
single line binaries can be achieved by splitting the total
integrations for each star into visits optimized in cadence to
identify the binaries with problematical barycentric velocities.
Given the expected instrument throughput, it was understood early
on that to reach distances of interest for studying a large fraction
of the Galaxy (and in particular, crossing the full extent of even
just the near side of the disk for the nominal Galactic plane
pointing) detector integrations of multiple-hour net length would be
needed.  However, because differential refraction limits the duration
of hour angle viability for any drilled fiber
plugplate\footnote{For definitions of this and other terms
specific to the fiber system of the 2.5-m SDSS telescope, see
\S \ref{sec:fibertrain}.}, it is necessary to break long exposures
into multiple observing stints --- either using plugplates drilled
for different hour angles (potentially observed on the same night)
or using the same plate observed on multiple nights.  It was most
efficient and natural to adopt the latter solution and exploit the
multi-visit strategy for the added purpose of binary star identification.

For effective identification of binaries, more velocity samples
over a longer time baseline is always preferable.  This desire must
be balanced against that of survey efficiency, which pushes in the
direction of breaking long exposures into the fewest possible number
of visits, to limit the fraction of time surrendered to overheads
of plugplate cartridge (\S \ref{sec:fibertrain}) changing 
and field acquisition.  While
mountain observing staff showed that this overhead can be as low
as 12-15 minutes per plugplate cartridge change, 15-20 minutes is a more
realistic ``typical" situation.  Under these circumstances, field
visits of less than 30-45 minutes accrue substantial overhead.
Moreover, frequent visits of such short duration place substantial
physical demands on the observers.  In any case, there were only
eight available ``bright time'' Sloan plugplate cartridges available
on which to put APOGEE fibers, so that no more than eight APOGEE
plugplates were observable on a given night.  Thus, given the
tradeoffs between observing efficiency and differential refraction
limits as well as the eight cartridge limit, it was decided that
the baseline APOGEE visit would include about an hour of integration
plus overhead (see \S \ref{sec:observing}) and that the ``nominal"
survey field integration of $\sim$ 3 hour length (see \S
\ref{sec:throughput}) would be divided into no less than three
visits.

With this basic multi-visit plan in place, one last requirement imposed is the adopted cadence for the visits.
To understand the potential effects of binary stars on measured APOGEE dynamical quantities, and to
assess the best way to distribute three visits over time to maximize the ability to detect ``problem'' binaries, 
a series of Monte Carlo simulations of stellar populations having 
nominal binary fractions and mass, 
period and orbital eccentricity properties was undertaken.  
The details of these models, wherein the parent sample of 
typical APOGEE targets had their radial velocities sampled 
over varying time intervals and net spans,
are given in Appendix C.

These simulations showed that the majority of binary systems ($\sim
74$~\%) are not expected to adversely affect the kinematical
measurements, where ``adversely affected'' had been defined as a
measured velocity of the primary star that is $>2$~km s$^{-1}$
different from the true systemic motion of the binary system.  Given
the above visit strategy, the most effective way of identifying
the remaining 26\% of binaries is by calculating the radial
velocity difference between every combination of paired measurements
and flagging stars showing a maximum velocity difference above a
certain threshold (we adopted for our modeling 4~km s$^{-1}$).
These simulations indicated that, for a set of at least three radial
velocity measurements of 0.5 km s$^{-1}$ precision, a temporal
baseline spanning at least one month was sufficient to make evident
at least a third 
of the remaining (26\%) binaries most likely to have a significant
impact on the APOGEE survey velocity distributions.  While longer
baselines improve detectability, that improvement is only marginally
better for baselines lengthened to a full season of typical object
visibility (Fig. \ref{fig:missed_binaries});
thus, a requirement of at least a 25 day span for the visits to a
single plugplate was adopted as a rule, with a minimum span between
epochs of 3 days.

In their CORAVEL study \citet{Famaey05} find 13.7\% of their local K giant sample to be in binaries and 
that with their strategy (two radial velocity measurements per star spanning 2-3 years)
and 0.3 km s$^{-1}$ velocity accuracy
``binaries are detected with an efficiency better than 50 percent \citep{Udry97}''.  
Famaey et al. actually find an even lower binary fraction of only
5.7\% for their M giant sample.  
These numbers suggest that one
might expect 27.4\% and 11.4\% binary fractions among K and M giants,
respectively.  If two-thirds of 26\% of {\it these} (i.e., 9\%)
slip through the APOGEE ability to detect them, then perhaps only
a few percent of APOGEE targets would remain kinematically
``problematical'', with measured velocities deviant from their
systemic motion by more than 2~km s$^{-1}$.  Even this fraction is
likely an upper limit because: (a) one month is the {\it minimum}
temporal baseline, whereas, at survey end, the median baseline for
all multi-visit fields
is almost two months
(see Fig.~\ref{fig:visits}b); (b) a
significant fraction of the primary APOGEE sample --- $\sim$32,600
stars, or 30\%, had more than three visits, by design or circumstance
(see Fig.\ref{fig:visits}a); and (c) the per-visit velocity
precision is substantially better than 0.5 km s$^{-1}$ (at
0.07 km s$^{-1}$; see \S 10.3 of \citealt{Nidever15}).  A more
complete assessment of the detected APOGEE binary fraction is
currently underway (Troup et al., in preparation).

\section{Survey Instrumentation}
\label{sec:instrumentoverview}

The APOGEE survey is made possible through the construction of 
the world's first high-resolution ($R$$\sim$22,500), heavily multiplexed
(300 fiber), infrared spectrograph (\citealt{Wilson10a},2015).
This cryogenic instrument (Fig.\ \ref{fig:instrument_schematic}), covering 
wavelengths from 
$1.51\mu$m $\le \lambda \le 1.70\mu$m, was
conceived, designed and fabricated in the University of Virginia (UVa)
astronomical instrumentation laboratory, but with considerable collaboration
on the design and fabrication of specific subcomponents from a number of 
SDSS-III institutions and private vendors.  
A full description of the instrument can be found in \citet{Wilson15}.
We provide here a broad overview of the instrument sufficient
to understand the format and character of the data it delivers.

\subsection{Fiber Train and Plugplate System}
\label{sec:fibertrain}

The APOGEE instrument leverages the wide-field (3 degree diameter
field-of-view) capability of the Sloan 2.5-m telescope \citep{Gunn06}
at Apache Point Observatory, New Mexico (USA), and the highly
efficient and proven survey infrastructure that has led to
the very successful SDSS-I and SDSS-II suites of experiments using
optical spectrographs \citep{Smee13}.
For the optical spectrographs, which are mounted to the telescope back end, 
short-length (1.8 m) fiber optic
bundles run directly from the telescope focal plane to the pseudo-slits
of the spectrographs.  In contrast, because of the sheer-size of
the APOGEE spectrograph, it is housed in a separate building adjacent
to the 2.5-m Sloan Telescope and fed light via a single, approximately
40-m ``long fiber'' run from the telescope.  This set of 300 ``long
fibers'' (called the ``fiber link'') is permanently attached to the
APOGEE instrument with one end of each fiber hermetically sealed
inside the cold, evacuated cryostat and terminating at the spectrograph
``pseudo-slit".  The warm end of the fiber link terminates at the
base of the telescope.

At the telescope
APOGEE employs the same plugplate system designed for use in the SDSS-I and SDSS-II surveys
\citep{Owen98,Siegmund98}, and, indeed, makes use of eight plugplate 
cartridges from those previous surveys that 
were converted to ``bright time" operations through the incorporation of 
distributed and mingled ``anchor 
blocks" of fibers linked to the MARVELS and APOGEE instruments.  
As with other Sloan spectrographic surveys, aluminum plugplates with 
precision-drilled holes matching the positions of APOGEE targets in a specific 
sky field are interchanged and manually plugged with the cartridge 
``short fibers" each day in preparation for the ensuing night time observing, 
The APOGEE fibers are step index, multi-mode, low-OH (i.e., ``dry") fibers
with a 120~$\mu$m diameter silica core 
that subtend 2 arcseconds of sky at the telescope focus.  
The sets of ``short fibers'' installed in the
fiber plugplate cartridges terminate in pluggable, stainless steel ferrules that impose
a fiber-to-fiber proximity limit (the so-called ``fiber collision limit'')
of 71 arc seconds. Each APOGEE anchor block of 6 fibers covers a
linear extent across the plugplate cartridge equivalent to a roughly circular
sky patrol area of about a 1.0 degree (22 cm) radius.  However, the
distribution of these anchor blocks is non-uniform, forming a ring around the 
central part of the field.  This arrangement allows either a uniform plugging across 
the entire plate or a higher central concentration
with all 300 fibers in a relatively narrow FOV.
The latter application is for
those plugplates that are drilled for high airmass (low declination)
fields, where all targets are selected within a restricted FOV (potentially
as small as a 1 deg diameter circle; see \S \ref{sec:scheduling}).

One primary difference in the plugging process for APOGEE plates
compared to previous SDSS projects is that APOGEE fiber plugging
imposes one level of fiber management by separating fibers into
three $H$-magnitude-defined groups that are plugged into holes
corresponding to the faintest, mid-brightness and brightest thirds
of targets on each plate.  This fiber management is accomplished
by color-coding the target holes on each plate either red, green
or blue by their brightness rank, and filling these holes with
fibers having matching colored sheathing. No other requirement is
imposed on which science fiber is plugged into which hole.  Each
anchor block has two fibers of each color, so that the ``bright'',
``medium'' and ``faint" fibers are evenly distributed across the
field.  At the spectrograph end, these different fibers are sorted
along the pseudo-slit into a repeating pattern of
faint-medium-bright-bright-medium-faint to ensure that bright spectra
are never placed next to faint spectra in the spectrograph focal
plane (this pattern of alternating spectrum brightness may be 
seen in Fig. \ref{fig:fibermanagement} below). 
This scheme minimizes the degree of contamination
of any given spectrum by 
overlapping wings of the PSF from the
adjacent spectrum of a brighter star.

During survey operations, the short-length fibers in each of the
eight plugplate cartridges are mated to the long-length fibers
approximately hourly (after each cartridge/plugplate change) via a
custom-built ``gang-connector" that simultaneously mates each of the 300
short fibers with its corresponding long fiber to within a few $\mu$m accuracy.  
Because of the frequency of this mating operation, the need for efficiency
of operations, and the sometimes dusty conditions at the site, no
index-matching gel is used in this fiber coupling operation; as a
result, there are some light losses at the connector, but they are
small enough (5\%) that the ease of operation without use of optical gels
more than makes up for the lost light.

An additional modification is required for the APOGEE fiber mapping
system.  In the case of the optical Sloan spectrographs fiber
mapping is undertaken after each plugplate is plugged
by running a laser directly up the pseudo-slit
(which is integrated as part of the cartridge) and recording which
plugged fibers light up on the plugged plate in succession.  In the
case of APOGEE, however, the true instrument pseudoslit is inaccesible,
as it lies within the cryostat.  Therefore, a warm copy of that
pseudoslit is mated to the gang connector for this operation.

\subsection{Spectrograph}
\label{sec:spectrograph}

\subsubsection{Technological Innovations}
\label{sec:innovations}

Although the APOGEE spectrograph design is simple in
concept, the sheer size of the optics and the need to feed 300 fibers
to a pseudo-slit inside a cryogenic instrument posed considerable
technological challenges.  In particular, the success of the instrument depended
on the development of five distinct technologies:
\begin{enumerate}
\item Implementation of the custom-made ``gang connector'', described above, which makes possible
the simultaneous high-efficiency coupling of 300 infrared transmissive fibers and
enables rapid swapping of telescope focal-plane plugplates.
\item Innovating hermetic fiber vacuum penetrations of the
cryostat stainless steel wall that simultaneously limit stress-induced 
fiber focal ratio degradation (FRD).  This was accomplished after
extensive testing of a wide range of materials and epoxies (\citealt{Brunner10}) for the seal.
\item Collaboration with Kaiser Optical Systems, Inc. (KOSI; Ann Arbor, MI) in the
design and fabrication of a volume phase holographic (VPH) grating 
larger than any previously deployed in an astronomical spectrograph
via innovation of a technique for precisely laying down multiple (three)
holographic exposures onto one glass substrate (\citealt{Arns10a}).
\item The design --- in collaboration with
New England Optical Systems, Inc. (NEOS; Marlborough, MA) ---
of a 6-element infrared transmitting camera that includes 
several unprecedentedly large diameter (40 cm) lenses of monocrystalline silicon.
\item The creation --- in collaboration with the James Webb Space Telescope (JWST)
Near Infrared Camera (NIRCam) team, Princeton University, and Johns Hopkins University ---
of a precision multi-array mount and translation stage for three Teledyne
HAWAII-2RG (2048$\times$2048) detectors. With this stage, the arrays 
can be ``dithered" together 
in the dispersion direction to $<$1~$\mu$m accuracy to mitigate the modest undersampling of the 
spectra as delivered to the instrument focal plane.
\end{enumerate}

\subsubsection{Basic Instrument Layout}
\label{sec:layout}

\begin{figure}[h]
\includegraphics[width=0.50\textwidth]{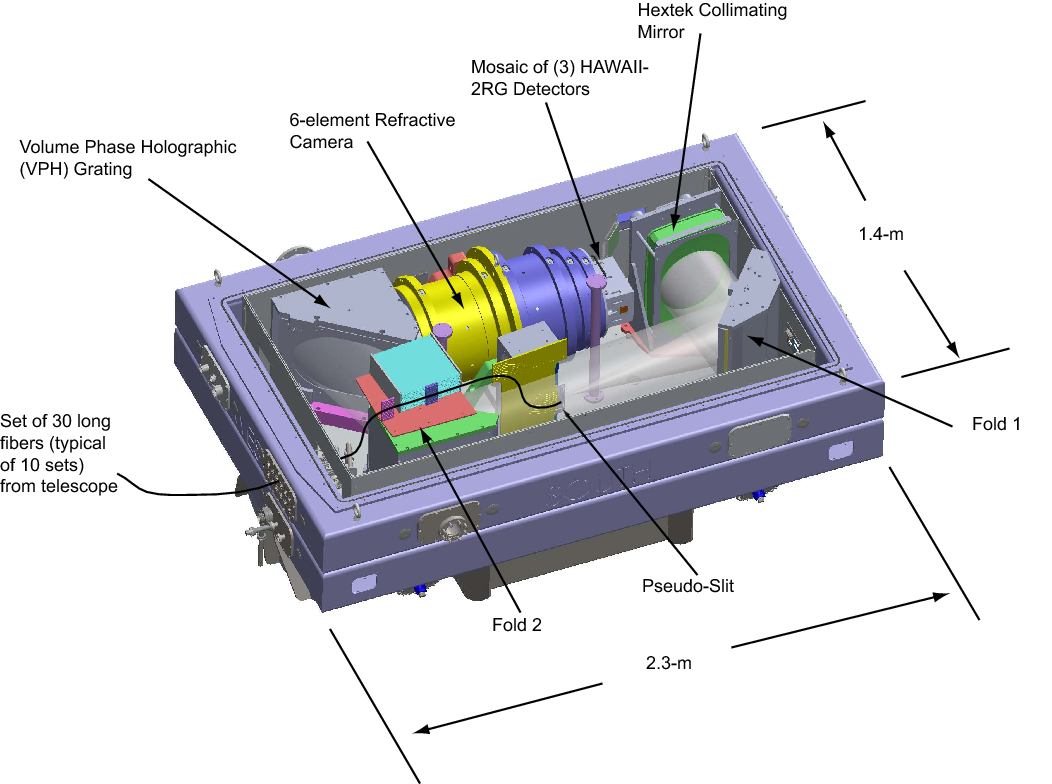}
\caption{\footnotesize Layout of the APOGEE spectrograph optical bench within the cryostat.  The
fiber train coming from the telescope enters the cryostat on the left.}
\label{fig:instrument_schematic}
\end{figure}

The basic optical design of the spectrograph leverages the successful optical design of the multifiber
optical SDSS spectrographs \citep{Smee13}, but modified as needed for high spectral resolution at infrared wavelengths.  The basic optical layout of the APOGEE instrument is illustrated in Figure \ref{fig:instrument_schematic}, and was built in a highly modularized fashion with distinct subcomponents:

{\it Cryostat:} Past the gang connector the long fibers route to
the spectrograph and enter the cryostat via vacuum feed-throughs
at the cryostat vacuum wall (keeping the fibers intact avoids 
the introduction of another optical junction and
thus minimizes throughput losses and focal ratio degradation).
The cryostat is a specially-designed, stainless steel, liquid
nitrogen-cooled vessel built by PulseRay Machining \& Design (Beaver
Dams, NY).  Together, the 1.4 $\times$ 2.3 $\times$ 1.3 m cryostat,
optical bench and instrument subcomponents weigh approximately 1.8
metric tons (2 U.S. tons).  The entire cryostat sits on four pneumatic
isolation stands to minimize vibration.  Within the steel container,
an aluminum cold radiation shield surrounds the entire cold volume;
this entire shield is surrounded by blankets consisting of 10 layers
of double-sided aluminized mylar interspersed with layers of tulle.

{\it Optical Bench:} The spectrograph optics are mounted to an
optical bench that is a single, 3-inch thick cold plate with extensive
underside lightweighting and suspended from the vacuum shell.  A
97 liter LN$_2$ tank is suspended from the bottom of the cold plate.
In the vicinity of the camera the cold plate maintains a temperature
of about 78K, and the entire cryostat experiences no more than a
35W heat load on the cold volume and has a 5.5 day hold time.  An
LN$_2$ liquid level sensor monitors the fill level, but, in any
case, an automatic filling system tops off the dewar every morning
after observing is over.  Although overall the bench-mounted,
fiber-fed spectrograph confers a distinct advantage in maintaining
a vibration-free, temperature-stable system with a constant gravity
vector that ultimately makes the APOGEE instrument deliver spectra
that are extremely stable and repeatable, the small nightly variation
in LN$_2$ liquid level was later found to induce slight variations in
mechanical flexure on the cold plate that can be observed as small,
$\sim0.1$ pixel shifts in the spatial position (and even smaller,
$\sim0.01$
pixel shifts in the spectral position) of the spectra on the
detector over the course of the night; fortunately these slowly
varying changes can be mapped and accounted for by regularly observed
flatfield exposures.

{\it Pseudoslit and Collimator:} The final 2~m of the long-length
fiber link trains are contained within the cold volume and terminate on
a curved pseudo-slit.  Fiber-to-fiber spacing at the pseudo-slit
is physically 350~$\mu$m between centers to yield 6.6 pixel spacings
between spectra on the detectors.  An ``uncorrected Schmidt'' camera,
used in reverse, collimates the light of each of the fibers.  Thus,
in keeping with the Schmidt design, the fiber tips are carefully
positioned to lie on, or close to, a curved surface with radius of
curvature approximately 1/2 that of the collimator and to emit light
in a direction orthogonal to that surface, so that they axially
point back as close as possible to the center of curvature of the
pseudo slit.  Moreover, the pseudo-slit and spherical collimator
mirror have a common center of curvature near the system pupil,
which is at the approximate position of the spectrograph grating.
In addition, the fiber ends are also aligned on a curved {\it
lateral} surface to ensure that at the detector the fiber ensemble
gives straight slit images; this lateral curve enables each fiber
to deliver the same rest wavelength range on the detectors.  As the
only active means to effect small changes in instrument focus, the
collimator is controlled by 3-axis (tip-tilt-piston) movement.  This
capability is also useful for implementing dithering in the spatial
dimension, an operational mode that is periodically activated 
for the creation of spatially-smoothed
instrument flatfields.  The resulting optical design is on-axis so
that the pseudo-slit is an obscuration in the collimated beam.

{\it Cold Shutter and Flat Field Illumination:} A ``swinging gate''
cold shutter with a light trap (not shown in Fig.
\ref{fig:instrument_schematic}) covers the pseudo slit to prevent
excessive light leaking into the cold volume when the spectrograph
is not taking observations.  This minimizes the accumulation of
unwanted charge that could contribute to detector ``persistence"
(see \S \ref{sec:inst_performance}).  This mechanism also contains a set of infrared
light-emitting diodes that can provide a diffuse illumination onto
the detectors useful for creating flatfield exposures.

{\it Fold Mirrors:}  
Two fold mirrors are used for efficient packaging of the optical train within the cryostat.
The second mirror flat is also a dichroic, which passes light longward of the APOGEE spectral range
into a trap behind the mirror; this assists in the mitigation of stray, thermal light (see below).

{\it VPH Grating:} The disperser is a three-segment mosaic VPH
grating, the first ever fabricated by KOSI.  Because the required
grating size exceeds that which can be recorded in a single VPH
exposure, the APOGEE VPH is made by recording, in close temporal
succession, three adjacent segments in gelatin on a common fused-silica
substrate. While prototype mosaic VPH's have been fabricated in the
past, none have been deployed in an astronomical instrument,
cryogenically cooled or not. The VPH has 1009.3 grooves/mm and
operates in first order with a 54$^{\circ}$ angle of incidence.
The grating is found to deliver peak efficiency of 90\% near the
center of the APOGEE spectral range, and 40\% at the edges.

{\it Camera:}
The wavelength-dispersed 
beams are focused by a six-element refractive camera designed and fabricated by NEOS.  The APOGEE camera is very large for an 
astronomical spectrograph: the largest element is 394 mm in diameter and the smallest element is 
237 mm in diameter.  Given such large camera elements, the variety of lens materials that can be 
considered is severely limited by economic and fabrication
limitations.
Fortunately the narrow wavelength range of APOGEE means that only two materials --- monocrystalline
silicon (for four of the lenses) and fused silica (for the other two) --- are necessary, and those two 
materials are also, coincidentally, very robust to thermal shock.  Overall the finished opto-mechanical 
camera assembly alone weighs 250 lbs.  When combined with (minimal) absorption through the fused 
silica and the performance of the anti-reflection coatings, the  
throughput for all six lenses is 93\% across the wavelength coverage.

{\it Detector Array:}
The spectra are recorded on three Teledyne Imaging Sensors H2RG,
 2.5~$\mu$m cut-off, 
$2048 \times 2048$ pixel detector arrays.  These are mounted in a detector mosaic opto-mechanical 
package similar to that used for the NIRCam instrument for the JWST, 
with the arrays tilted to approximate the field curvature of the optical system within a 
tolerance of 15~$\mu$m through precise shimming of piston, tip and tilt. 
 These detectors are operated in sample-up-the-ramp mode, with readouts
every 10.7 seconds; thus, ``images" of the spectra are in datacubes for each of the three arrays.
As mentioned above, the arrays lie on a movable stage which is used for 
``dithering" translation of the entire assembly in the dispersion direction; in practice, images
are taken in pairs with half-pixel shifts, which, in the data processing, can be used to recover the
full sampling of the spectral line-spread-profile (\S \ref{sec:1dred}).

{\it Baffling and Other Stray Light Mitigation:}
Mitigation of stray light is an important consideration for achieving the required 
$S/N$ because the APOGEE wavelength range is small compared to the 
wavelength sensitivity range of the 
detectors; of particular concern is thermal light, because of the use of
2.5~$\mu$m cut off arrays.
Zero'th order light transmitted through the grating is intercepted with a blackened panel behind the VPH.
Of more concern are first order wavelengths outside of the nominal
APOGEE wavelength range.
Light shortward of 1.0~$\mu$m is absorbed by the four silicon elements in the camera.
Thermal light is mitigated in several ways: (1) The fibers are cooled over 2~m of travel within the cryostat
before reaching the pseudo-slit.  (2) The ``long-pass'' dichroic on the front face of the second fold mirror 
and a broadband anti-reflection coating on the backside creates a light-dump that intercepts some
95\% of the residual thermal light ($\lambda > 2$~$\mu$m) before if gets to the grating.  (3) The silicon lenses in the camera
have antireflection coatings that, together, permit transmission of only 9\% (3\%) or the thermal
light to the detectors at 2.3~$\mu$m (2.5~$\mu$m).

{\it Calibration Box:}  Unlike the optical SDSS spectrographs, which
take wavelength calibration images by illuminating the closed
telescope covering petals, APOGEE employs a separate, off-telescope
calibration module that enables access to calibration lamps at any
time.  When not attached to a bright time plugplate cartridge, the
APOGEE gang connector can be connected to separate fiber runs that
terminate at an integrating sphere that can illuminate the fibers
with nominal $f/5$ light (to mimic the telescope) with either a
ThArNe hollow-cathode lamp, a UNe hollow-cathode lamp, or a tungsten
halogen light source.  During commissioning and testing, the sphere
also at times was illuminated with a precision-controlled blackbody
source.  Two possible fiber links to the integrating sphere are
available: (a) a ``DensePak'' bundle with a full set of 300 fibers,
or (b) a ``SparsePak'' bundle that sends light to every sixth fiber
in the spectrograph focal plane.  The latter is particularly useful
for evaluating the wings of the point spread function and the effects
of scattered light.

{\it Instrument Control:} At the observer level, operation
of the APOGEE instrument takes place through scripted sequences in
the Sloan Telescope User Interface (STUI; \S \ref{sec:observing}).
For manipulation of the spectrograph detectors, the STUI interfaces
with a Digital Signal Processor (DSP) based
controller that provides both clocking and bias/power supply voltages
to the three arrays.  All three share a common clock and most of
their bias lines, with just a few power supply voltages unique to
the individual arrays.  This ensures common timing for the three
arrays as they are read out.  The read out scheme utilizes
``sampling up the ramp'' (SUTR), where the arrays are clocked and
read out continuously and non-destructively with a period of 10.6
seconds.  The data are formatted as a single output image containing
the data for all three arrays, and including the three H2RG reference
outputs.  Because the array clocking is DSP based, the interval
between reads is very stable, which allows for accurate curve of
growth analysis of the developing signal in each pixel (\S
\ref{sec:datahandle}).  This is facilitated by the rearrangement
of the flat, three-array data frames into time series data 
cubes for each array during post processing of the data for each observation.

\subsection{Instrument Development and Operations Timeline}
\label{sec:timeline}

A white paper describing the potential of high throughput, multifiber,
near-IR spectroscopy on the SDSS 2.5-m telescope was presented to
the Astrophysical Research Corporation (ARC) Futures Committee by
\citep{Skrutskie15} in August 2005.
The APOGEE project, refining the concept to a focus
on high resolution spectroscopy of Milky Way stars, was proposed
as an SDSS-III\footnote{At the time, the SDSS-III project was called
the ``After Sloan-II" project, but, for clarity, we use ``SDSS-III''
throughout this paper.} project in August 2006 and was officially
approved by the ARC Board as one of the four SDSS-III projects in
November 2006.  The APOGEE instrument Conceptual Design
Review (CoDR) was held in April 2008, with the goal of having the
spectrograph collecting data on the Sloan Telescope
by 2011Q2.  The instrument Preliminary Design Review (PDR) took place
in May 2009, with approval to start fabrication given at a Critical
Design Review (CDR) held in August 2009.  Despite the technical
challenges enumerated in \S \ref{sec:innovations}, the primary
APOGEE hardware construction phase 
spanned only 16 months from CDR to obtaining spectra of bright stars
in February 2011.\footnote{This starlight was delivered to the
APOGEE instrument while still in the UVa instrument lab by way of
a 10-inch Newtonian reflector with the diagonal flat replaced by a
``hot mirror" dichroic that directed optical light to the nominal
Newtonian port for eyepiece acquisition and guiding, but passed the
$H$-band light to a sparsely packed grid of fibers linked to the
APOGEE instrument.}
The instrument was delivered to APO in April 2011 and on-site first
light occurred on the evening following deployment of the fiber train,
on 6 May 2011 --- consistent with the original instrument
development schedule.

Testing/commissioning observations of the instrument commenced
immediately.  It was soon realized that while 
instrument performance met, or exceeded, the original requirements,
it was also suffering from significant
astigmatism that made it impossible to achieve simultaneous focus
in the spatial and spectral directions.  In addition, the placement
of the arrays, particularly the array recording the reddest
wavelengths, was non-optimal.  While the source of the astigmatism
has yet to be identified confidently, it was possible to mitigate
its effect by introducing a slight cylindrical bend on the first
fold mirror using a specially made fixture that induces a calculated
axial force along the center of the mirror backside.   On the other
hand, the realignment of the focal plane arrays, which required
shipping the entire detector assembly package to the University of
Arizona, was not effected until the APO ``summer shutdown" in July
2011.  Thus, from May-July observations with the APOGEE instrument
were taken without parfocal arrays, and this resulted in data being
taken with a reduced resolution of
$R \sim 15,000$ across the
``red'' detector array.
The observations collected during this phase of operations are
commonly referred to as ``pre-shutdown'' or ``commissioning'' data;
although these data are being released publicly, application of APOGEE
data reduction and analysis pipelines to those data does not lead to data
products within the science quality requirements,
and any results from them are not of survey quality.  Moreover, all
but a few plugplates observed with this instrument configuration
were eventually repeated (see Fig. \ref{fig:field_plan}).  Nevertheless,
the data are of some interest, for example, in providing additional
epochs for the study of time series phenomena.

Official APOGEE survey data collection commenced after summer
shutdown, August 2011, with all three detectors in
focus.
The instrument parameters given in Table \ref{tab:Inst_Char} pertain
to this configuration of the APOGEE spectrograph, which was maintained
throughout the remainder of SDSS-III operations (which concluded
July 2014).  Throughout this period, APOGEE observations were
smoothly carried out by the SDSS observers with minimal daily
oversight by the APOGEE team and no loss of time due to instrument
problems.

\subsection{Basic Instrument Performance and Properties}
\label{sec:inst_performance}

The overall instrument performance is obtained from a variety of test data taken on-site.
Table \ref{tab:Inst_Char} summarizes the instrument characteristics.  Much greater
detail on the instrument performance can be found in \citet{Wilson15}, \citet{Nidever15}
and \citet{Holtzman15}.

\begin{figure}[h]
\vspace{-82mm}
\includegraphics[width=1.00\textwidth]{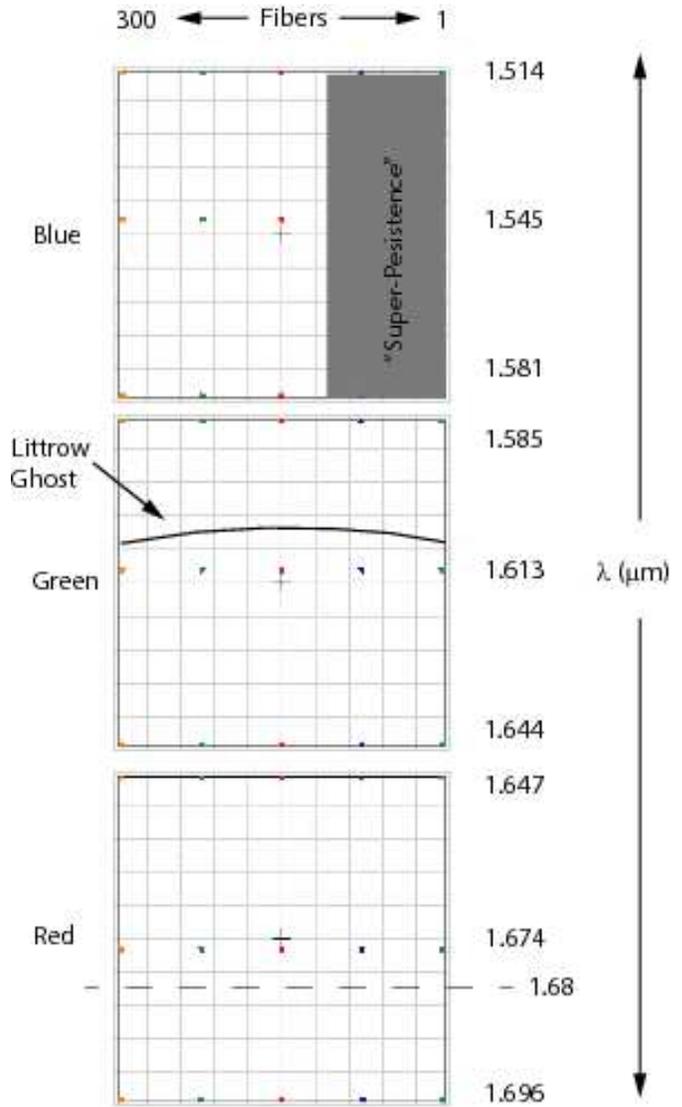}
\caption{\footnotesize Schematic figure showing the arrangement of fibers and wavelengths  
across the three APOGEE detectors.  The wavelengths indicate the edges of 
the arrays as well as fiducial wavelength positions (indicated by grey dots) 
corresponding to the ``mid-chip" properties 
given in Table \ref{tab:Inst_Char}.  The location of the Littrow ghost (curved line) and the super-persistent 
region (grey area) are also indicated.   The dashed line at 1.68~$\mu$m shows the red limit of 
the wavelength coverage for which the technical performance of the instrument 
was specified by the science requirements, but the instrument performance is
still good redward of this.
}
\label{fig:array_schematic}
\end{figure}

{\it Wavelength Ranges:}
While the APOGEE spectrograph was designed to meet technical specifications that included
the specific wavelength limits set by the 15160\AA\  potassium line and the 16720-16770\AA\ 
aluminum lines (\S \ref{sec:resolution}), the spectral range recorded by the detectors extends 
almost to 1.7~$\mu$m (Fig. \ref{fig:array_schematic}), 
although these ``extra'' wavelengths were not designed to meet the
resolution, throughput and other technical specifications and did not drive design considerations.
Of course, because of the physical limitations of butting detectors together, the spectral coverage
is interrupted by inter-chip gaps.
The exact wavelength region falling onto the array ensemble can be controlled by micro-positioning of the 
dithering stage, but the default position of the instrument delivers the wavelength regions on each
chip as shown in Figure \ref{fig:array_schematic} and given in Table \ref{tab:Inst_Char}.

{\it PSF, LSF and Resolution:}
Image quality can be judged from the delivered line-spread function
(LSF) and point-spread function (PSF) across the arrays.
The PSF has a FWHM of typically 2.16, 2.14, and 2.24 pixels at the
fiducial centers (1.54, 1.61 and 1.66~$\mu$m)\footnote{For the ``red''
array, the fiducial wavelength lies at the midpoint of the blue
edge of that array and the 16770\AA\  red limit of the technical
specification (Fig. \ref{fig:array_schematic}).} of each of the
three array wavelength spans (Fig. \ref{fig:array_schematic}). The
wings of the fiber PSFs 
reach far enough from the peak that there is 
a small amount of overlap
between the PSFs of adjacent fibers on the detector focal plane.  
When the magnitude difference between stars on adjacent fibers is large, 
contamination of the spectrum of the fainter one by the wings of the brighter
spectrum can become important.
The amount of contamination varies across the three arrays, 
but analysis of commissioning data showed that between
$\sim$0.1 and 0.2\% of the total power of the PSF is located within
3 pixels of the central pixel of the adjacent PSF.  It is for this 
reason that the fiber management scheme described in 
\S \ref{sec:fibertrain} was implemented.  The LSF also varies across 
the arrays, both as a function of wavelength and fiber,
as discussed by \citet[][see their Figures 14-16]{Nidever15}.  In
particular, it is slightly undersampled in the blue part of the
spectrum, which is
what motivated our use of the detector array dithering mechanism
during observations (\S\ref{sec:resolution} and \S\ref{sec:innovations}).
The resulting resolution in the properly
sampled spectra varies by $\sim$ 25\%, peak to peak, being higher
at shorter wavelengths.  Typical values at 1.55, 1.61 and 1.66~$\mu$m are
$R=23,500$, 23,400 and 22,600 \citep[for details, see][]{Nidever15}.

{\it Instrument Throughput:} Observations of 
stars with well known 2MASS magnitudes make possible empirical estimates of the 
throughput of the APOGEE instrumental apparatus.  
The end-to-end (i.e., from primary mirror to detector) measured
throughput has a peak of 20 $\pm$ 2\% at $\lambda\sim$~1.61 $\mu$m.
This number is somewhat higher than expected from predictions based
on the product of the component-by-component (measured or
manufacturer-supplied) wavelength-dependent throughputs
(see Table~\ref{tab:Inst_Char}).
These throughput measurements have obvious implications for the
$S/N$ achieved under survey conditions; those are discussed 
in \S\ref{sec:perf_specpar}.

{\it Array Persistence:}
As with most Teledyne infrared detector arrays, those installed in the 
APOGEE instrument have a small degree of image persistence, 
which results in the carryover of latent charge from exposure to exposure.
This typically does not affect most APOGEE data.  However, roughly one
third (in the spatial direction; see Fig. \ref{fig:array_schematic}) of the
detector used for the bluest wavelengths is affected by excessive
and long-lasting 
``superpersistence", which appears to behave like normal persistence, but 
with significantly greater accumulated charge and
a very long time constant (see \S5 of \citealt{Nidever15}).  Thus, intensely exposed
pixels on one image can yield inordinately ``hot" pixels in subsequent
exposures.  A small portion of the ``green'' array (a thin ``frame''
around the edges) is also affected.
The seriousness of this phenomenon has had a strong influence on
our observing procedures --- e.g., the timing and strength
of calibration exposures and the imposition of a bright limit to 
targeted sources --- with the goal of limiting unnecessary
overexposure whenever possible.  This problem also motivated the installation of the 
cold shutter (\S \ref{sec:layout})
to prevent stray light to enter the instrument when not in use.
While there is hope that the effect of the superpersistence on the data 
may be correctable in software, it is a complex hysteresis problem that we 
currently have not fully resolved and no mitigation is currently implemented
up to, and including, Data Release 12.

{\it Ghosts:} Despite mitigation efforts, there remain two in-band sources of stray
light in the form of ghosts: (1) A Littrow ghost of each fiber
(created by light reflected off the detector surface, 
recollimated by the camera, recombined by, and reflected from, the grating, and reimaged by the 
camera onto the detector; \citealt{Burgh07}) forms on the detector at 0.4\%
the intensity of all of the recorded spectral light in each fiber
near the spectrograph Littrow position at 1.604~$\mu$m.  Because
the pseudo-slit is actually curved, the Littrow ghost centers this
excess light at a slightly different wavelength for each fiber,
ranging from 1.6056-1.6067~$\mu$m ($\sim$35 pixels; Fig.
\ref{fig:instrument_schematic}).  This spectral region was
chosen through an optimization procedure aimed at minimizing the
impact of loss of absorption lines due to ghost overlap on the
quality and diversity of the final APOGEE abundances.  Optimal ghost
positions were identified for which only very few interesting lines
are lost, and for which in all cases there are other lines for the
same element making up for the missing ones.  The final position,
was selected so as to minimize any additional grating tilts that
could lead to a substantial change in spectral resolution.  The
resulting spectral interval happens to coincide with the natural
position of the Littrow ghost for the nominal APOGEE grating with
no fringe tilt. The FHWM of the ghost in the wavelength dimension
is about 9 Angstroms ($\sim$32 pixels).
(2) Fiber tip ghosts occur from light that reflects off the detector
face, transits through the entire instrument in reverse, reflects
off the fiber face (or v-groove block area adjacent to the fiber) and
returns through the instrument a third time, back to the detector.  While ghost
intensity varies with wavelength and fiber position, stray light
analysis of the optical design predicts the ghost images will have spot
size RMS radii approximately 1.5-4.5 times larger and intensity $< 1/1000$
compared to the primary images at the detector.  Moreover, ghost images should
arrive within 1 pixel of the primary image positions.

\subsection{Example Spectra}
\label{sec:example}

\begin{figure}
\includegraphics[width=0.50\textwidth]{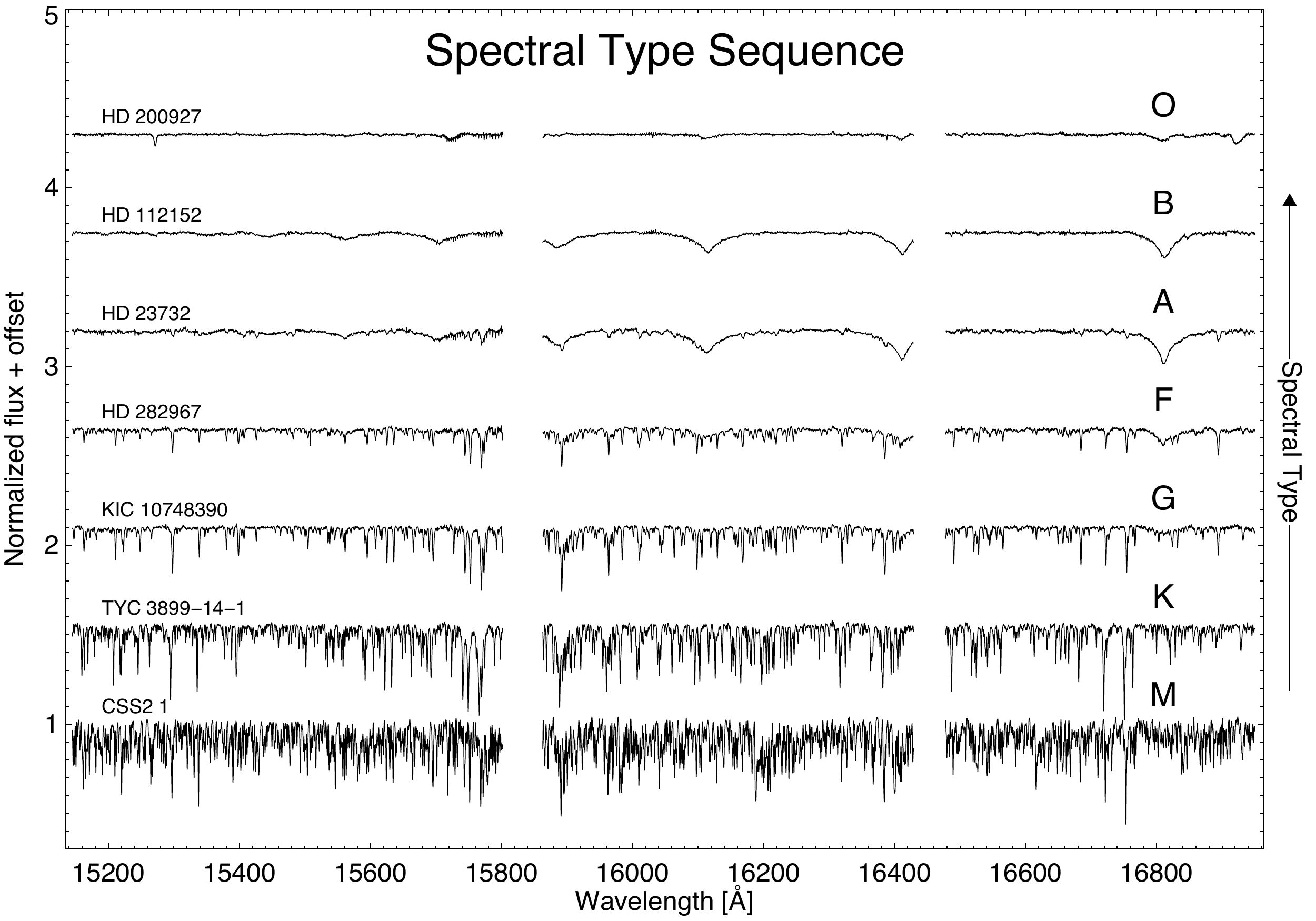}
\caption{\footnotesize An image showing continuum normalized APOGEE spectra as a function of 
stellar spectral type.  The earlier spectral types are representative of those
seen among the telluric standards, whereas the later types are typical of
those seen among the main APOGEE survey.}
\label{fig:spec_sequence}
\end{figure}

\begin{figure}
\includegraphics[width=0.50\textwidth]{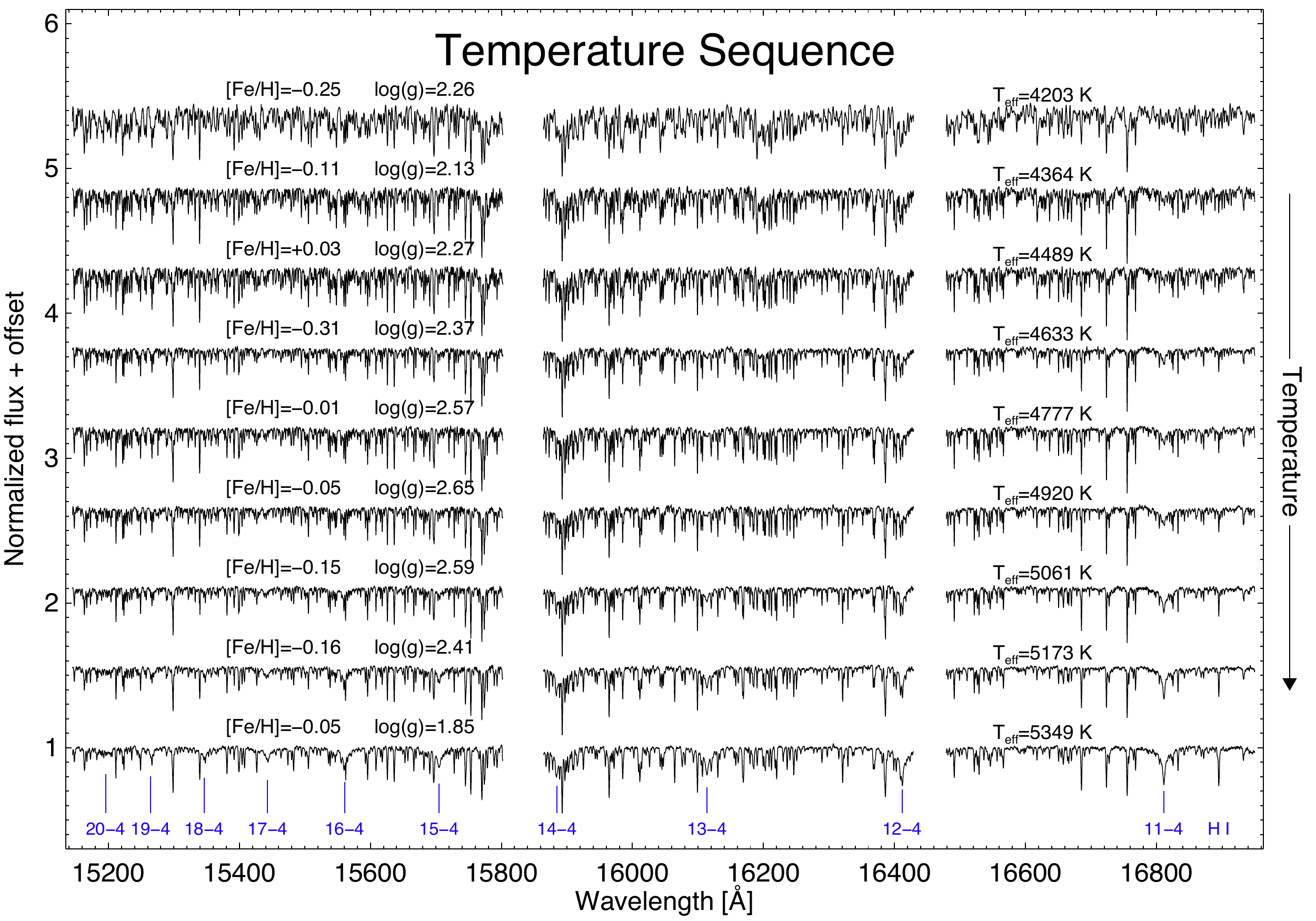}
\caption{\footnotesize An image showing continuum normalized
APOGEE spectra as a function of 
stellar surface temperature for typical APOGEE main survey RGB stars of 
near-solar abundance.
At the bottom, Bracket hydrogen lines are identified; these lines show the clear trend of increasing 
strength for increasing temperature.}
\label{fig:temp_sequence}
\end{figure}

\begin{figure}
\includegraphics[width=0.50\textwidth]{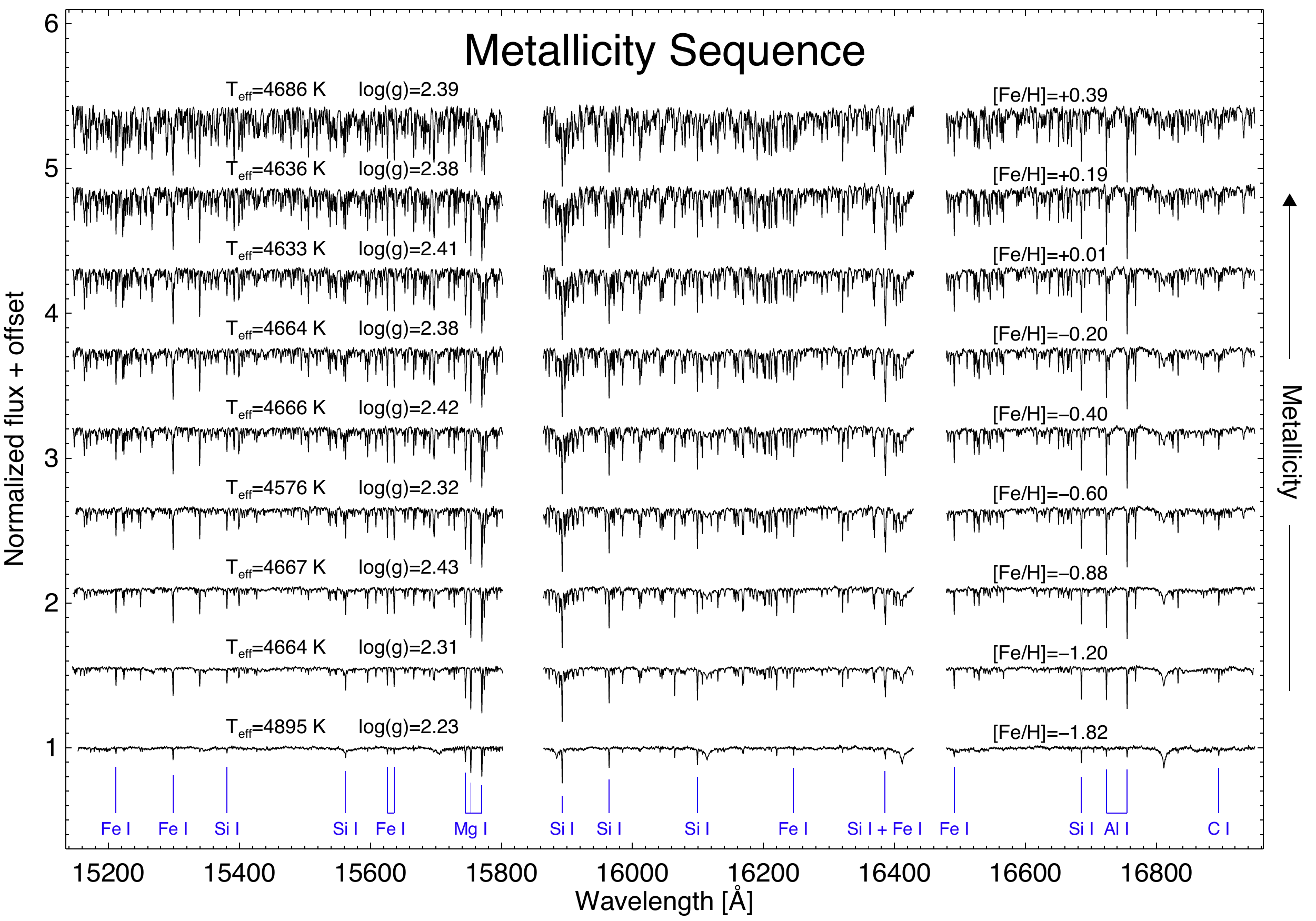}
\caption{\footnotesize An image showing continuum normalized APOGEE spectra as a function of metallicity for giant stars
of similar temperature.  Some of the strongest metal lines seen are identified at the bottom of the figure. }
\label{fig:abund_sequence}
\end{figure}

\begin{figure}
\vspace{-15mm}
\hspace{-10mm}
\includegraphics[width=0.50\textwidth]{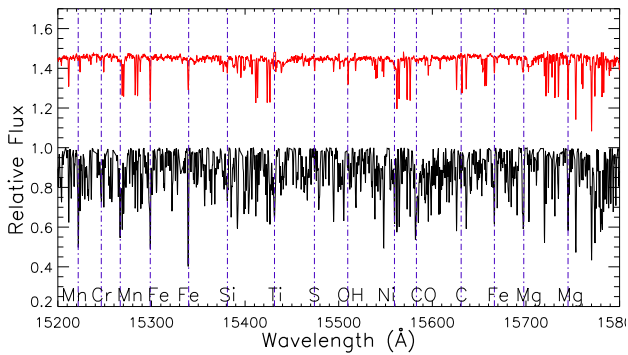}
\vspace{-50mm}
\caption{\footnotesize Comparison of a section of the APOGEE spectra for two stars of the same temperature
(approximately 4060 K) with about a 100$\times$
ratio in abundance of iron. The red spectrum is for a star that has [Fe/H] = -1.8 and $\log{g} = 0.158$.
The black spectrum is for a star that has [Fe/H] = 0.365 and $\log{g} = 1.5$.  }
\label{fig:abund_zoom}
\end{figure}

Examples of the appearance of stellar spectra as obtained by the APOGEE spectrograph are 
shown in Figures \ref{fig:spec_sequence}-\ref{fig:abund_zoom}.
Figure \ref{fig:spec_sequence} shows stars ranging from spectral
type O to M; the primary APOGEE science targets are of type G and
K, whereas most of the early spectral types were observed
as telluric standards and some M types are selected by the random
sampling of the parent distribution (\S \ref{sec:targeting}).  Across
the temperature range of the primary survey target types (G-K stars),
it is still possible to discern line strength variations, as shown
in Figure \ref{fig:temp_sequence}.
A primary driver of the APOGEE
project is the exploration of chemical abundance variations among its late type stellar sample; 
Figure \ref{fig:abund_sequence} demonstrates the appearance of RGB stars of similar temperature
but a 2.2 dex metallicity spread.
To show greater detail and a broader array of chemical species, 
Figure \ref{fig:abund_zoom} highlights the blue array
spectra for two giant stars separated by about 2.2 dex in [Fe/H].

\begin{table*}[tdp]
\caption{Summary of APOGEE Instrument Characteristics}
\begin{center}
\begin{tabular}{l l }
\hline 
Property  & Performance \\
\hline 
On-sky field of view (typical declinations)			& 3.0 deg diameter circle\\
On-sky field of view (high airmass)				& 1.5 deg diameter circle\\
Total number of spectrograph fibers				& 300\\
Fiber center-to-center collision limit on plugplate 	& 70 arcseconds \\
Fiber scale on sky (diameter)					&  2.0 arc seconds \\
Detectors  								& 2.5~$\mu$m cut-off, $2048^2$ pixel, Teledyne  H2RG Imaging Sensors  \\
Detector pixel size							& 18~$\mu$m \\
Detector wavelength regions					&  1.514-1.581, 1.585-1.644, 1.647-1.696~$\mu$m \\
Littrow ghost position 						& 1.6056-1.6067~$\mu$m \\               
Littrow ghost intensity						& 0.150\% of full fiber intensity \\    
Dispersion (at 1.54, 1.61, 1.66~$\mu$m)			& 0.326, 0.282, 0.235 \AA /pixel \\ 
Point Spread Function (spatial FWHM) (at 1.54, 1.61, 1.66~$\mu$m)			& 2.16, 2.14, 2.24 pixels		\\
Line Spread Function (resolution element) FWHM (1.54, 1.61 1.66~$\mu$m)	& 2.01 ,  2.44 ,  3.14 pixels \\
Median native ($\lambda$/FWHM) resolution (at 1.54, 1.61, 1.66~$\mu$m)	& 23,500, 23,400, 22,600 \\
Predicted\tablenote{Calculated as the product of the
wavelength-dependent transmittance or reflectivity for all components of the
as-built telescope+instrument design.} throughput  (1.54, 1.61 1.66~$\mu$m) & 14, 15, 10\% 				\\
Measured\tablenote{Based on measured flux for stars
of known $H$
magnitude.  Error bars reflect uncertainties regarding extinction by
Earth's atmosphere and (seeing-induced) fiber losses.} throughput  (1.61$\mu$m) &  20 $\pm$ 2\%  \\
$S/N$ for $H$=$12.2$ K0III star in an 8$\times$500 sec visit (1.61~$\mu$m)  &  105 \\
Specific fiber numbers most affected by excessive persistence		 		&   1-100 \\
\hline 	
\end{tabular}
\end{center}
\label{tab:Inst_Char}
\end{table*}%


\section{Survey Design}
\label{sec:surveydesign}

\subsection{Field Selection}
\label{sec:fieldselection}

\subsubsection{Field Selection Principles}
\label{sec:fieldselectionprinciples}

The APOGEE field targeting strategy was designed around several
motivations and requirements:

\begin{itemize}

\item A desire to sample, with minimal bias, all stellar populations
of the Galaxy, from the bulge, across the disk, and into the halo.

\item The need to probe fields to a variety of magnitude limits to
access stars over a wide range of distance in all parts of the
Galaxy.

\item The ability to calibrate efficiently against stars with
well-established physical properties, such as the chemical abundances
and radial velocities that are often well established for
star cluster members, or the masses and gravities that can be derived
for asteroseismology targets.

\item The need to coordinate with the other SDSS-III bright time
program, MARVELS, which relied on frequent visits to a relatively
limited number of fields.

\end{itemize}
In the end, changes in the latter two requirements as well as the realities of the
actual distribution of clear weather and several other considerations
led to the evolution of the APOGEE target selection over the three
year observing campaign.

\subsubsection{Field Selection Evolution}
\label{sec:fieldselectionevolution}

{\it Initial Survey Design:} 
For its expansion into bright time
observing the SDSS-III collaboration planned to capitalize
on the existence of two new fiber-fed instruments that
could operate simultaneously from shared plugplates, thereby doubling
the effectiveness of the Sloan Telescope.
Because the MARVELS project required many visits to each of its
target fields, whereas APOGEE had always planned at least some deep
field probes, the original SDSS-III plan was for 75\% of the bright
time to be in co-observing mode, whereas the remaining 25\% of
bright time would be given to APOGEE to observe fields of no interest
to MARVELS and to fill out its sky coverage.
Moreover, because 
MARVELS targets were relatively bright, relatively nearby stars,
both surveys could make good use of many visits to fields
at high latitude (in APOGEE's case, for accumulating signal on faint, distant
halo stars) as well as in the disk (where APOGEE could
{\it both} accumulate flux on highly dust-extinguished stars across
the disk as well as cycle through large numbers of brighter stars).

The baseline for co-observed fields 
was to accumulate a total of 24, approximately one hour visits.
Under these overriding restrictions, the initial APOGEE field
selection plan focused on fulfilling the other principles described
in \S\ref{sec:fieldselectionprinciples}.  The 75\% shared
survey time was distributed in a series of 24- and 12-visit fields
across the disk and halo (the latter used for fields that MARVELS
began observing before APOGEE came on line).
The disk plan included, a
regular ``picket fence'' Galactic longitude distribution of these
deep fields, and with multiple visits at each picket broken up into a series
of plate designs that enable stars of different magnitudes (i.e.,
mean distances) to be cycled through for different numbers of total
visits.
The adopted distributions of Galactic latitude and cycling of stars 
were based on modeling stellar population distributions using both
the Trilegal \citep{Girardi05} and Besan\c{c}on \citep{Robin03}
Galaxy models.
A major concern addressed by this modeling and that
drove the specific latitude distributions chosen, was ensuring ample
representation of stars from the Intermediate Population II, ``thick
disk".  Further information about this modeling is given in Appendix
D; an example of the results are given in Figure \ref{fig:XYApogeePlot}.

For the shared halo pointings, the APOGEE team focused on fields
containing globular clusters, which serve as both science and
calibration targets.  A number of globular cluster stars having
high resolution spectroscopy in the literature are faint enough to
require deep APOGEE observations.  Moreover, the multiple globular
cluster visits make it possible to increase the number of
globular cluster stars to be sampled, given the limitations posed
by fiber collisions (Table \ref{tab:Inst_Char}).  Additional high
latitude long fields were placed in fields known to be traversed
by halo substructures, such as the Sagittarius stream (with field
placement guided, e.g., by the results of \citealt{Majewski03}) or
the Virgo Overdensity \citep[e.g.,][]{Vivas01,Newberg07,Juric08}.

With this basic structure in place for 75\% of the planned observing
time, the remaining bright time time was distributed to various classes
of ``APOGEE-only'' fields: (1) fields across the bulge, a primary
region of the Galaxy sought for our primary science goals (\S
\ref{sec:sciencegoals}); (2) fields at low declinations that are
not viable for MARVELS work; (3) a number of fields across the disk,
filling in the relatively large gaps between the long field
``pickets''; and (4) additional disk fields that include open
clusters useful for further calibration of APOGEE spectroscopy.
There are two important considerations relevant to fields of class
(1) and (2):  First, the limited accessibility for these fields
resulted in them typically being reduced to having single 1-hour visits
which, at constant $S/N$, mandated a brighter magnitude limit
($H=11.1$, see \S\ref{sec:maintargets}) there.  This
was needed to ensure that statistically significant samples would
be obtained at the end of the survey, particularly 
in high value regions such as the Galactic bulge, where
good spatial coverage was also desired.  Second, the sizes for these
fields had to be reduced to only a 1-2$^{\circ}$ diameter field
because of the severe differential refraction experienced over the course of a
1 hour visit at high air masses.  Fortunately these reduced
field-of-view fields occur in high stellar density environments,
so that there is no shortage of targets from which to choose.

This overall plan was in place by early 2011 --- as needed to begin
drilling plugplates in preparation for 2011Q2 instrument commissioning
and 2011Q3 survey operations --- and  thus dictated the early survey
observing plan.

{\it Reconfiguration at MARVELS Descope:}
During the 2011 summer shutdown, and just prior to the commencement
of the formal APOGEE survey operations, a decision was made to
gradually curtail the MARVELS program over the course of the following
year.
A select number of MARVELS fields that had already obtained at least
12 pre-APOGEE epochs of MARVELS observation were chosen for completion, 
but reduced to either 6 or 12 hour
APOGEE fields.
Thus, in the end, only a handful of the original 24-hour fields
were preserved (primarily the ``deep disk mid-plane spokes'' at
$l=30, 60$ and $90^{\circ}$ and a few globular cluster
fields:
see Figs. \ref{fig:field_plan} and \ref{fig:XYApogeePlot}).

The sudden, substantial increase in the 
share of ``APOGEE-only'' bright time observing  
allowed a number of new pointings to be added to the baseline APOGEE
field placement plan:

\begin{itemize}

\item 
more bulge pointings, including fields useful for
cross-calibration to the BRAVA \citep{Rich07} and ARGOS \citep{Freeman13}
surveys;

\item additional calibration open and globular clusters;

\item numerous 3 hour fields to give a finer angular sampling at
latitudes of $b=0,\pm4,\pm8$ and $\pm12^{\circ}$) between the preserved
12/24-hour pickets at $l=30,60,90,120,150,180$ and $210^{\circ}$;

\item rings of high latitude fields at $b=+30,\pm45,+60$ and
$+75^{\circ}$.

\end{itemize}
The greater flexibility afforded by the increased control over field placement also 
aided in the implementation of the initial set of Ancillary Science programs.

{\it Incorporation of the Kepler Field:}
The success of ESA's {\it CoRoT} mission \citep{Auvergne09} and
NASA's {\it Kepler} mission \citep{Borucki10}, and, in particular, the 
asteroseismology programs for each \citep{Michel08,Chaplin10,Gilliland10}
--- through which non-radial oscillations were detected and characterized
for a substantial 
sample of RGB stars and subgiants 
\citep{Mosser10,Hekker11} --- presented a special opportunity for the
APOGEE program.  The asteroseismic frequencies are sensitive
probes of stellar masses and radii \citep{Chaplin13}.
Apart from providing invaluable independent measurement of 
stellar gravities for testing and calibrating the APOGEE stellar
parameters pipeline (\S \ref{sec:stelpar}), when combined with
precision abundance measurements of the quality that APOGEE could
provide, asteroseismically measured stellar masses can
provide reliable age estimates, at the level of 15\% \citep{Gai11}.
The opportunity to obtain such reliable age data for a large number
of {\it field stars} is unprecedented, and provides pivotal
temporal benchmarks for a survey of Galactic chemical evolution,
the primary mission of APOGEE.  Moreover, the APOGEE instrument
presents the only practical means to obtain high resolution
spectroscopic assays for a large fraction of this {\it Kepler}
sample, which is distributed over a relatively large area of sky;
serendipitously,
the FOV of the SDSS plugplates is nicely matched to the
size of a {\it Kepler} tile.

With formally established collaborations --- the APOGEE-{\it Kepler}
Asteroseismic Science Collaboration (APOKASC) and the {\it CoRoT}-APOGEE
Collaboration (COROGEE) --- plans were established to target a large
fraction of the KASC giant/subgiant sample as well as CoRoT giants
in the direction of the Galactic center and anticenter; however,
practically, this meant non-negligible reorganization of the APOGEE
targeting scheme.  The two {\it Kepler} tiles containing the star clusters 
NGC 6791 and NGC 6819 already were planned to have long pointings, but 
the remaining 19 {\it Kepler} tiles were now included
with two 1-hour visits each, and with each visit focusing on a unique set of
targets.\footnote{Because chemistry was a primary goal of the APOGEE
visits, and the amount of available observing time was greatly
limited, the normal three-visit cadence for binary detection was
not implemented for the APOKASC program.}
This resulted in observations of (a) some 8,000 APOKASC giant stars, 
along with (b) about 600 subgiant and dwarf stars, whose ages could be
determined using gyrochronology \citep[e.g., ][]{vanSaders13},
as well as (c) targets for other ancillary science programs (e.g.,
eclipsing binaries).  Further information on the {\it Kepler} field
targeting can be found in \citet{Zasowski13} and \citet{Pinsonneault14}.
Meanwhile, in the CoRoT fields APOGEE targeted 121 giant star
candidates in one plate designed for the CoRoT LRa01 (``anticenter'')
field and 363 giant candidates on 3 plates designed for the LRc01
(``center'') field.  Unfortunately, incorporation of the Kepler
field observing required thinning out the APOGEE survey of the
Galactic plane at similar longitudes (though see below).

{\it Survey Year Three Modifications:} Several circumstances led
to further modifications in the third year of APOGEE observations,
fortunately in the sense of allowing expansion of the APOGEE
footprint.  First, overall clearer than average winters put the
surveying of the anticenter disk ahead of schedule.  This enabled
an expansion of the Galactic anticenter grid with broader
latitude coverage and more finely sampled longitude coverage at all
latitudes, both an advantage for exploring the properties of the
disk warp, disk flare and the presence of low latitude
substructure in the outer Galaxy, such as the Monoceros and TriAnd
structures.  In addition, several Ancillary Science programs that
could take advantage of the relevant LSTs were slightly expanded.

Meanwhile, our somewhat lagging spring and summer field schedule
was greatly aided in the final year by both the twilight and dark
time observing campaigns (\S \ref{sec:specialobs}).
With this extra telescope time the APOGEE program not only was able
to catch up on observations of spring and summer fields (including
Kepler field pointings), but to restore previously removed disk
grid pointings near the Kepler field.

\subsubsection{Final Field Plan}
\label{sec:fieldplan}

\begin{figure}[htbp]
\begin{center}
\includegraphics[width=0.47\textwidth]{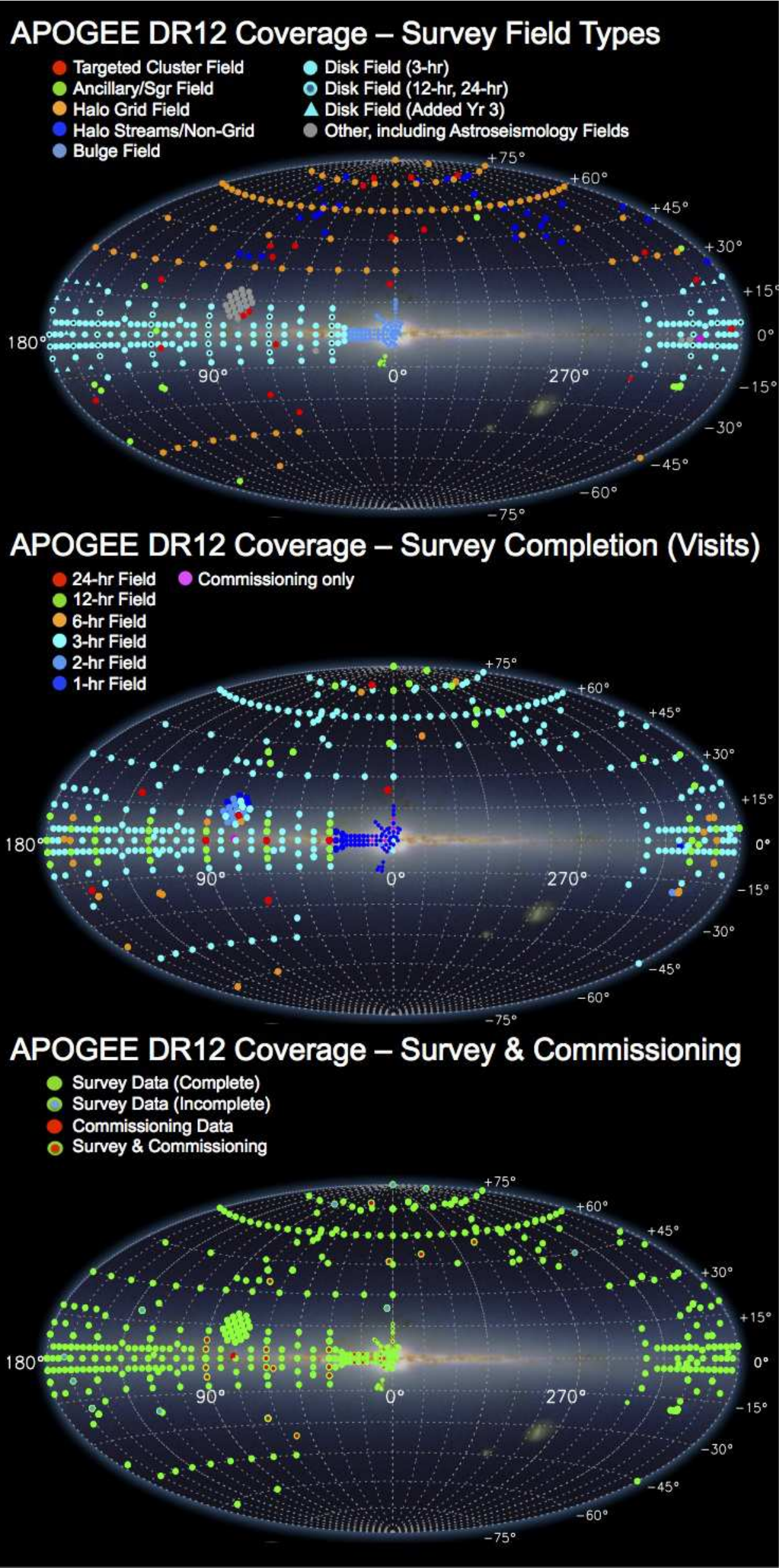}
\caption{
(Top) The final APOGEE field targeting plan, the product of the
evolving strategy described in \S\ref{sec:fieldselectionevolution}.
Fields are color-coded by their primary purpose or sought-after
target class.  The grey fields include both Kepler and CoRoT asteroseismology
targets in the Kepler and CoRoT databases, as well as MARVELS Calibration fields.
(Middle) Distribution of observed APOGEE fields,
color-coded by the number of approximately 1-hour visits.  
(Bottom) Distribution of APOGEE survey and commissioning fields,
and, for the former, whether the survey observations were completed.
Most commissioning observations were repeated during the main survey
with the spectrograph in its survey configuration.}
\label{fig:field_plan}
\end{center}
\end{figure}

\begin{figure}
\includegraphics[width=0.50\textwidth]{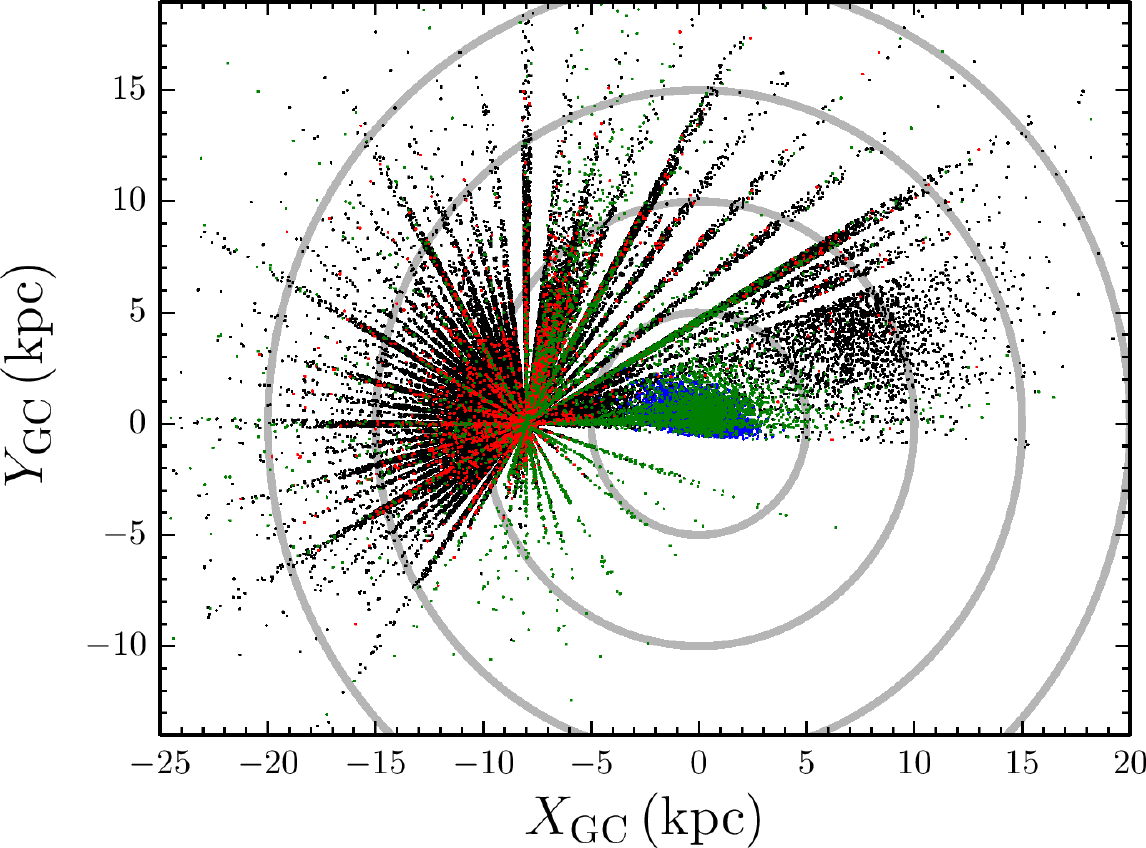}
\caption{\footnotesize The expected Galactic distribution of APOGEE targets as projected
on the Galactic plane, as predicted by the Trilegal model for the field
plan prior to the final, survey year three modifications.  
Stars are color-coded by
expected stellar population:  blue = bulge, green=halo, red=thick disk, black=thin disk.
}
\label{fig:XYApogeePlot}
\end{figure}

The final APOGEE targeting footprint is thus the product of
the evolving plan described in \S\ref{sec:fieldselectionevolution}.
Figure~\ref{fig:field_plan}, which supersedes the previously published
APOGEE field targeting plan in \citet{Zasowski13}, shows the final
implemented survey plan (\S\ref{sec:fieldselectionevolution}) with
the targeted fields color-coded according to different criteria.
In the top panel, the fields are color-coded according to the
intended primary purpose.  The middle panel shows the same fields
color-coded by number of visits.  Finally, in the bottom panel the
fields are broken up by formal survey versus commissioning fields
and, for the former, by completion status. 
Few ``commissioning-only" fields remain because most commissioning
observations were repeated during the main survey with the spectrograph
in its proper survey configuration (\S \ref{sec:timeline}).

\begin{figure}
\includegraphics[width=0.50\textwidth]{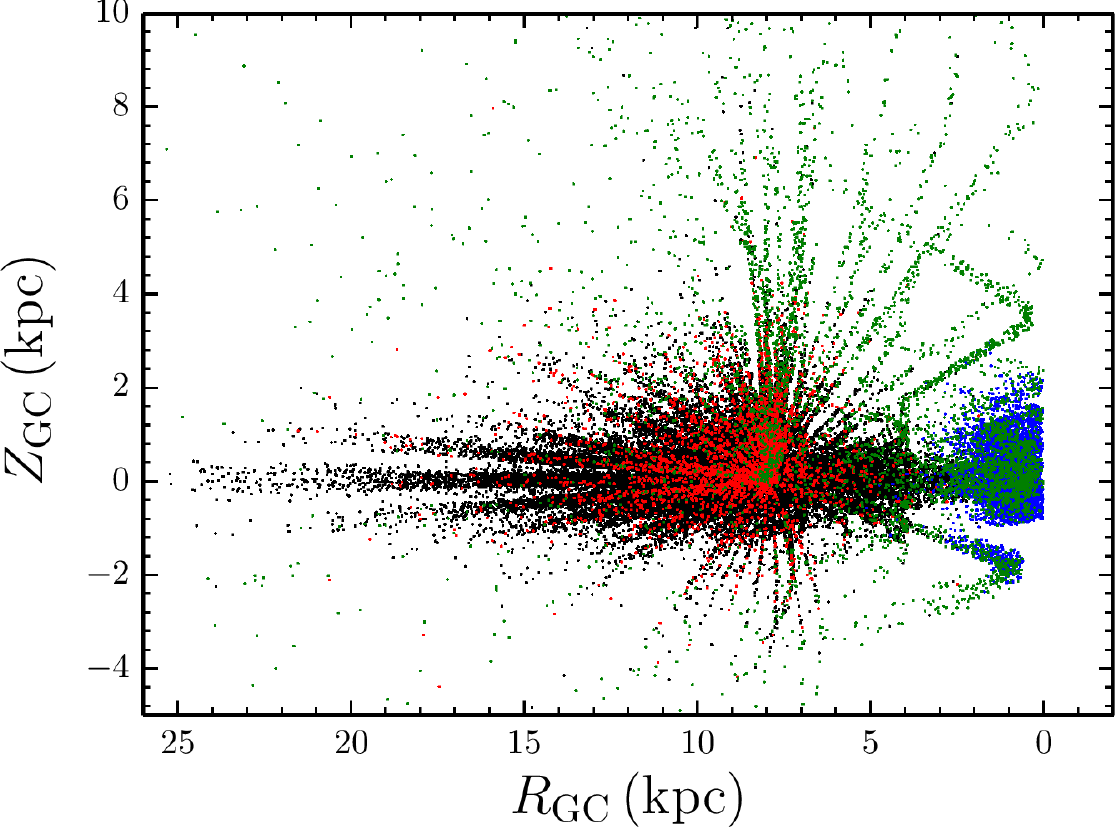}
\caption{\footnotesize Same as above Fig. \ref{fig:XYApogeePlot} for the expected 
Galactic azimuthally-averaged $R_{GC}-Z_{GC}$ 
distribution of APOGEE targets 
 as predicted by the Trilegal model. 
}
\label{fig:RgZApogeePlot}
\end{figure}

\subsection{Target Selection}
\label{sec:targeting}

APOGEE targeting consists of (1) the ``main sample'' or ``normal
science targets'', (2) ``special targets'', which include (among
others) calibration stars with measured stellar parameters and
abundances from other spectroscopic studies, star cluster members,
and targets submitted by one of APOGEE's Ancillary Science programs,
and (3) a sample of early-type stars observed as telluric absorption
monitors for each exposure.  A complete and detailed discussion on
how each of these targets is selected, and how they are identified
within the publicly released databases, is given by \citet{Zasowski13}.
We only give a broad overview here, with an emphasis on motivations
for the overall procedures followed.

\subsubsection{``Minimum Criteria'' Philosophy}
\label{sec:mincriteria}

From the start of APOGEE survey planning there was a strong desire
to maintain the utmost simplicity in the rules for target selection 
for the main sample of normal science targets.
As the first large spectroscopic project to truly survey all major
components of the Milky Way, questions related to interface and
overlap of these components 
are central to the APOGEE mission.  
To see these signatures with clarity 
a homogeneous sample and a well understood
selection function are both critical.  Moreover, a first exploration
of uncharted territory mandates a prudent attitude, curbing a natural
temptation towards forcing overrepresentation of certain populations
in any given position in the sky.  As a consequence, however, the resulting
sample strongly favors the most common stellar types (e.g., metal-rich disk
stars), with rare populations (e.g., metal-poor stars) constituting
a small---even negligible---fraction of the whole.  To some extent,
this situation is mitigated by the field distribution, which naturally leads to 
variable relative sampling of the bulge, thin disk, thick disk, and halo by
Galactic line-of-sight
(\S\ref{sec:fieldselection}, Fig.~\ref{fig:field_plan}).  In addition,
the emphasis of APOGEE's targeting on a stellar color range dominated
by RGB star candidates (\S\ref{sec:motivations}) enhances
the representation of more distant populations, despite the 
relatively bright magnitude limits of the survey.

Nevertheless, nearly every 
APOGEE field has many more objects in it than APOGEE can reasonably observe, 
and the strategy for selecting targets
from the available parent population 
inherently imposes additional
biases in the selection function.  In particular, the adopted schemes for
selecting stars across the magnitude distribution (see
\S\ref{sec:maintargets}) have been designed to achieve large spreads
in distance representation along each line of sight.  Moreover,
additional photometric criteria were adopted in the halo
fields to favor the targeting of halo giants and minimize
foreground dwarf star contamination (\S\ref{sec:maintargets}).
Despite these concessions, which were meant solely to improve
the {\it spatial sampling} of the Galaxy, 
a goal of maintaining the simplest
and most consistent selection function {\it at each position}
was central to the survey
design.

\subsubsection{Source Catalogs and Supplemental Data Used}
\label{sec:sources}

Target selection for APOGEE was made primarily using the Point
Source Catalog (PSC) of the Two Micron All-Sky Survey (2MASS;
\citealt{Skrutskie06}), which is complete
to $H<15.1$, and
therefore more than sufficient for our primary selection of targets
with $H<12.2$.  In effect, APOGEE represents the first comprehensive
stellar spectroscopic follow-up survey of 2MASS.

These data were supplemented, where available, with {\it Spitzer} IRAC data 
taken from the GLIMPSE I, II, and 3-D
surveys \citep[e.g.,][]{Churchwell09}.  The addition of IRAC data
in the Galactic mid-plane, where extinction is greatest, allows us
to take advantage of star-by-star dereddening techniques exploiting
$JHK_s[3.6][4.5]$ data \citep{Majewski11}.  In the vast majority
of fields falling outside of the {\it Spitzer} footprint, we made
use of the mid-IR data from NASA's {\it WISE} mission \citep{Wright10}.
Finally, to enhance our efficiency in identifying stars from the
distant halo, we also made use of an ad hoc Washington $M,T_2$,$DDO51$
filter observing campaign in high latitude fields using the Array
Camera on the U.S. Naval Observatory 1.3-m reflector; this filter
system has been shown to be effective in photometrically distinguishing
dwarf from giant stars \citep{Geisler84,Majewski00,Morrison00}.

\subsubsection{Main Survey Targets}
\label{sec:maintargets}

{\it Color Selection Criterion:}
A primary driver of the APOGEE survey was the desire to exploit 
luminous, evolved (RGB, RSG, AGB, RC) 
stars as our primary tracer of the Galaxy because they allow
access to large distances, even in regions of high extinction,
at magnitudes reachable with the Sloan Telescope.
Moreover, these post-main-sequence stars are found in stellar
populations of almost all ages and metallicities and so do not
impose a strong bias in this regard.  Finally, it is possible to
generate relatively pure samples of these 
stars with simple
color criteria applied to the 2MASS PSC.  It is well known that the red 
side of the typical $(J-K_s, H)_0$
color-magnitude diagram (CMD) produced from the 2MASS PSC is dominated
by red giant and red clump stars, so that a simple red color selection
suffices to 
generate a target catalog dominated by such evolved stars.

Choice of an optimal blue $(J-K_s)_0$ limit entails a trade
between (1) increasing dwarf star contamination towards the
blue, (2) increased fractional representation of fainter (and
therefore typically closer) red clump versus RGB stars towards
the blue, and (3) increasing bias against metal-poor giant and red
clump stars towards the red.  Comparison to stellar atmospheric
models, Galactic stellar population models and theoretical isochrones
indicate that within APOGEE's typical magnitude range, a color limit
of $(J-K_s)_0 \ge 0.5$ produces a sample that substantially reduces
the dwarf star contamination in the final sample while imposing a
minimal bias against metal-poor giants \citep[see \S 4.3 of][]{Zasowski13},
and this limit was adopted for the main APOGEE survey. 

{\it Correction for Extinction:} To obtain the extinction-corrected
CMDs we applied a correction to each potential target based on its
$E(H-4.5\mu$m) color excess according to the Rayleigh-Jeans Color
Excess Method (``RJCE"; \citealt{Majewski11}), if 4.5~$\mu$m
photometry is available from {\it Spitzer} or {\it WISE}, with the
former preferred because of its better resolution. Unfortunately,
most of the {\it Spitzer} data derive from the GLIMPSE or other
programs that are tightly confined (generally to within 1 deg, and
at most 4 deg) to the Galactic mid-plane.  Fortunately,
these are the latitudes where image crowding is worst and the need
for  {\it Spitzer}'s better
spatial resolution is greatest.  For halo fields, it was found
that a slightly more sophisticated, ``hybrid" dereddening method,
invoking limits from the \citet{Schlegel98} maps, proved more
effective \citep[see \S4.3.1 of][]{Zasowski13}.

{\it Magnitude Ranges:}
Given the requirement of $S/N = 100$/pixel for the faintest targets
in any field, the magnitude limits are set by the number
of visits (thus integration time) to each field.
Thus, because of the variable numbers of visits across the
survey (\S\ref{sec:fieldplan}, Figure~\ref{fig:field_plan}), different lines of
sight probe to different magnitude limits, and, consequently, distances. 
The nominal 3-visit survey field
is limited to $H \le 12.2$, but across the survey magnitude limits range from
$H \le 11.0$ to $H \le 13.8$ for fields ranging from 1 to 24 visits
(see Table 4 of \citealt{Zasowski13}).  
A universal bright magnitude limit of $H=7.0$ prevents saturation of the
detectors and minimizes
scattered light contamination of adjacent spectra.

However, only a fraction of the stars in a particular FOV
require the full integration delivered by all visits to that field.  
Moreover, as described in \S\ref{sec:fieldselectionevolution},
numerous visits to the same field
afford the opportunity to sample discrete groups of stars and
accumulate a much larger stellar sample. Therefore, a
``cohort'' scheme was developed to divide the parent target sample
into groups of stars that could be successfully observed in only a
fraction of the visits and then rotated out and replaced with new
targets.  The details of the breakdown on the number of fibers per
plate design delegated to each cohort and the magnitude ranges
assigned to each cohort are detailed in \S4.4 of \citet[][]{Zasowski13}.

{\it Magnitude Distribution Function:}
With magnitude limits established for each cohort in a field, stars
within the relevant color and magnitude limits are then 
sampled randomly within each cohort.
Consequently,
the final magnitude distribution of spectroscopic targets in a field
may differ significantly from the distribution of candidates, because
the former also depends on (a) the number of each type of cohort in the
field, (b) the fraction of APOGEE's science fibers allocated
to each cohort, and (c) the vagaries of which targets may be rejected
during the actual plate design phase due to fiber collisions
\citep[see \S4.5 of][for details]{Zasowski13}.  

{\it Halo Field Considerations:} 
A larger fraction of available stars can be targeted in the
halo fields than at lower latitudes, because of the lower target
density.  However, because of the steep density fall-off, the
nominal dwarf:giant ratio in the standard survey color and magnitude
range is substantially higher in the halo.  Therefore, to
ensure access to the smaller fraction of giant stars available
per field, in many halo fields we used combined Washington
($M$ and $T_2$)  and $DDO51$ photometry to classify stars as likely
dwarfs or giants prior to their selection as spectroscopic targets
(see \citealt{Zasowski13} for details).
In some halo fields the number of targets brighter than the nominal
magnitude limit was too small to employ all APOGEE fibers. 
Unused fibers were assigned to stars that either lacked $DDO51$
classification or were classified as dwarfs, or on stars classified as
giants, but fainter than the magnitude limit, with the expectation
of getting at least some useful data from the resulting lower $S/N$
spectra (see \S3.3 and 7.1 in \citealt{Zasowski13}).

\subsubsection{Calibration Fibers}
\label{sec:calfibers}

Despite the great multiplexing advantage afforded by a 300 fiber instrument, observing
in the near-infrared means that unfortunately
a non-negligible fraction of these fibers must be surrendered to real-time calibration.
APOGEE spectra are affected (see Figs. \ref{fig:wavelengths} and \ref{fig:element_maps}) by
both airglow (OH emission) and telluric absorption 
(by CO$_2$, H$_2$O and CH$_4$),\footnote{A detailed breakdown of this telluric absorption
by molecular species is shown in Fig. 17 of \citet{Nidever15}.} and both phenomena vary on short 
enough timescales that they must be monitored simultaneously with science observations.  
Moreover, these atmospheric effects
vary on angular scales comparable to the APOGEE/SDSS FOV
(see, e.g., Fig. 19 of \citealt{Nidever15}).  
Thus, large numbers of broadly distributed calibration fibers are needed
for the derivation of two-dimensional airglow and telluric absorption
corrections across the same FOV as the science fibers.
For airglow correction, 35 APOGEE fibers are assigned 
(by the plate design algorithm --- see \S \ref{sec:platedesign}) to an evenly
distributed selection of blank sky positions.  

To monitor the telluric absorption it is most useful to depend on
the spectra of hot stars, which are characterized by very few
and very broad atomic lines that can be easily distinguished from
telluric lines.  Thirty-five of the bluest and brightest stars
evenly distributed across the field are chosen for telluric absorption
calibration
(see \S5 and Fig. 8 of \citealt{Zasowski13}). 
Although not originally envisioned as part of the primary science
focus of APOGEE, the number of hot stars targeted and the ample
spectral time series collected for many has turned out to yield a
number of interesting science results, particularly in the study
of emission line (B[e]) stars and other, non-emission stars with
circumstellar disks \citep[][see \S\ref{sec:timeseriesspec} and 
Fig. \ref{fig:BeStar}]{Chojnowski15}, and including the discovery
of rare stellar types \citep{Eikenberry14}.

\subsection{Ancillary Science Program}
\label{sec:ancillary}

Several motivations led to the inclusion of an ancillary science
program in the APOGEE survey plan: (1) The APOGEE spectrograph, its
mating to the very large FOV Sloan 2.5-m Telescope, and
the extremely effective multifiber optical interface between the
two represents a unique, state-of-the-art capability applicable to
a broad range of groundbreaking Galactic science applications that
may not fall within the purview of the primary APOGEE mission.  (2)
Not all interesting and relevant Galactic science could be included
in the primary survey program, but could be addressed in a limited
way through an ancillary science program.  (3) Some science programs
that might be worth pursuing 
as main survey science
require some verification and testing in pilot programs.  (4) Leaving
some amount of survey time in reserve allows the opportunity to
respond to new developments in the field or to incorporate originally
unanticipated science of great value.

Given these motivations, 5\% of the total fiber-hours\footnote{The
``fiber-hour" metric is defined so that one fiber-hour represents
the allocation of one fiber for one visit, which is about one hour
long.} of the APOGEE survey were made available for a formal APOGEE
Ancillary Science Program.
The main criteria for selecting such proposals was that the ancillary
observations result in novel and compelling scientific contributions
and that they not impact negatively the primary
objectives of the APOGEE survey.  Especially compelling were proposed
programs that could enhance the productivity and impact of the
primary APOGEE survey.  Several of the most meritorious proposals
in the APOGEE calls for ancillary science had as primary goals the
improved calibration of the APOGEE database.  A few approved ancillary
science programs served as the basis for a major redefinition of
APOGEE targeting to include significant attention to Kepler mission
targets (\S \ref{sec:fieldselectionevolution}).

Three calls for proposals to the APOGEE Ancillary Science Program
were solicited: September 2010, March 2012,  and March 2013.  
Two flavors of ancillary science targeting were implemented:
(a) sets of individual fibers placed on specific targets in
already-existing APOGEE survey pointings, and (b) use of up to all
$\sim$230 APOGEE ``science" fibers in a new pointing not already
within the general APOGEE survey plan.  The selected Ancillary
Science programs are described in detail in Appendix C of
\citet{Zasowski13}.  Note that, while all collected APOGEE spectra
are automatically processed through the data reduction and analysis
pipelines, for some of the programs focused on targets
significantly different from those in the main survey, there is no
guarantee that the automatically generated data products are
optimal, or even reliable.  All special processing and analysis of Ancillary Science
Program data are the responsibility of the principal investigators
of each selected project.

\begin{figure}
\vspace{-75mm}
\includegraphics[width=0.70\textwidth]{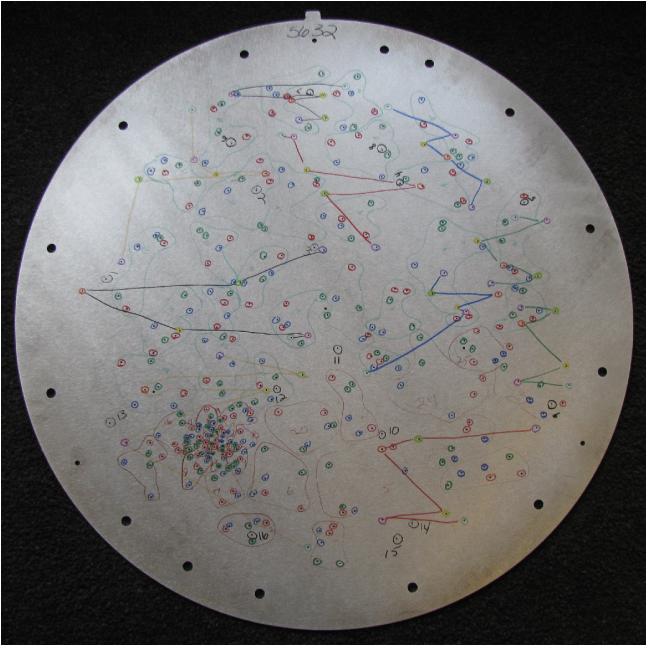}
\caption{\footnotesize Photo of a shared APOGEE/MARVELS plugplate
(plate \#5632), as marked for plate plugging. This particular plate
is for a field featuring the globular cluster M3, whose 
position on
the plate can be identified by the concentration of fiber holes to the lower left.
The holes connected by the zigzagging tracings mark those
associated with MARVELS, whereas the red, green and blue circled
holes show those intended for the bright, medium and faint APOGEE
fibers, respectively.  The latter holes are grouped into small
``zones" (indicated by the irregularly-shaped, closed loops) by the
pluggers as a way to organize areas on the plate anticipated to be
serviced by fibers in a single anchor block (having two
red, two green and two blue fibers each). (Photo by W.
Richardson.)} \label{fig:plugplate} \end{figure}

\subsection{Plate Design and Drilling}  
\label{sec:platedesign}

Once prioritized lists
of selected targets (science, telluric calibrators, sky positions)
have been generated for each plate design, they are fed to standardized
SDSS plate design software.  This software takes the input targets'
celestial coordinates 
and generates the final linear $(x,y)$ plug plate drill pattern for
the plate design.  The software accounts for potential fiber collisions
between all fibers from both APOGEE and MARVELS, as well as
collisions between science fibers and acquisition or guide fiber
bundles.  The algorithms also take into account the field curvature of
the Sloan 2.5-m Telescope (to which the plugplates are bent during 
observing) and the differential
refraction expected for the nominal hour angle at which each plate
of a given declination might be observed.
In some cases, due to the uncertainty in scheduling, multiple plates
might be generated from the same plate design input files, differing
only in the potential hour angle of observation.

\begin{figure}[h]
\vspace{-42mm}
\includegraphics[width=0.50\textwidth]{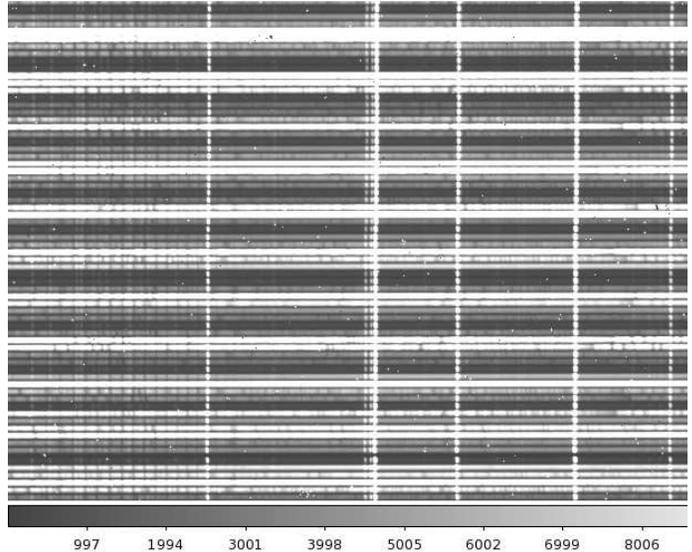}
\caption{\footnotesize 
A portion of a 
raw 2-D 
APOGEE image from observations of
a bulge field.
The horizontal stripes correspond to individual stellar spectra.  Vertical
bright bands correspond to airglow features at the same rest
wavelength in each spectrum, whereas
absorption features at the same horizontal position from spectrum to spectrum
correspond to telluric absorption features.  Also obvious are variations in
the expression of stellar atmospheric absorption features from star to star, 
evidenced by their varying strengths due to temperature and chemical composition 
differences, as well as changing relative positions due to Doppler shifts.
Fiber assignments were managed by color-coding the fiber jackets at
the telescope end for stars in each field sorted into three brightness groups 
(bright, medium, faint).
These fibers were sorted at the spectrograph slit head into a repeating pattern
of faint-medium-bright-bright-medium-faint  
to minimize the contamination
of any given spectrum by the PSF wings of a much brighter spectrum in 
an adjacent fiber.  
This management scheme gives rise to the brightness modulation pattern
apparent in this image.  
}
\label{fig:fibermanagement}
\end{figure}

In addition to establishing the precise coordinates for each star
based on refraction considerations, the plate design code also sorts
the intended targets into three magnitude bins of 100 stars each.
The stars in each magnitude bin are assigned to fibers of a given
sheathing color (red, green or blue), by which fiber management is
achieved in the telescope focal plane to separate the brightest
spectra from the faintest spectra in the spectrograph focal plane
(see \S \ref{sec:fibertrain}); this separation is needed 
to minimize contamination of any spectrum by the PSF wings of adjacent spectra.
Figure~\ref{fig:fibermanagement} illustrates how this
fiber management scheme creates a repeating pattern of variable 
spectrum brightness
as a function of fiber pseudoslit position as projected onto the 
spectrograph focal plane.

The plates are drilled on a 6-axis, computerized (CNC) milling
machine at the University of Washington, and then shipped to APO.
At APO, the plates are manually marked to identify which holes
correspond to stars designated to red, green or blue-sheathed fibers
by way of an overhead projection onto the aluminum plate of the
fiber plugging color scheme (Fig.~\ref{fig:plugplate}).  
Note that the red/green/blue = bright/medium/faint division of stars in each
plate design are not directly correlated to any designated cohort
divisions, except as the sorting by magnitudes of stars in the
cohorts places them into an appropriate fiber color by default.

\section{Survey Operations}
\label{sec:operations}

\subsection{Standard Observing Procedures}
\label{sec:observing}

As with all SDSS observing, APOGEE observing was typically conducted with the use
of a package of standard operating scripts that orchestrate
nightly activities through the observatory STUI (\S \ref{sec:layout}).

APOGEE science observing was based on standardized ``visits'' (\S
\ref{sec:binaries}) to a scheduled set of fields using corresponding
plugplates designed and drilled for specific hour angles (\S
\ref{sec:platedesign}), and plugged with fibers in advance.  
Each standard visit consisted of eight 500
second exposures taken at two array dither positions (``A''
and ``B''; \S \ref{sec:innovations}) in two ABBA sequences.  A 500
second exposure consists of a sequence of 47 detector readouts,
performed in intervals of 10.7 seconds, which generates a
sample-up-the-ramp data-cube (see \S \ref{sec:layout}).
This $\sim$67 minute exposure sequence plus two dark exposures taken
during the change of the plugplate cartridges yields a typical visit
length of 75 minutes.  A plugplate was typically revisited
on multiple nights to build up the required $S/N$ according to the
cadence rules described in \S\ref{sec:binaries}. 

To operate usefully in less than ideal weather conditions and to
take full advantage of extra pockets of observing time, guidelines
had to be established for maximizing the usefulness of ``non-standard
visits''.  Therefore, a minimum data quality to count a visit as
successful was set at
at least one AB dither pair with each 500 second exposure having a
$S/N \ge 10$, the minimum needed to derive the stellar radial
velocity at the required survey precision.\footnote{For reference,
the typical visit of eight 500 second exposures  for a ``3-hour''
plate reached a $S/N \sim 63$ for the faintest stars.} To aid in
the assessment of exposure quality, the observers had access to
``quick look'' reductions (simplified versions of the data reduction
pipeline; \S \ref{sec:reductions}-\ref{sec:1dred}) of the data in
near real time that produced plots of accumulated $S/N$ as a function
of magnitude.  Over the course of a night, the available APOGEE
time would be divided into standard field visits, with any additional
observing time allocated to either gathering extra $S/N$ on a
particular plate or creating a ``short visit'' with a new plate,
at the discretion of the observing staff to maximize observing
efficiency.

Because telescope guiding is done at optical wavelengths, APOGEE
plates were observed with the guiding software making refraction
corrections to keep 1.6~$\mu$m light in the fibers.
In the case of fields observed jointly with MARVELS 
the guiding wavelength was set to a compromise wavelength of
1.1~$\mu$m.

Stability of the APOGEE instrument limits the amount of
calibration needed on a nightly basis.  At the beginning and
end of each night with potential APOGEE observations the gang
connector is connected to the calibration box to collect a standard
calibration sequence that includes long dark frames as well as
exposures of the tungsten halogen, ThArNe and UNe lamps at both
dither positions (\S \ref{sec:layout}).  At the end of the
night we also take a set of internal flat fields. In addition,
once each night 4 $\times$ ABBA exposures are taken with all fibers
on sky; the resulting airglow spectra are used for monitoring the
LSF and PSF of the instrument.

A full observing night can generate $\sim$100 
GB of data, which are then compressed and transferred from the
mountain to the {\it Science Archive Server} (SAS) (see
\S\ref{sec:access}), where they are stored in disk.  These raw
data consist of large data cubes containing all the 47 readouts
making up every single 500 second exposure.  The subsequent processing
and reduction of these data are described in \S\ref{sec:datahandle}.

\begin{figure}[h]
\includegraphics[width=0.48\textwidth]{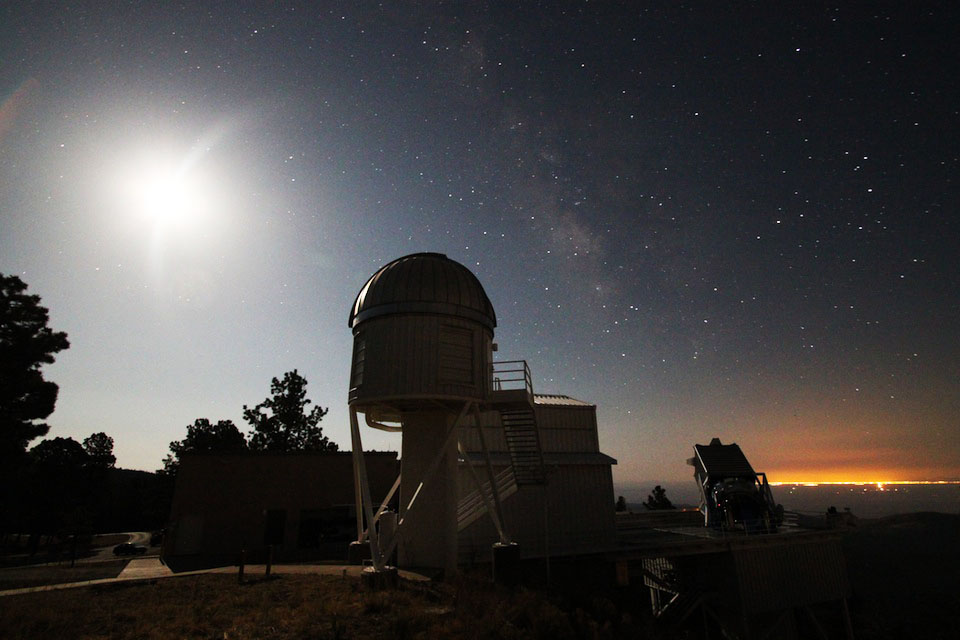}
\caption{\footnotesize Photo of Apache Point Observatory during the APOGEE first light
observing run showing the Sloan Telescope (right of center) pushed to new observing regimes ---
pointed to the Galactic center at extreme airmass, near the full moon, and near the 
light pollution from El Paso (which affects near-infrared bright time observations less
than dark time optical observations). The constellations of Sagittarius and
Scorpio are obvious on the right hand side of the image.  (Photo by S. R.
Majewski.)
}
\label{fig:firstlight}
\end{figure}

\subsection{Observing Constraints, Strategies and Scheduling}
\label{sec:scheduling}

From 2011Q2 to 2014Q2  APOGEE (and initially, MARVELS in
parallel) operated during all bright time
(lunar phase $< 39\%$), as well as all ``grey" time (lunar phase
39-56\%) for LSTs when the North Galactic Cap was not visible.
APOGEE observations pushed the Sloan Telescope to several new
observing regimes and limits --- e.g., with respect to lunar phase,
airmass, twilight, cadencing and 
sharing of the focal
plane by two different instruments (Fig.~\ref{fig:firstlight}).
With a number of observing constraints different than those required
by the optical programs, integrating the APOGEE program into SDSS
operations added new layers of complexity to telescope scheduling
and plugplate cartridge organization, especially on nights shared
between all three operating surveys (BOSS, MARVELS and APOGEE).
Within the APOGEE portions of nights, internal scheduling software
was developed to organize the nightly observing for efficiency, and
to account for APOGEE observing constraints, as well as those for
MARVELS during joint operations.  These APOGEE scheduling constraints
included:

\begin{itemize} \item {\it Moon avoidance:} Observations were not allowed
within $15^{\circ}$ of the moon ($30^{\circ}$ for MARVELS shared
observations).  However, because the ecliptic passes directly through
the Galactic bulge, this limit was loosened to $10^{\circ}$ for bulge
observations; without this adjustment the amount of potential bulge observations
would have been reduced by 50\%.

\item {\it Airmass limits:} The central regions of the Galaxy,
containing highly prized APOGEE targets, transit at very high airmass
at APO.  Compared to the optical, near infrared observations
benefit from reduced differential refraction and atmospheric
extinction, which made it possible to undertake the
desired extreme airmass
observations.  One important limitation, however, is the still 
significant {\it differential}
atmospheric refraction at low elevation, which forced the adoption
of more limited drilled areas (1-2 degrees) on the plugplates.
Despite the smaller angular coverage it was easy to fill all the
science fibers in these fields, due to the high stellar density of
the central regions of the Galaxy.  Those fortunate advantages made
it possible for APOGEE to probe the Galactic bulge, the Galactic
center and even further south (to $\delta = -32^{\circ}$).
Hardware limits set the maximum APOGEE airmass ($X$)
to $X$ $<$ $3.2$, but a limit of $X$ $<$ $1.7$ was necessary for
MARVELS co-observed plates.  APOGEE utilized the standard $X>1.01$
limit of the Altitude-Azimuth mounted Sloan Telescope.

\item {\it Hour angle:} 
The APOGEE windows of opportunity were set
so that plates had to be observed with no more than 0.5 arc seconds of
differential refraction across the plate.
However, the reduced $H$-band differential refraction also allowed greater
APOGEE flexibility in observing plugplates farther from their nominally drilled 
hour angles than is possible for optical observations. 

\item {\it Plate cadence:}  As discussed in \S \ref{sec:binaries},
the nominal survey plates 
had to be observed over at least three visits each meeting the minimum $S/N$ 
per visit requirement (\S \ref{sec:observing})
with separations of at least 3 days between the two closest observations 
and at least 25 days between the first and last observation.

\end{itemize}

Survey plate scheduling was done by a module originally designed to optimize
cadence observations for the MARVELS survey that was later adapted to
account for both cadence and $S/N$ constraints of the APOGEE survey.
Beyond accounting for the above constraints, the scheduling software invoked 
several additional rules to optimize efficiency.

For example, bulge plates and other plates with limited observability
windows were given highest priority.  Other plates were given
relative priorities that accounted for their individual cadence
histories and net accumulated $S/N$.
Special attention was needed for
scheduling of ``non-standard"
visits to take advantage of occasional extra pockets of observing
time.  For example, 
standard visits for
the eight bright time cartridges were insufficient to fill the available time
on long winter nights; 
in this case, longer than standard visits could be applied to, e.g., (a) halo
plates that have been designed with fainter than main survey stars
(see \S \ref{sec:maintargets}), (b) plates that --- due to poor weather
or prematurely ended previous visits --- were behind on $S/N$ accumulation
despite satisfying cadence constraints, or (c) plates that could,
conversely, be ``pre-loaded" with extra $S/N$ allowing useful, but
shorter than standard visits on other 
(e.g., shorter) nights.  
In the interest of steady progress on the completion of fields,
another, albeit more
loosely followed, scheduling strategy was that the full set of
observations for nominal, three-visit plugplates, if at all possible,
not stretch beyond one observing season.

On long nights, when the full eight fiber plugplate setups
could be observed, APOGEE was able to record spectra for 1840 target stars,
along with 280 hot telluric star calibrators and 280 sky fibers (\S \ref{sec:calfibers}).

\subsection{Special Observing Strategies and Campaigns}
\label{sec:specialobs}

\subsubsection{Twilight Observing}
\label{sec:twilight}

Another advantage of near-infrared over optical spectroscopy is the
ability to work deeper into twilight.  By the second year of the
APOGEE campaign it became clear that above average poor weather at
certain LSTs was going to make it challenging to complete the planned
observations of the bulge and {\it Kepler} field plates. 
In view of this situation, the BOSS team and SDSS observing staff
graciously agreed
to allow APOGEE to
make use of the small windows of the dark and grey time morning twilight
not useful for BOSS observing.  Fortunately, the LSTs of greatest
need could be serviced in spring and summer, so this special twilight
observing was conducted only between the vernal and autumnal
equinoctes to limit the impact on the observers.  BOSS observing
is limited to 15$^{\circ}$ twilight, but in cases where a standard
BOSS observation concluded by 20$^{\circ}$ twilight there was
insufficient time for a new BOSS observation, but enough time for
APOGEE to observe a plate to 8$^{\circ}$ twilight. This was sufficient to
collect, at minimum, an AB dithered pair of exposures and as much as
an ABBAAB sequence.  These short visits --- useful for accumulating
$S/N$ for the 1-hour bulge and {\it Kepler} field plates, as well
as cadence visits for main survey plates that compete for the same
LSTs --- were found to be essential to the completion of the APOGEE
survey plan.

\subsubsection{Year 3 and Dark Time Campaign}
\label{sec:year3}

In the final half-year of SDSS-III it became evident that the BOSS survey
was ahead of schedule and likely to finish early; thus some dark time was made
available to the collaboration for additional projects.
At this point, though on pace to reach the required number of stars, APOGEE was
significantly behind schedule on completing plates in the inner Galaxy and Kepler regions,
due to atypically poor summer weather.\footnote{
APOGEE remained on pace to complete
the 100,000
star goal primarily because it was ahead of schedule in the Galactic anticenter region
due to atypically good {\it winter} weather.  As discussed in \S 
\ref{sec:fieldselectionevolution},
this enabled a significant expansion of the anticenter program.}
Through access to significant portions of that dark time, not only did the main
APOGEE survey manage to complete virtually its entire field plan, but a number 
of APOGEE bulge plates that had only lower quality commissioning observations
could be reobserved for survey quality data (Fig. \ref{fig:field_plan}c).  In addition, two new APOGEE ancillary
science programs\footnote{``Infrared Spectroscopy of Young Nebulous Clusters (INSYNC)'' 
ONC clusters'' (e.g., \citealt{Cottaar14})
and  ``Probing Binarity, Elemental Abundances, and False
Positives Among the Kepler Planet Hosts'' \citep[e.g.,][]{Fleming15}}
 were added beyond those described in \citet{Zasowski13}.

\subsubsection{Bright Standard Star Calibration}
\label{sec:standards}

Calibration of the APOGEE velocity, stellar parameter, and chemical
abundance data relied, to a large extent, on data obtained
from special targeting of numerous open and globular clusters as
well as the asteroseismology targets in the {\it Kepler}
and {\it CoRoT} fields (\S \ref{sec:surveydesign}).  
In addition a large range of bright, previously well-studied ``standard
stars'' were also observed for calibration purposes.   A compiled 
target catalog of 
such stars 
included an
assortment of stellar types meant to calibrate specific regions of
stellar parameter space.  Especially useful were
stars not well represented in clusters (e.g., carbon stars) and
subsamples designed to address specific issues, such as, e.g., S
class stars, which aided
the search for lines due to neutron
capture species in the APOGEE wavelength window.  
Two targets
critical to calibration efforts were the well-studied metal-deficient
K giant ``reference'' standard Arcturus (e.g., \citealt{Hinkle95})
as well as the asteroid Vesta 
(providing a reference solar spectrum).  

To obtain spectra of these bright sources
is a challenge for the Sloan 2.5-m telescope
and not practical through drilling and observing specialized
plugplates.  Initially these spectra were obtained using an observing
script (``Any Star Down Any Fiber'' or ``ASDAF'') that enabled the
observers to put the bright standards down an APOGEE fiber on any
currently loaded plugplate, a procedure implemented only during
moderately cloudy nights when main survey observing was not practical.
Subsequently, this rather labor-intensive strategy was
replaced by use of
New Mexico State University's
(NMSU's) 1-m telescope, to which a fiber optic link was run that
can be connected to the APOGEE long fibers.  Through a time-sharing
agreement with NMSU, a fraction of the dark time was reserved for
1-m bright star calibration observations with APOGEE, made even
more efficient by it being robotized (the 1-m program is described
further in \citealt{Holtzman15}).

\subsection{Survey Timeline}
\label{sec:surveytimeline}

The APOGEE program consists of two distinct observing campaigns ---
``commissioning" (May-July 2011) and ``survey" (August 2011-July 2014) ---
divided by the change in spectrograph optical configuration during the
shutdown in Summer 2011 (see \S \ref{sec:timeline}).
``Commissioning" observations consisted primarily of 1-visit and 3-visit
fields to test instrument performance, calibration, and limitations.
The ``survey" observations were conducted over the originally intended 
three year APOGEE campaign from August 2011 to July 2014 and produced
acceptable quality survey data during 520 days spanning over 1900 individual
field visits.
The entire three year survey campaign was conducted uninterrupted, with the 
instrument continuously sealed and cold in the same optical state to provide
an extremely uniform data set.

\section{Data Handling and Processing}
\label{sec:handling}

The software chain used to convert the raw APOGEE data to final
data products is divided into three primary programs: (1) real or
near-real time codes to pre-process, bundle and archive the raw
data (\S\ref{sec:datahandle}); (2) the data reduction pipeline,
which converts the collected data cubes into extracted, 1-dimensional,
calibrated spectra, and, along the way, derives radial velocity
information (\S\S\ref{sec:reductions}, \ref{sec:1dred}, and
\ref{sec:rvs}); and (3) the APOGEE Stellar Parameters and Chemical
Abundances Pipeline (ASPCAP), which aims at achieving the unprecedented
feat of determining stellar parameters and up to 15 elemental abundances
through the automatic analysis of APOGEE's high-resolution $H$-band
spectra (\S\ref{sec:stelpar}).  Steps (1) and (2) are performed
by the {\tt apred} software \citep{Nidever15},
whereas step (3) is performed by {\tt ASPCAP} \citep{GarciaPerez15}.

Because of the APOGEE observing strategy, the subsequent reduction
routines generate a number of intermediate files.  For the following
discussion, a few terms need a clear definition.  Each final {\it
combined spectrum} (1D) consists of the combination of a number
(NVISITS) of {\it visit spectra} (1D).  In turn, each normal visit spectrum
results nominally from the combination of 4 (AB-BA-AB-BA) pairs of {\it dither
spectra} (1D), obtained at two different dither positions (i.e., 8 distinct spectra).
Each 1D dither spectrum is extracted from a bias-subtracted,
flat-field and cosmic-ray corrected {\it 2D array}, which in turn is created 
by pixel-by-pixel fits to   
the numerous detector readouts that constitute the raw
{\it data cubes} (\S \ref{sec:layout}).  Each data cube consists of a
time series of 47 up-the-ramp readouts of all three detectors,
performed every 10.7 seconds along the exposure (\S
\ref{sec:observing}).

\subsection{Basic Reductions: From Data Cubes to 2D Arrays} 
\label{sec:datahandle}

At the end of every observing night, APOGEE data are compressed and
transferred to the SAS (\S \ref{sec:observing}), and all data reduction is done
subsequently off the mountain.
In the following, we briefly describe the steps leading
to the generation of a final APOGEE combined spectrum.  In this
first processing stage each data cube is corrected for standard
detector systematic effects and converted into a 2D array.  Every individual
readout is corrected for bias variations in the detectors and
electronics.  Bias measurements are performed on a combination of
pixels generated by the readout electronics and a set of reference
pixels around the edge of each detector.  Next, a dark frame resulting
from combination of multiple individual exposures is subtracted
from each individual readout.
The 2D arrays are then generated through linear fits to the time
series of SUTR
readouts for each pixel, and the best fitting
slope is multiplied by the exposure time to generate the final pixel
counts.
The process allows for detection, correction, and flagging of pixels
affected by cosmic rays.  Finally, 2D arrays are corrected for
pixel-to-pixel sensitivity variations through division by a normalized
flat field frame.
The output of this reduction step for one visit is eight calibrated 2D
arrays, four for each dither position.

\subsection{From 2D Arrays to 1D Dither Spectra} \label{sec:reductions}

As a next step, spectral extraction and wavelength calibration are performed
on each 2D array.  
Spectra are extracted through modeling of the spatial PSF of all
300 fibers as a function of wavelength in a way that accounts for
the overlapping of the PSFs between adjacent spectra.  The model
is fit to a high $S/N$ flat-field frame obtained immediately after
each science exposure.

Wavelength calibration is the next stage of the reduction, and
as usual, is based on arc lamp exposures.  Because each
fiber occupies a different position in the pseudo-slit, fiber-to-fiber
wavelength scale variations exist, so that individual calibrations
for each fiber are necessary.  The APOGEE spectrograph is stable
enough that a single polynomial relation is adopted for each fiber, with
zero point corrections applied on the basis of measurements of central
wavelengths of airglow lines.
In conformity with previous SDSS standards,
APOGEE adopts vacuum wavelengths.  For details of the adopted
conversion between vacuum and air, see \citet{Nidever15}.

The overall wavelength scale suffers drifts linearly over time, due
to a slowly varying flexure in the instrument optical bench as the
liquid nitrogen tank depletes over time (\S \ref{sec:layout}).
Every time the tank is refilled, the scale undergoes a large ``reset''
shift, which brings it back to the original scale.  These shifts
are measured using a set of bright airglow lines and the wavelength
scale is corrected accordingly.  The accuracy of the resulting
wavelength solution at any given pixel of an APOGEE spectrum is of
order 0.1 pixel or 0.03-0.04~${\rm\AA}$ \citep{Nidever15}.  
The outputs of this reduction stage for one
visit are 8 wavelength-calibrated 1D dither spectra, 4 for each dither
position.  

\subsection{Dither Combination, Sky Subtraction, Telluric Correction, and
Flux Calibration}
\label{sec:1dred}

In the next reduction stage, dither pairs are combined into well sampled 1D
spectra, sky subtraction is performed, and the signature of telluric
absorption is removed.

The shift 
between the spectra in each dither pair is determined to high
accuracy through cross correlation of the two spectra.  Before
combination, each dither spectrum is subject to sky subtraction,
which is critical due to the presence of strong OH emission lines
and a faint continuum, which is stronger in the presence of clouds
and moonlight.  
The contribution of sky
background to the spectrum of any science fiber is determined through
interpolation of the spectra of the four closest fibers from among the 35
sky fibers 
distributed across
the APOGEE FOV (\S \ref{sec:calfibers}).  
Because of fiber to fiber LSF differences, subtraction of sky lines is not
perfect, and can result in the presence of significant residuals in pixels
situated at or near the positions of very strong lines that renders
these pixels useless for science.  While future improvements in the
reduction pipeline may ameliorate the situation, the $S/N$ in those
pixels will nevertheless be substantially deteriorated due to high
Poisson noise.

Telluric line absorption in the APOGEE spectral region 
due to the rovibrational transitions of the H$_2$O, CO$_2$, and CH$_4$ molecules
are removed through the fitting of telluric absorption
models to observations of the 35 telluric standards distributed
across the field (\S \ref{sec:calfibers}).
For each telluric standard, synthetic telluric spectra based on
model atmospheres by \cite{Clough05} are fitted to the full family of
absorption lines from each molecule separately to 
generate scaling factors
to the model spectrum of each
molecule at the position of each telluric calibration fiber.
Polynomial surfaces are then fitted to describe the spatial
variation of the scaling factors, and the correct scaled model
is determined for each science fiber through interpolation within those
surfaces.  For each science fiber, models are then convolved with
fiber-specific LSFs, and divided into the science spectrum

Although the above telluric correction method works well, it has
shortcomings related to errors in the wavelength solution, and uncertainties
in both the telluric absorption model and the adopted LSFs.
Because a large fraction of APOGEE pixels are affected by telluric
absorption, improvements in telluric correction are a high priority
for future pipeline improvements.

Each sky-subtracted, telluric-corrected pair of dither spectra are
then combined into a single better-sampled spectrum, using the
shifts determined as described above.  Each of these resulting
spectra are then coadded to generate a single visit spectrum.

Flux calibration consists of two steps.  First, an approximate relative flux
calibration is applied to dither spectra to remove the spectral signature of 
instrumental response; this response function was determined
through observation of the black body spectrum from a calibration
source (\S \ref{sec:layout}).  Later on, after dither spectra are
combined to generate visit spectra, the latter are scaled to match
the object's cataloged $H$-band magnitude.  
Because the spectra are later reshaped through
polynomial fits to the pseudo-continuum prior to performance of
stellar parameter and abundance analysis, flux calibration is not
a critical aspect of data processing.

\subsection{Radial Velocities and Generation of Combined Spectrum}
\label{sec:rvs}

Radial velocities (RVs) are one of APOGEE's key data products.
There are two main steps related to the RV determination within
APOGEE.  One step determines relative RVs
between different visits, and the other fixes these measures to an
absolute scale.

In both steps, RVs are determined via a cross correlation between
the object spectrum and a particular template.  Observed and template
spectra are initially both in a log-linear 
wavelength scale, so
that a Doppler correction can be performed by shifting all
pixels by the same value.  Before cross correlation, bad pixels
are flagged and the 
pseudo-continuum is normalized through
a low-order polynomial fit to the spectrum of each detector separately.
A Gaussian fit is performed to the cross-correlation distribution
and the position of the peak and its error are converted into a
velocity shift and uncertainty.

Visit RVs are determined through an iterative process via cross
correlation with the 
combined spectrum.  
Initial relative RVs, obtained from cross correlation with the
highest $S/N$ visit spectrum, are used to bring all visit spectra to
a common velocity scale, and making possible the production of an
initial combined spectrum.  The process is then iterated by adopting
the most recently created combined spectrum as a template.

Absolute RVs are obtained through cross-correlation of the combined (and
visit) spectra with synthetic spectra from an ``RV mini-grid'', which
is a subset of the APOGEE spectral grid (\S\ref{sec:grid}),
and consists of 538 spectra over a wide range of stellar parameters
and chemical compositions.  The numbers resulting from this cross
correlation are further adjusted by the
barycentric correction, to
produce heliocentric RVs.

\subsection{Stellar Atmospheric Parameters and Elemental Abundances}
\label{sec:stelpar}

Elemental abundances are another primary data product of the APOGEE
survey.  Stellar parameters --- effective temperature ($T_{\rm eff}$),
surface gravity ($\log g$), metallicity ([M/H]), and microturbulence
($\xi_t$) --- are also 
necessary stepping stones
towards elemental abundances and spectroscopic parallaxes.  
An understanding of the possible systematic effects on the derived
elemental abundances and distances inferred from APOGEE spectra requires a
good grasp of the procedures followed for the derivation of atmospheric
parameters and metallicities.  In this section, a brief
description of those procedures is provided, but the reader is
referred to \citet{GarciaPerez15} for 
details.

The APOGEE Stellar Parameters and Chemical Abundances Pipeline
(ASPCAP) implements a two-step process: first, the determination of
stellar parameters from a fit of the entire APOGEE spectrum to model
spectra, and second,
adoption of these parameters as inputs for a fit to
small windows of the
spectrum 
containing spectral features associated with each particular
element to derive its abundance.  In the following subsections
we describe each of the main ASPCAP processing steps.

\subsubsection{Grid of Synthetic Spectra} 
\label{sec:grid}

Stellar parameters are obtained through determination of the best
fitting synthetic spectrum from across an extensive grid spanning
six stellar atmospheric parameter dimensions ($T_{\rm eff}$, $\log
g$, [M/H], [$\alpha$/M], [C/M], and [N/M])
by $\chi^2$ minimization (\S \ref{sec:ASPCAP}).  The accuracy of
the results is fundamentally dependent on the fidelity with which
spectra from the synthetic grid reproduce real stellar spectra.  We
briefly describe the main ingredients entering the calculation of
this spectral grid, and refer the reader to \citet{Zamora15} for
further details.

Synthetic spectra were calculated using the Advanced Spectrum
Synthesis 3D Tool (ASS$\varepsilon$T) code \citep{Koesterke09},
adopting 1D model atmospheres calculated in local thermodynamic
equilibrium (LTE) by \citet{Meszaros12} and a line list customized
for the analysis of APOGEE spectra (\citealt{Shetrone15}; Appendix
\ref{sec:linelists}).  The adopted model atmospheres were calculated
using the ATLAS9 code \citep{Kurucz93}, adopting newly computed
opacity distribution functions as described by  \citet{Meszaros12}
and the solar abundance pattern of \citet{Asplund05}, as well as
variations in the abundances of carbon and $\alpha$ elements.  
Spectra were calculated 
over a range of [M/H], [$\alpha$/M] (where
all $\alpha$ elements are assumed to vary in lockstep), [C/M], and
[N/M].  The chemical compositions adopted matched 
those used in the generation of the model photospheres, except for the
case of nitrogen, whose variation was not seen to affect the
photospheric structure in an important way.

The line list resulted from an initial implementation of the Kurucz
line list, improved by introduction of both theoretical and laboratory
transition probabilities ({\it gf} values) following an exhaustive
critical search of the existing literature, and further supplemented
by laboratory values of key transitions obtained by our collaborators
\citep[e.g.,][]{Wood14} by request (see Appendix \ref{sec:linelists}).
Further refinement of {\it gf} values and damping constants was
achieved through spectral synthesis of the solar and Arcturus spectra
\citep[see][for details]{Shetrone15}, where departures from laboratory
values were capped at no more than twice the nominal uncertainties.

The synthetic spectra are smoothed to the APOGEE resolution ($R$$=$$22,500$)
by convolution with a single, empirically-determined, average APOGEE LSF 
\citep{Nidever15,Holtzman15} and sampled
into a logarithmic scale to match the sampling of the APOGEE data
($\sim 10^4$ wavelengths). 
Synthetic spectra are further normalized
through fitting of a polynomial to the upper envelope of the spectrum,
for comparison with observed spectra treated in the same
way (see below).

Efficient computation would require storage of the entire spectral
grid in memory, which is currently not practical.  Therefore, fluxes
are compressed using Principal Component Analysis (PCA) and it is
the PCA-compressed grid that is compared with the observed spectra
for atmospheric parameter determination.  To expedite calculations
further, the grid is split into two distinct sub-grids, with $T_{\rm
eff}$ spanning  ranges approximating those of GK (3500-6000 K) and
F (5500-8000 K) spectral types \citep[see][]{Zamora15}.

Each synthetic spectrum is characterized by seven parameters, namely
$T_{\rm eff}$, $\log g$, [M/H], [$\alpha$/M], [C/M], [N/M], and
$\xi_t$ (microturbulence).  With multiple nodes in each parameter,
the final 7-dimensional spectral sub-grids consist of about 1.7
million (GK stars) and 1.4 million (F stars) spectra covering the
entire range of expected atmospheric parameters and chemical
compositions.

Abundances of individual elements are defined as follows:
\begin{equation}
[X/H]\, =\, \log_{10}(n_X/n_H)\, -\, \log_{10} (n_X/n_H)_\odot
\end{equation}
where $n_X$ and $n_H$ are the number, per unit volume of the stellar
photosphere, of atoms of element X and hydrogen, respectively.  
The
metallicity [M/H] is defined as an overall scaling of metal abundances for
a solar abundance pattern, while [X/M] is the deviation of element X from
that pattern:
\begin{equation}
[X/M]\, =\, [X/H]\, -\, [M/H]
\end{equation}
Because the search for the best fitting spectrum within a 7-D space
is considerably slow at present, the library dimensionality has been reduced to 6 
(thereby reducing the overall size of the libraries by a factor of 5)
by constraining microturbulent velocities through the adoption of
a relation with surface gravity (see details in \citealt{Holtzman15} and \citealt{GarciaPerez15}).  
For $T_{\rm eff} > 8000\,K$, where molecular lines
are entirely absent, the grid is described by only three parameters,
$T_{\rm eff}$, $\log g$, and [M/H].

\subsubsection{Pre-processing of Observed Spectra}
\label{sec:preprocessing}

A few additional processing steps are taken to prepare the observed
spectra for comparison with the synthetic grid.  First, to optimize
the fitting process and increase the robustness of the $\chi^2$
statistic, pixels affected by cosmic rays, saturation, cosmetic
problems, or strong airglow lines are flagged and 
ignored during spectral normalization and $\chi^2$ minimization.  Moreover,
to account for small systematic errors in spectral calibration, we
set a minimum flux error 
of 0.5 percent for all remaining pixels.

Next, to minimize uncertainties due to interstellar
reddening, atmospheric extinction, and errors in relative fluxing,
spectra are flattened and normalized through the fit of a polynomial
to their upper flux envelopes.  Fits are performed through a
$\sigma$-clipping algorithm to the spectra on each of the
three detector arrays independently.  An identical normalization is performed
on the grid of synthetic spectra, using the same spectral regions
with the same $\sigma$-clipping and polynomial form.

This process does not necessarily produce a normalization to the
true stellar continuum, but rather to a ``pseudo-continuum''.
This is because, at the APOGEE resolution, it is impossible to
resolve spectral regions that are unaffected by any line opacity
(i.e., true continuum regions) in the spectra of the coolest and/or
most metal-rich stars.  This fact alone largely dictates our
methodological choice for normalized fluxes over equivalent
widths as APOGEE's fundamental observable 
for atmospheric parameter and elemental abundance determination 
through comparison with model predictions.
This choice is predicated on the notion that normalized fluxes are
less strongly affected by continuum placement uncertainties than
equivalent widths, especially
if synthetic and observed spectra are normalized identically.

\subsubsection{Stellar Atmospheric Parameter and Abundance Determinations}
\label{sec:ASPCAP}

Stellar atmospheric parameters and the relative abundances of 
C, N and the $\alpha$ elements are determined by the 
FORTRAN90 code FERRE
\citep{AllendePrieto06}, which searches within the 6-D grid of
synthetic spectra for the best match to each observed APOGEE spectrum.
The code uses a $\chi^2$ criterion as the merit function, and the
searching method is based on the Nelder-Mead algorithm
\citep{Nelder65}.  The search is run 12 times starting from
different grid locations: three positions in $T_{\rm eff}$ and two each in
$\log g$ and [M/H].  Two points symmetrically located around the grid center are adopted
for $\log g$, [M/H], and $T_{\rm eff}$, whereas for the latter a starting point at the
central grid value is also adopted.  A single (solar) starting value is
adopted for [C/M] [N/M], and [$\alpha$/M].
The code returns the best matching spectrum, obtained through
cubic B\'ezier interpolation
within the grid, as well as the parameters
associated with that spectrum ($T_{\rm eff}$, $\log g$, [M/H], and
[C/M], [N/M] and [$\alpha$/M] abundance ratios), the covariance
matrix of these parameters, and the $\chi^2$ value for the best-matching
spectrum.

The analysis described above delivers an overall metallicity
and a mean $\alpha$-element relative abundance, as well as
relative abundances of carbon, and nitrogen.  Based on fits of the
entire spectrum, these numbers can only be considered as preliminary
values.  A subsequent, more refined analysis takes place that 
directly and more accurately evaluates the abundances 
of carbon and nitrogen, and also derives the abundances for all remaining 
target elements 
(O, Na, Mg, Al, Si, S, K, Ca, Ti, V, Mn, Fe, and Ni). 
This is accomplished by re-running FERRE on each spectrum,
this time restricting the search to a more limited area of parameter
space, where $T_{\rm eff}$, $\log g$, and $\xi_t$ are held fixed.
For each element, spectral windows are defined that maximize the
sensitivity to that particular element's abundance, while minimizing
sensitivity to the abundances of all other elements \citep[see][for
details]{Smith13,Cunha15}.
After the first iteration of FERRE determines the stellar
parameters, a series of new FERRE runs are performed, one for each
element. In each of them, all the dimensions remain fixed except
that used for the abundance of the element of interest fitting only
the specific spectral windows for that element.  Each pixel is
weighted on the basis of sensitivity to the elemental
abundance being fitted, and also according to the quality of the
fit of the Arcturus spectrum at that pixel 
by APOGEE synthetic models \citep{Shetrone15}.  

Elemental abundances are thus obtained by searching for the best
fit of each spectral window.  For any given element, the search is
performed in only 1-D, where the only varying parameter is the
abundance of that element while the stellar parameters and all other
elemental abundances are held fixed.  For further details we refer
the reader to \citet{GarciaPerez15}.

\section{Achieved Performance}
\label{sec:performance}

In this section, we briefly examine the performance of the APOGEE
survey, understood as the result of the combination of instrument,
survey strategy, operations, and data processing and analysis tools
and procedures, contrasting it with the requirements described in
\S\ref{sec:topleveltechnical}.

\subsection{Final Sample Statistics and Galactic Distributions}
\label{sec:surveystats}

Upon concluding three years of operation, the APOGEE survey obtained
over a half million spectra of 163,278 stars.  Of these,
12,140 were obtained during commissioning but were 
never reobserved; due to the issues discussed in \S\ref{sec:timeline}, these 
commissioning data do not meet the survey science requirements.  Therefore, 
the total number of {\it survey quality} targets
delivered in DR12 is 151,138.  Of these, 14,692 are telluric
standards, which leaves a net total of {\it 136,446 survey quality
science targets.}
Thus, APOGEE exceeded by more than 35\% the original technical
requirement on sample size.
Such a substantial increase over the required performance was
achieved due to the combination of factors described in \S
\ref{sec:fieldselectionevolution} and \S \ref{sec:specialobs}.

Determining the sample breakdown according to Galactic
component is not simple, because many of the fields present a substantial
overlap of components, particularly those fields below
$|b| = 20^\circ$.
Nevertheless, we provide below a simple
breakdown of the sampled survey stars by rough field type, with the 
caveat that of course not all targets in a field belong to the Galactic 
component defining these types, which, for simplicity, we categorize 
broadly by Galactic coordinates: 
\begin{itemize}
\item 13,473 stars in bulge fields ($|b|<16^\circ$, $-10^\circ<l<11^\circ$);
\item 54,988 stars in halo fields ($|b|>16^\circ$);
\item 82,677 stars in disk fields ($|b|<16^\circ$, $11^\circ<l<350^\circ$).
\end{itemize}
The halo numbers are inflated by inclusion of fields targeting
nearby stars at relatively large Galactic latitudes (some examples are the
Kepler fields and Ancillary Science fields focused on nearby clusters or
associations).  Accounting for those, the total number of halo field
targets drops 
to fewer than 40,000.  

Along these lines, some specialized target classes of particular interest include:
\begin{itemize}
\item 12,443 stars in {\it Kepler}/{\it CoRoT} fields;
\item 2,035 stars in Sagittarius dSph core fields;
\item 3,782 stars in fields in the direction of other known halo
substructure, including streams associated with the Sagittarius dSph;
\item 7,291 stars in fields placed on suspected halo overdensities from the
Grid Giant Star Survey \citep[GGSS,][]{Bizyaev06,Majewski12};
\item 8,112 stars in star cluster fields (fields specifically targeting open or 
globular clusters, but not counting disk fields in which open clusters 
serendipitously were observed);
\item 12,115 objects in Ancillary Science fields;
\item 880 bright stars observed with the NMSU 1-m telescope + the APOGEE spectrograph. 
\end{itemize}

\begin{figure}[h]
\includegraphics[width=0.50\textwidth]{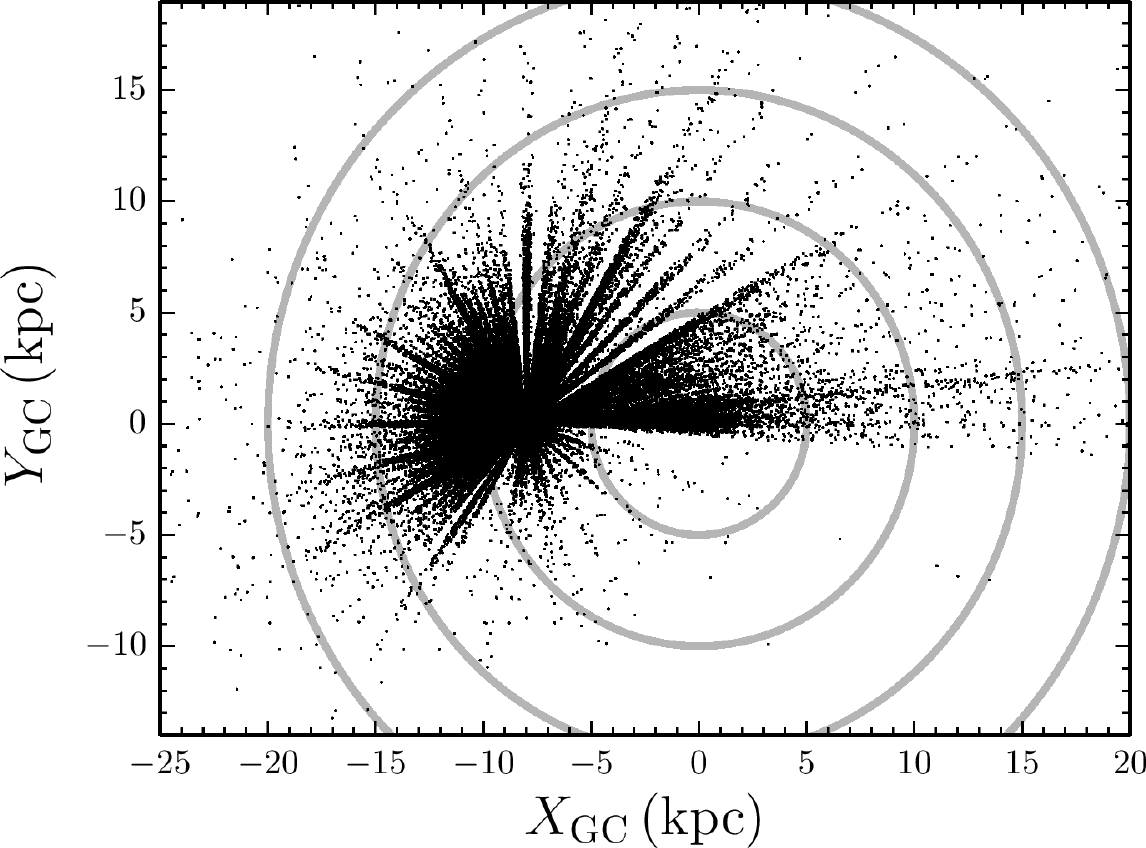}
\caption{\footnotesize 
The computed Galactic distribution of APOGEE targets as projected
on the Galactic plane.  Comparison
to Figure \ref{fig:XYApogeePlot} shows that the anticipated spatial coverage of
the Milky Way has been achieved.
}
\label{fig:finalXYdistribution}
\end{figure}

\begin{figure}[h]
\includegraphics[width=0.50\textwidth]{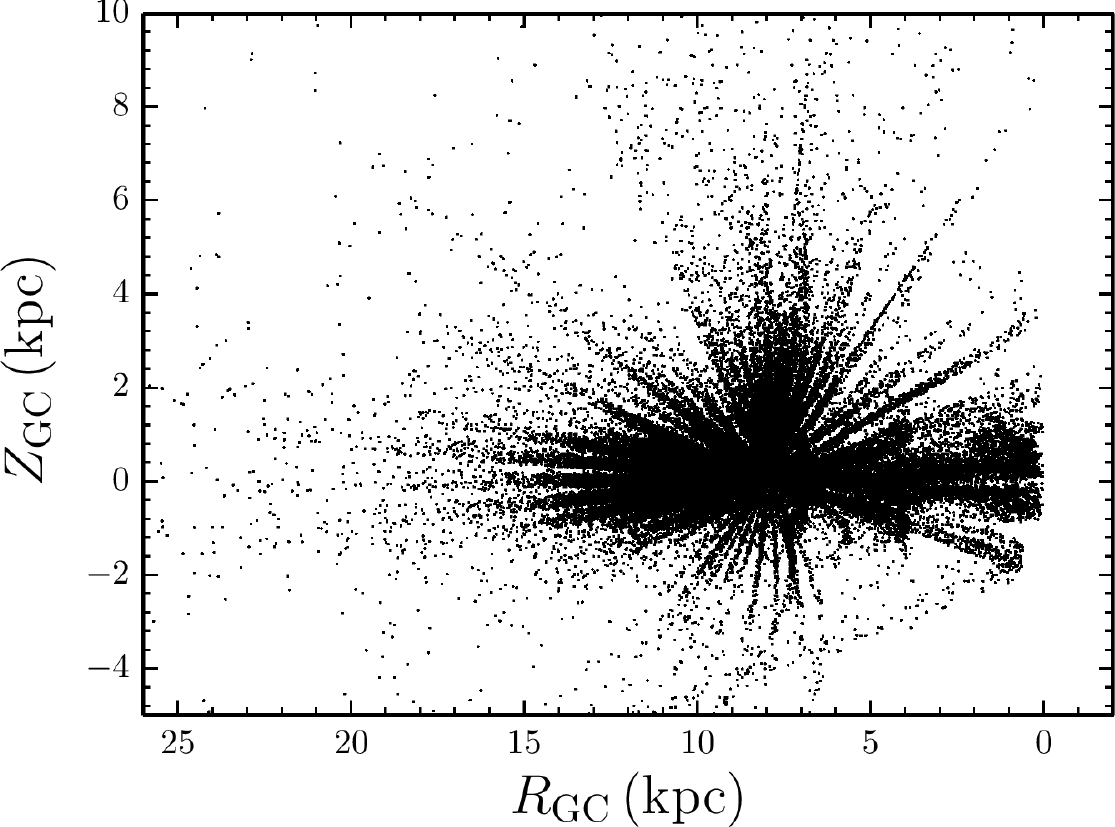}
\caption{\footnotesize Same as Fig. \ref{fig:finalXYdistribution}
for the computed azimuthally-averaged $R_{GC}$-$Z_{GC}$ distribution
of main APOGEE survey targets. Compare to pre-survey prediction in
Fig.\ref{fig:RgZApogeePlot}.}
\label{fig:finalRZdistribution}
\end{figure}

Figures
\ref{fig:finalXYdistribution} and \ref{fig:finalRZdistribution}
show the computed\footnote{Distances are calculated using the method
of \citet{Santiago15a} applied to stars in DR12 with signal-to-noise
ratio larger than 70, a temperature range of $3500\,\mathrm{K}
< T_\mathrm{eff} < 5500\,\mathrm{K}$, positive extinctions with
$A_K < 3$, good 2MASS photometry, ASPCAP analysis $\chi^2 < 50$,
and metallicity errors less than 0.3 dex. The distance
inference uses PARSEC isochrones \citep{Bressan12a} applied to the
spectro-photometric data from APOGEE and 2MASS with the overall
metallicity of each star determined as the $[\mathrm{Fe/H}]$ measured
from iron lines plus the overall $[\alpha/\mathrm{Fe}]$ abundance
determined from the synthetic fit to the whole APOGEE spectrum (see
\citealt{Santiago15a} for further details).}  Galactic spatial 
distributions for the main APOGEE survey targets and
demonstrate that the targeting plan as implemented achieved its
general goals for Milky Way coverage (compare to Figs.
\ref{fig:XYApogeePlot} and \ref{fig:RgZApogeePlot}).  A full
description of the data is given in \citet{Holtzman15}.

\begin{figure}[h]
\vspace{-7mm}
\includegraphics[width=0.50\textwidth]{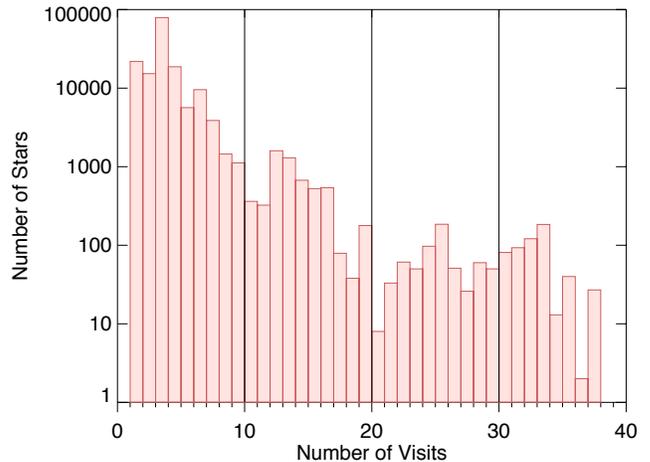}
\includegraphics[width=0.50\textwidth]{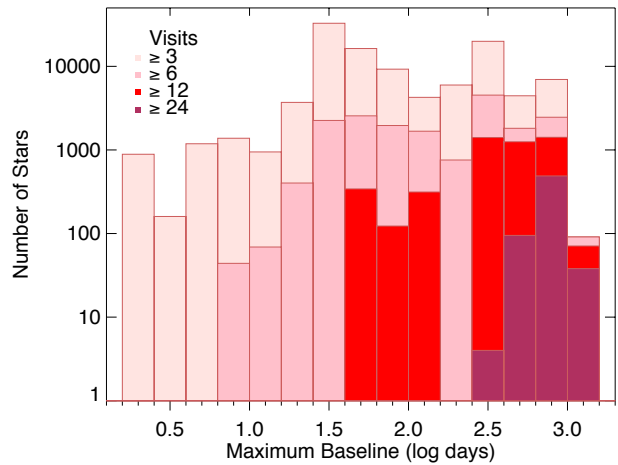}
\caption{\footnotesize 
({\it Top}) Distribution of the number of stars having a given number of visits over the
course of the survey.  
({\it Bottom}) Maximum time baseline for stars having at least 3, 6, 12, or 24 visits.
}
\label{fig:visits}
\end{figure}

A unique aspect of the APOGEE survey 
is that the majority of the stars are visited multiple times.
Figure \ref{fig:visits}a shows the distribution of numbers of stars having any particular
number of visits over the course of the survey.  
As may be seen, the vast majority of stars have three visits, but a significant 
tail of stars have been visited many times more.  For example, of order 1,000
stars have more than two dozen visits.  In many cases, these stars were observed
for over a year, and some as long as three years (Figure \ref{fig:visits}b).
Such time series
data enable a variety of scientific applications,
including the search for stellar companions
(\S \ref{sec:RVperformance}) and stars with spectral variability
(\S \ref{sec:timeseriesspec}).

\subsection{Signal-to-Noise Ratio}
\label{sec:perf_specpar}

\begin{figure}[h]
\vspace{-60mm}
\includegraphics[width=0.60\textwidth]{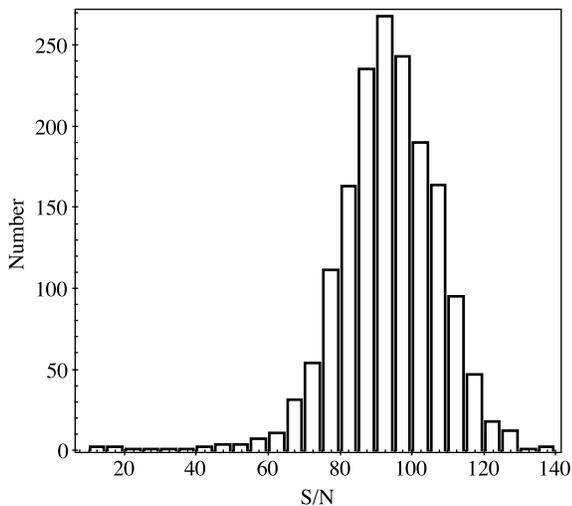}
\caption{\footnotesize Distribution of $S/N$ for stars with $12.15 < H < 12.20$
for stars on 3-visit plugplates.  
}
\label{fig:snhistogram}
\end{figure}

Achievement of survey requirements in depth would only be possible
if the instrument throughput met the original specifications of
$S/N$ = 100 pixel$^{-1}$, at the specified resolution and sampling,
at $H=12.2$ in 3 hours of integration time, which translates into
an expected $S/N = 105$ for a combination of three visits
with eight exposures of 500 s duration each.
Overall survey performance is, of course, also impacted by observing
conditions (transparency, seeing) and procedures adopted to track
and ensure accumulated signal over multiple visits.  Delivered
performance in this area is demonstrated by
Figure~\ref{fig:snhistogram}, which shows the distribution of $S/N$
measured in 3-visit spectra of all stars (main science and ancillary
targets --- $\sim$ 1,600 stars) in a narrow range of (undereddened)
magnitudes (12.15 $<$ $H$ $<$ 12.20) at the 3-visit plate limit of
$H=12.2$. The distribution is nearly Gaussian (despite the
long tail towards very low $S/N$), with a peak at $S/N$ $\sim$ 93.
The dispersion in $S/N$ is about 30 (FWHM), which
is consistent with a $\pm$ 0.2 mag (1$\sigma$) dispersion in flux.
Some of this spread can be ascribed to photometric uncertainties in the 2MASS
catalog ($\sigma_H\sim$~0.02 mag at $H$=12.2)\footnote{
www.ipac.caltech.edu/2mass/releases/allsky/doc/figures/secii2f9.gif}, 
but most of it
reflects the realities of accommodating variable weather conditions
within a semi-rigid observing plan structured on gathering set
numbers of paired 500 second standard exposures for each plugplate
(nominally a dozen pairs for each 3-visit plugplate). 
Assuming that the whole of the scatter is due to weather and seeing
variations, one finds that for the best observing conditions, 
represented by those observations more than
1$\sigma$ (2$\sigma$) above the median of the distribution,
the $S/N$ achieved was about 106
(118), indicating that the spectrograph delivered, and probably
exceeded, the throughput needed to meet the
$S/N$ requirement established at the survey outset.

\subsection{Derived Parameters}
\label{sec:perf_stelpar}

More important than achieving a nominal $S/N$ requirement is the ability
to achieve the desired uncertainties on higher level survey data products,
and in this regard APOGEE has generally succeeded.
A good estimate of RV precision can be obtained from analysis of repeat
measurements of single stars.  In DR12, the RV residuals for stars
with multiple visits and $S/N$ $\geq$ 20
peak at $\sim$ 70 m s$^{-1}$
\citep{Nidever15}.  Uncertainties in the cross correlation
method tend to be smaller for spectra with lots of sharp absorption
lines.  Therefore, RVs are more accurate for cooler and/or metal-rich
giants than for metal-poor and/or hotter/high surface gravity stars,
whose spectra are characterized by fewer and/or broader absorption
lines, respectively.  In any case, the precision achieved 
surpasses the stated requirement (\S\ref{sec:kinematicalprecision}).
The absolute RV zeropoint of APOGEE is estimated to be 
$\sim$350 $\pm$ 30 m s$^{-1}$, from comparison with data from
other sources for stars in common \citep{Nidever15}.

\begin{figure}[h]
\vspace{-115mm}
\includegraphics[width=0.80\textwidth]{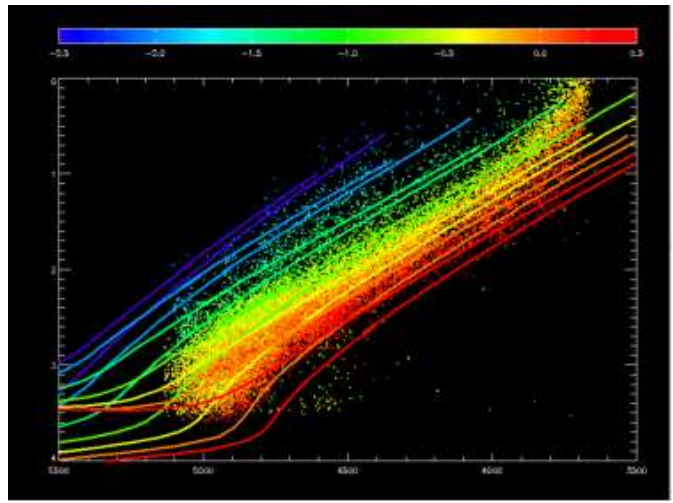}
\caption{\footnotesize 
Final, calibrated, ASCAP-derived 
parameters for main survey stars in the nominal temperature range of the APOGEE main sample.
Colors show the metallicities derived for the stars according to the scale at the top.
The lines show Padova PARSEC
\citep{Bressan12a} isochrones
for 4 Gyr populations at various metallicities, 
against which the calibrated spectral metallicities
may be compared.
}
\label{fig:HR}
\end{figure}

Even though no technical requirements were specified
for the stellar parameters $T_{\rm eff}$ and $\log g$, they drive
the accuracy
of derived chemical
abundances and spectroscopic parallaxes.
Figure~\ref{fig:HR} shows final DR12 values \citep{Holtzman15} 
for APOGEE stars in a ``spectroscopist's HR diagram''.  The final stellar
parameters adopted and shown in this plot result from calibration of the raw
parameters delivered by ASPCAP against photometry and abundance
data from the literature, and asteroseismological surface gravities from
APOKASC \citep{Pinsonneault14},
as described by \cite{Holtzman15}.
There is very good agreement between stellar parameters from APOGEE spectra
and independent theoretical predictions, shown by the isochrones.
Of course, some scatter of the data about the isochrones is expected given the 
dispersion of the sample in age and [$\alpha$/Fe] at any metallicity.  
Moreover, some unresolved issues remain. The most important
example happens for red clump stars, which should be distinctly
separated from the red giant branch.  Stellar evolution models, which
are confirmed independently by accurate gravities from
asteroseismology for a sub-sample of APOGEE stars in the Kepler field
\citep[e.g., see Fig. 18 of][]{Pinsonneault14}, predict 
a narrow range of surface gravities for red clump stars, which are not seen
in APOGEE data, which instead show a correlation between $T_{\rm eff}$ and
$\log g$.
Another unresolved issue 
is the slightly too warm APOGEE temperatures at 
the cool end.  These are areas
where refinements in ASPCAP will bring improvements in the
future. 
For a further discussion of these issues, see \citet{Holtzman15}
and \citet{GarciaPerez15}.

The precision and accuracy of the derived elemental abundances
is reviewed in detail by \citet{Holtzman15}.
To achieve the goals in probing Galactic chemical evolution specified in
\S \ref{sec:sciencegoals}, APOGEE had a specification for an internal
precision in measured chemical abundances of 0.1 dex, and an external 
accuracy of 0.2 dex.
The precision of the actually derived
APOGEE elemental abundances was established by measuring the dispersion of
the measurements at fixed $T_{\rm eff}$ in clusters stars, where the
abundances are expected to be constant.  Precisions were found to be better
than 0.1 dex for all top and medium priority elements, thus formally
meeting the original requirement.  For many elements, precisions are
better than 0.05 dex and for only one low priority element (vanadium) is the
precision worse than 0.1 dex.  

Comparison of APOGEE [Fe/H] with mean literature
abundances of cluster stars based on high resolution optical
spectroscopy, shows that the 0.2 dex external accuracy 
requirement is met everywhere except
perhaps for [Fe/H] lower than $\sim -1.5$.  In view of the presence
of clear systematic differences between ``raw'' derived APOGEE
values via ASPCAP and literature values  
(where APOGEE values are higher by up to 0.2 dex) a calibration to convert
APOGEE [Fe/H] into the literature scale was derived
\citep{Holtzman15}.
For some of the other elements, abundance {\it ratios} --- i.e., [X/Fe],
computed adopting the non-calibrated [Fe/H] --- compare well with the
literature, in spite of some scatter, 
which is likely due to the
heterogeneity and uncertainties in the literature data.  For a few
elements, such as Ca and Ti, trends with $T_{\rm eff}$ are found.
In the absence of a statistically significant homogeneous sample
of optical data from the literature, it is difficult at this stage
to draw firm conclusions regarding the APOGEE accuracy for
some of the elemental abundances.
Nevertheless, while the original requirements are met for most 
of the elements, more work is required to, on one
hand, refine the elemental abundances delivered by ASPCAP and, on
the other, more firmly establish the systematic differences
that will inevitably be present between the APOGEE
abundance scales and those of other large observational programs
(typically performed at optical wavelengths).

\subsection{Science Demonstrations of Data Capabilities}
\label{sec:science_demo}
The achieved performance and capabilities of APOGEE are perhaps
most eloquently demonstrated by examples of how the collected data
may be used in a wide variety of science applications.  In this
Section we briefly discuss a few examples of published and in
preparation research based on APOGEE data, with an eye towards
illustrating the breadth of potential APOGEE science, both within
the confines of Galactic astronomy and beyond.  The order
of presentation of these science examples roughly tracks the degree
of processing of the APOGEE data, as described in \S\ref{sec:handling}
--- i.e., from direct analyses of the spectral character of sources,
to analyses of derived stellar velocities, to explorations of bulk
metallicity and then more detailed chemical abundance patter data,
to even higher level results made possible by inclusion of derived
stellar ages, and concluding with analyses incorporating the former
information for the analysis of star clusters and the interstellar
medium.

\begin{figure}[h]
\vspace{-5mm}
\hspace{-9mm}
\includegraphics[width=0.60\textwidth]{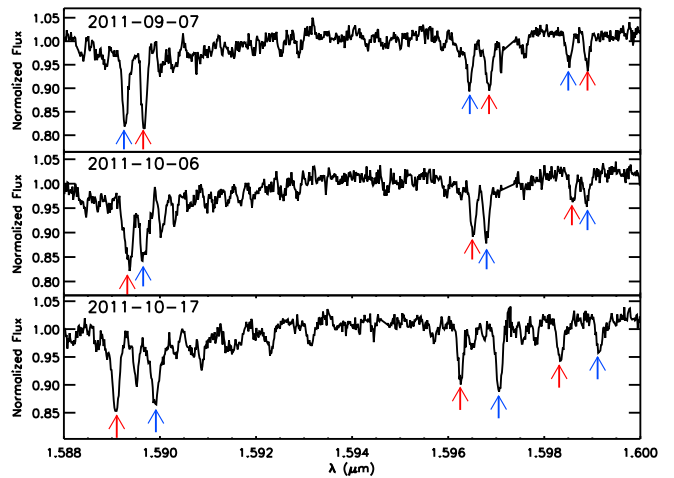}
\vspace{-65mm}
\caption{\footnotesize 
Time series data of a double-line
spectroscopic binary, illustrating the variation in the position of the lines
for both the primary (lines marked by blue arrows) and secondary (red arrows) stars in the system.}
\label{fig:SB2}
\end{figure}

\begin{figure}[h]
\vspace{-50mm}
\includegraphics[width=0.50\textwidth]{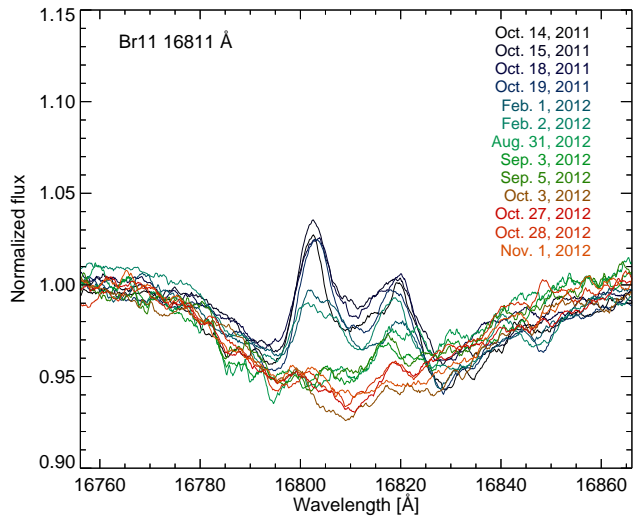}
\caption{\footnotesize 
Close-up view of the variation in the Brackett 11 (16811 \AA\ )
line in one of the Be stars discovered by APOGEE
(HD 232940, with $V=9.45$, $H=8.62$).
}
\label{fig:BeStar}
\end{figure}

\subsubsection{Time Series Spectral Data}
\label{sec:timeseriesspec}

A unique aspect of APOGEE in the context of Galactic archaeology
surveys is that for a large fraction of the targeted sources
multi-epoch observations are obtained (e.g., Fig. \ref{fig:visits}).
These time series data lend
themselves to a number of useful applications, including the 
identification of binary stars that partly motivated this 
survey strategy (\S \ref{sec:binaries}), and beyond.
For instance, Figure \ref{fig:SB2} shows an example
of a double lined spectroscopic binary made evident through the multi-epoch
spectroscopy).
As another example of the interesting phenomena that
a detailed exploration of APOGEE's time series spectral data might uncover, 
Figure \ref{fig:BeStar} shows the strong variation in a Brackett series
line for one of the 
numerous Be stars 
discovered by APOGEE
\citep{Chojnowski15}
within the sample of hot stars observed for telluric absorption
correction.  

\begin{figure}[h]
\vspace{-10mm}
\hspace{-10mm}
\includegraphics[width=0.55\textwidth]{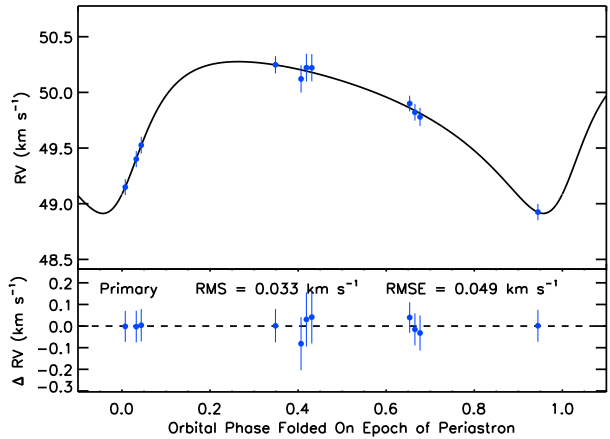}
\vspace{-60mm}
\caption{\footnotesize (Re-)discovery of the hot Jupiter around HD 114762 using eleven epochs
of APOGEE data.  }
\label{fig:planet}
\end{figure}

\subsubsection{Time Series Precision Velocity Data}
\label{sec:RVperformance}

The high precision of the APOGEE radial velocity measurements, made
possible by the tightly controlled environmental conditions (gravity
vector, temperature, vacuum) of the APOGEE spectrograph combined
with the availability of multi-epoch visits, makes the survey
sensitive to not only stellar, but substellar mass companions.  The
opportunity for large-scale statistical explorations of brown dwarfs
and even hot Jupiters is illustrated by Figure \ref{fig:planet},
which shows one of a number of exoplanet candidates identified
within the APOGEE database.  In this case, the star, HD 114762, has
a known 11$M_{\rm Jup}$ companion in an 83.9 day period \citep{Latham89}.
Before recognizing that this was a previously known exoplanet, the
APOGEE team, through a project to automatically fit multi-epoch
APOGEE velocity data, determined this source to have a 14$M_{\rm
Jup}$ companion in a 77.9 day period.  This result, based on only
11 APOGEE spectra, is not far from that already determined by Latham
et al.  The RMS of the best fit to this relatively bright star
system yields a residual of only 33 m~s$^{-1}$; further refinements
in the velocity measurement pipeline may bring similar RMS precisions
to fainter stars and perhaps enable further improvements in velocity
precision for brighter stars.

\begin{figure}[h]
\vspace{-30mm}
\includegraphics[width=0.50\textwidth]{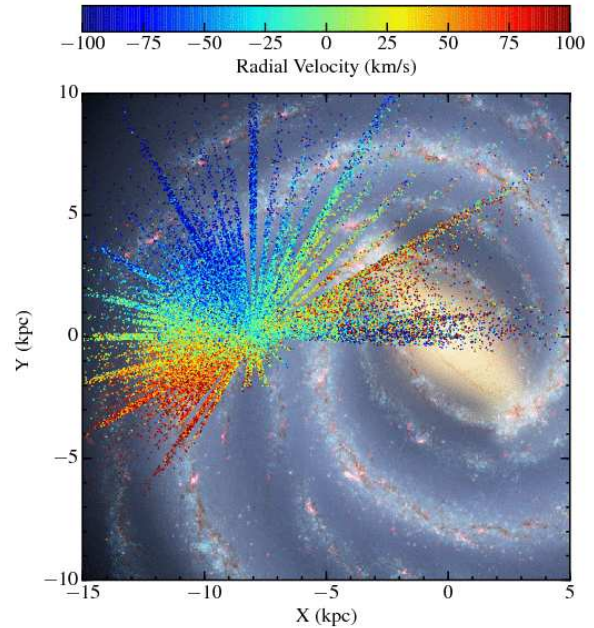}
\caption{\footnotesize 
Star-by-star APOGEE heliocentric velocities
as a function of Galactic $X$-$Y$ position and projected on
an artist's conception image of the Milky Way.
The points represent main APOGEE
survey stars (i.e., excluding those in the Kepler field) having projected $|Z_{GC}|<2$ kpc
and $\log{g} \ge 1$.  Only stars with no ``bad'' flags in the database are used.  
(Background image credit, R. Hurt, NASA/JPL-Caltech.)
}
\label{fig:rvgradients}
\end{figure}

\subsubsection{Radial Velocities Across the Galaxy}
\label{sec:rvgradients}

While not all stars in the APOGEE database will have more than
several radial velocity measurements (see Fig.\ref{fig:visits}),
almost all of them will have mean radial velocities measured with
an accuracy greater than is typically found in previous large
spectroscopic surveys of Milky Way stars.  Combined with its
systematic probe of all stellar populations, APOGEE's radial velocity
database makes possible new large-scale explorations of Galactic
dynamics (e.g., see Fig. \ref{fig:rvgradients}).  Because APOGEE
was designed to include extensive and deep probes of the low latitude
Milky Way, including the highly extinguished regions of the disk
and bulge, previously poorly studied regions of the Galaxy have now
been opened up to extensive kinematical canvassing.  As a result,
a number of Galactic dynamical phenomena have already been discovered
or reexamined with unprecedented statistical significance, including:
the discovery of a kinematically extreme population in the bulge
\citep{Nidever12}, likely associated with the Galactic bar
\citep{Molloy15,Aumer15}; a comprehensive analysis of the
dynamics of the bulge \citep{Ness15}; a new derivation of the
circular velocity of the Sun afforded by the more global APOGEE
view of disk kinematics \citep[Fig.~\ref{fig:rvgradients};][]{Bovy12,Bovy15};
and finally an assessment of the power spectrum of non-axisymmetric velocity
perturbations of the disk, of which most can be attributed to the
action of the central bar \citep{Bovy15}.

\begin{figure}[h]
\vspace{-30mm}
\includegraphics[width=0.50\textwidth]{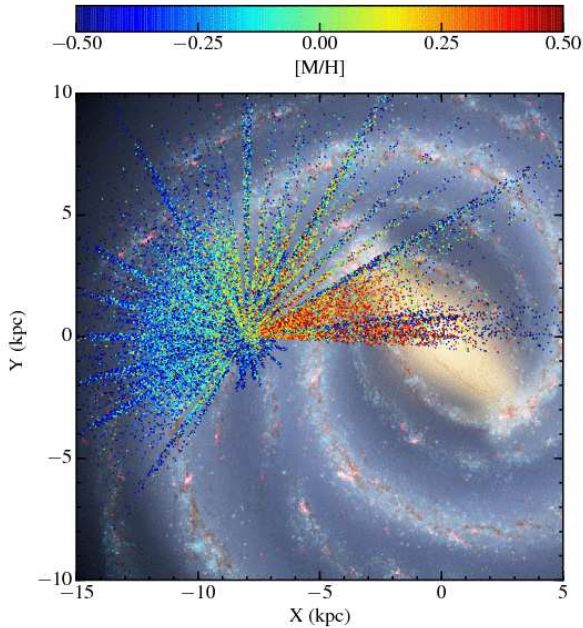}
\caption{\footnotesize 
Same as Fig. \ref{fig:rvgradients}, but with points color-coded by 
metallicities [M/H]
and using stars with projected $|Z_{GC}|<2$ kpc.
(Background image credit, R. Hurt, NASA/JPL-Caltech.)
}
\label{fig:metalgradients}
\end{figure}

\subsubsection{Metallicities Across the Galaxy}
\label{sec:gradients}

A primary goal of the APOGEE survey is to probe the large-scale
distribution of stellar chemical compositions across the Galaxy.
Typical results are demonstrated by the [M/H] metallicity gradients shown in
Figure~\ref{fig:metalgradients}, where the distribution of APOGEE
targets projected onto the XY plane is
displayed.  Accurate measurements of gradients across a range of
distances, both Galactocentric and away from the Galactic plane,
provide strong constraints on models for the formation of the thick
and thin Galactic disks (for further details on the application
of APOGEE to these problems, see \citealt{Bovy14}, \citealt{Nidever14},
\citealt{Hayden14},2015, \citealt{Anders14}).

\begin{figure*}
\vspace{-85mm}
\includegraphics[width=1.00\textwidth]{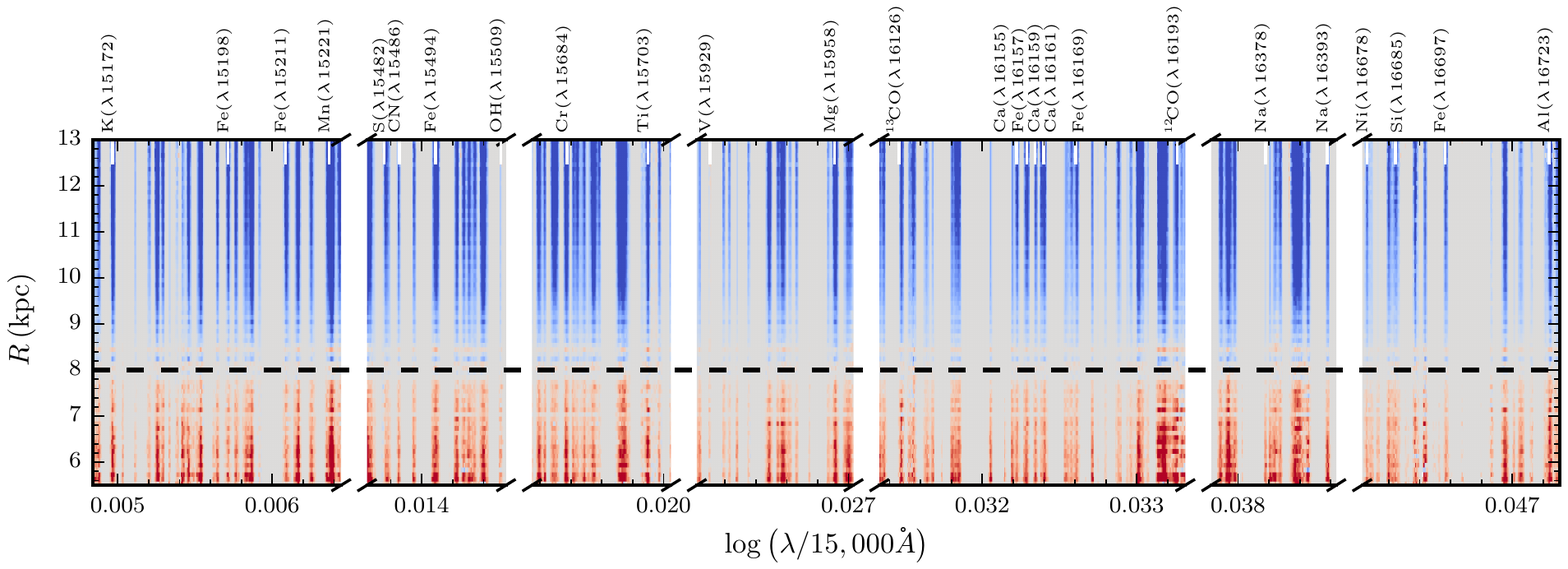}
\vspace{-90mm}
\caption{\footnotesize Direct spectral representation of the Milky Way disk's
  metallicity gradient. This figure displays median-stacked
  normalized APOGEE spectra of RC stars in bins of
  Galactocentric radius of width $0.1\,\mathrm{kpc}$ and normalized
  by the spectrum at the solar circle (assumed to be $R = 8\,\mathrm{kpc}$,
  indicated by the dashed line). Redder/bluer colors represent
  more/less
  absorption compared to that at the solar
  circle.  Only about 1/6th of the full APOGEE spectral range is shown
  here, primarily regions containing strong, clean lines and representing all of
  the elements whose abundances can be determined from APOGEE 
  spectra. Some of the interesting lines of neutral-atomic and
  molecular species as well as their central wavelengths are labeled
  at the top. Because RC stars span a narrow range of stellar
  parameters, absorption line strengths translate into rough elemental
  abundances directly.  All elements display a clear abundance gradient from the
  inner to the outer disk.  The only exception perhaps is vanadium, which
is based on a very weak line at 15929\AA\ that becomes vanishingly weak
 beyond the solar circle.
}
\label{fig:bovystack}
\end{figure*}

\subsubsection{Multi-Element Abundance Variations}
\label{sec:multi_gradients}

The power of APOGEE to deliver similar information for {\it multiple}
chemical elements and with great statistical power is illustrated
by Figure~\ref{fig:bovystack}, a representation of the metallicity
gradient in the Milky Way's disk using APOGEE spectra directly. We
stack the pseudo-normalized spectra of red clump (RC) stars in the
DR12 APOGEE-RC catalog \citep{Bovy14} as a function of Galactocentric
radius in bins of width $0.1\,\mathrm{kpc}$ and normalize by the
stacked spectrum at the solar circle ($R = 8\,\mathrm{kpc}$). Because
the RC occupies a small region in $(\log g$, $T_{\mathrm{eff}})$, the
$T_{\rm eff}$ and $\log{g}$ values for RC stars are all very similar, so
that spectrum-to-spectrum differences in absorption line strengths
are mostly due to star-to-star elemental abundance variations.  This
figure clearly demonstrates that metal absorption increases toward
the center of the disk and decreases toward the outskirts
for every element with transitions
in the APOGEE spectral region (except for vanadium, which has such
a weak line that absorption disappears almost completely around the
solar circle).  A quantitative measurement of the overall radial 
[M/H] gradient of the RC sample is
$-0.09\pm0.01\,\mathrm{dex}\,\mathrm{kpc}^{-1}$ \citep{Bovy14}.\footnote{
While some radii do not follow the overall radial trend (e.g., the stacked
spectrum at $R = 7\,\mathrm{kpc}$ is almost exactly that at the
solar circle), this is most likely due to first-ascent RGB
contamination in the APOGEE-RC catalog.}

\begin{figure}[h]
\includegraphics[width=0.50\textwidth]{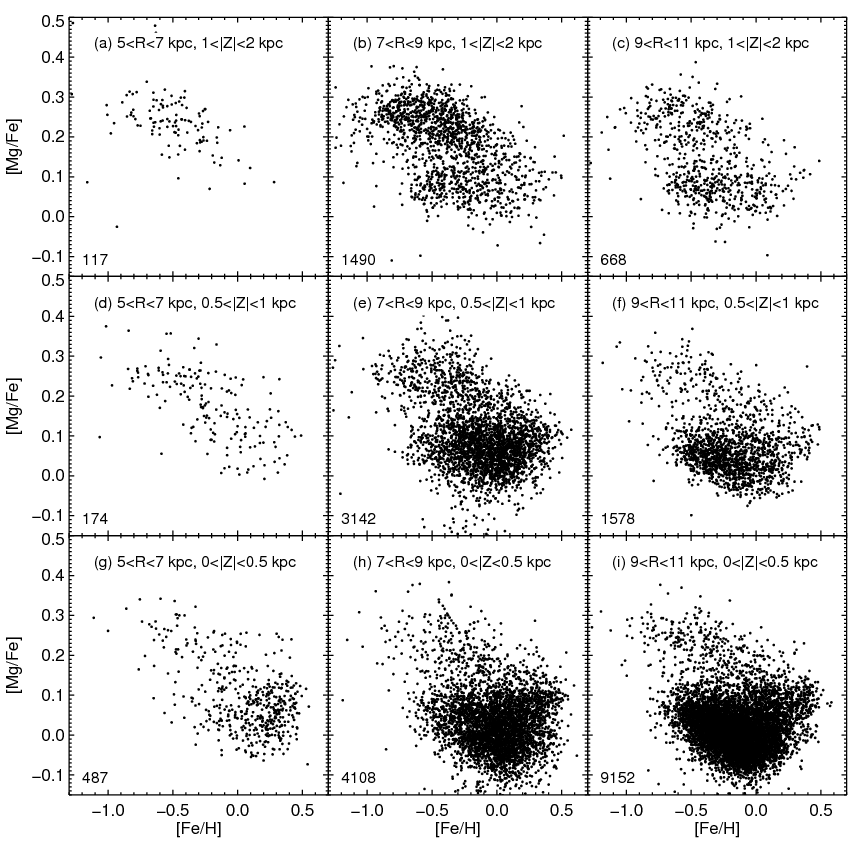}
\caption{\footnotesize Spatial variation of [Mg/Fe] vs. [Fe/H] distributions across the
Galactic disk, from \cite{Nidever14}}
\label{fig:alphaFe}
\end{figure}

\subsubsection{Analysis of Abundance Patterns}
\label{sec:alpha}

A central rationale for APOGEE's sensitivity to the abundances of multiple
chemical elements is to access the information contained in variations
of chemical abundance {\it patterns}, particularly for patterns containing 
elements synthesized on different evolutionary timescales within populations.
The power of APOGEE to probe Galactic chemical evolution in this way
is demonstrated by Figure~\ref{fig:alphaFe}, which shows the 
spatial distribution of [Mg/Fe] with [Fe/H]
in different spatial zones
of the Galactic disk. Figure~\ref{fig:alphaFe} 
is a more elemental-specific example of the more general
[$\alpha$/Fe] distributions explored in \citet{Anders14}, \citet{Nidever14}, and
\citet{Hayden15}, and shows that APOGEE not only confirms  
the existence of a bimodal distribution of [$\alpha$/Fe] at fixed
[Fe/H], but that this bimodality exists over a large extent of the disk, 
including for previously uncharted Galactocentric distances for 
$|Z| < 0.5$~kpc
(see \citealt{Hayden15} for an even more extensive coverage of the disk).  
These
results pose important constraints on disk formation models, in
particular calling for the presence of two stellar populations with
different chemical enrichment histories in the outer Galactic disk.
That APOGEE data can provide such data across a large
multi-dimensional space of elements probing different nucleosynthetic 
pathways is further demonstrated by the results shown
in Figure \ref{fig:clusterabund} below.

\subsubsection{Stellar Ages}
\label{sec:ages}

\begin{figure}
\includegraphics[width=0.50\textwidth]{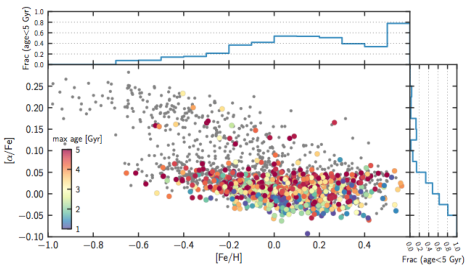}
\caption{\footnotesize Chemical abundance patterns informed by
individual stellar ages, derived from the combination of {\it Kepler}
asteroseimology and APOGEE chemistry (see \citealt{Martig15}).}
\label{fig:ages}
\end{figure}

Collaboration between APOGEE and the {\it Kepler} and {\it CoRoT}
asteroseismology teams has greatly increased the
power of
APOGEE data to shed light on Galaxy formation scenarios.  The combination
of APOGEE chemical compositions with asteroseismological data makes
possible the determination of accurate masses, thus ages, for a
large sample of APOGEE {\it field} giants \citep{Pinsonneault14}.
Demonstrations of the power of 
analyzing combined detailed chemical
compositions and ages from these APOGEE subsamples are starting to
emerge \citep[e.g.,][]{Epstein14,Martig15,Chiappini15}; one example
is demonstrated in Figure~\ref{fig:ages}, which shows the unexpected
discovery \citep{Martig15} of a small sample of stars with
high [$\alpha$/Fe] and relatively low ages ($\sim$ 3--5 Gyr).  The
mere existence of intermediate-age stars with such high [$\alpha$/Fe]
constitutes a serious challenge to existing chemical evolution
models and is, as yet, not fully explained.

\begin{figure}[h]
\vspace{-37mm}
\hspace{-5mm}
\includegraphics[width=0.55\textwidth]{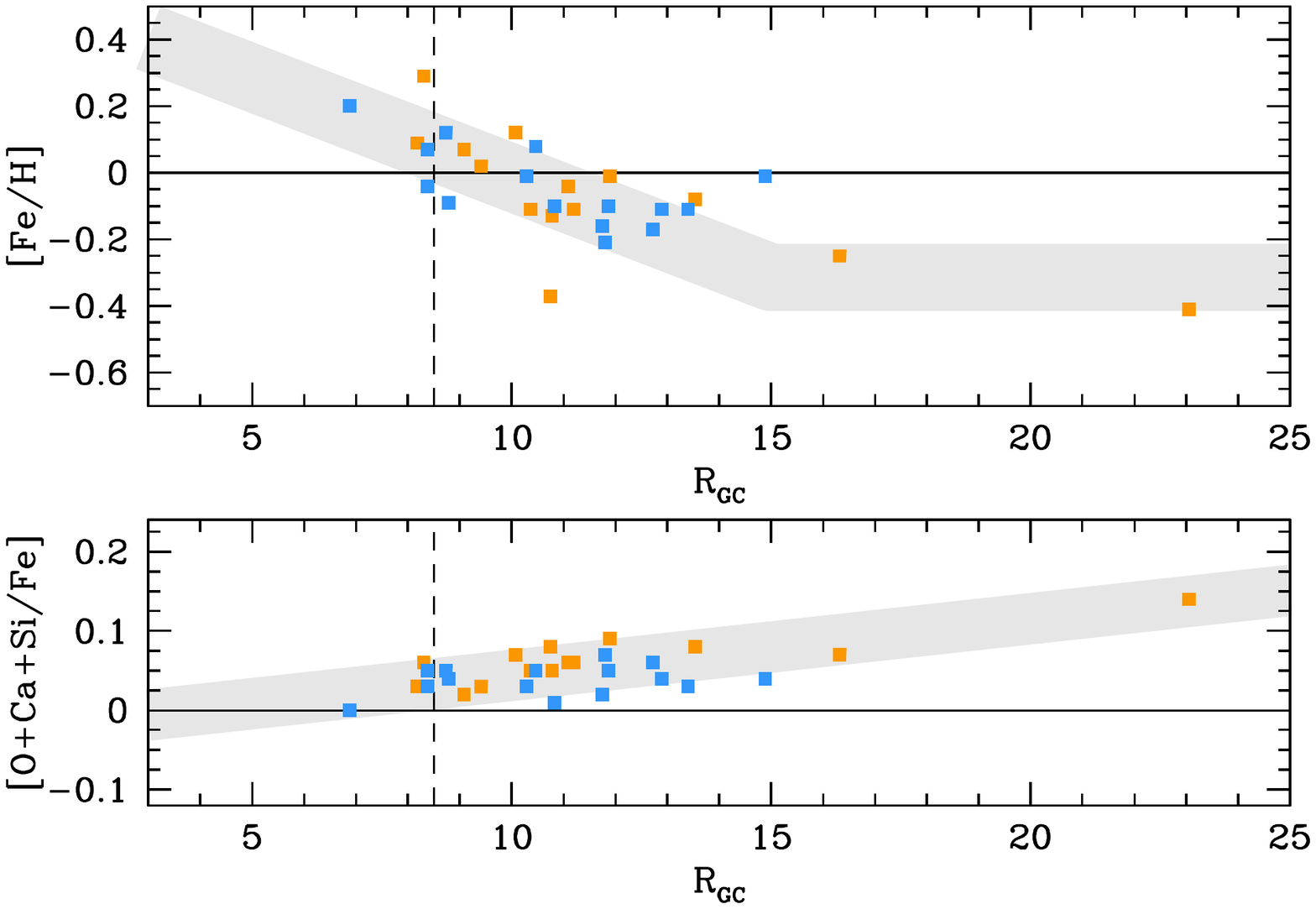}
\vspace{-25mm}
\caption{\footnotesize  
APOGEE-derived metallicity and [$\alpha$/Fe] (using the reliable
individual DR12 elements O+Ca+Si as $\alpha$) as a function of 
Galactocentric radius for 29 
clusters in two age regimes: 
Blue (orange) points denote clusters younger (older) than 1 Gyr.
The light grey region in the upper panel shows a gradient in [Fe/H] of $-0.06$ dex kpc$^{-1}$
within $R_{GC} = 15$ kpc and hints of a flat distribution (admittedly based
on very few data points) beyond with a spread of 0.2 dex.  A shallower, 
positive gradient, of $+0.007$ kpc$^{-1}$dex, is seen in
[$\alpha$/Fe] (lower panel) consistent with previous literature-based open cluster
studies \citep[e.g.,][]{Yong12}.  Additional individual element abundance gradients
are  explored in \cite{Frinchaboy15}.}
\label{fig:open_cluster_metallicity}
\end{figure}

\subsubsection{Open Clusters}
\label{sec:OCs}

The APOGEE database has made the targeting of star clusters a priority, not
only because they provide reliable abundance calibration references, but because
they are extremely useful stellar population components of great interest 
in their own right.  Open clusters, in particular, provide a powerful, independent
means by which to explore the chemistry, kinematics and extinction of the Galactic disk
because clusters distances can be precisely gauged via isochrone fitting,
made all the more accurate when the variable of metallicity can be removed using 
spectroscopic metallicities.
An important output of such isochrone fits are cluster
ages, which provide critical and confident 
timestamps for benchmarking evolutionary studies
across the disk.
Fortunately, apart from a set of key open clusters that 
drove specific plugplate pointings, the vast majority of the large APOGEE sample
of over 150 sampled open clusters was a natural product of its extensive
canvassing of the Galactic plane, and only required attention to the allocation
of fibers on already overlapping plugplates from the APOGEE grid to known
or suspected members of these clusters.  As part of the Open
Cluster Chemical Analysis and Mapping (OCCAM) survey
\citep{Frinchaboy13}, these data have already been used to study 
disk metallicity gradients and abundance patterns as a function of 
both space {\it and} time;
an example of the opportunities opened up with this kind of analysis 
is demonstrated in Figure \ref{fig:open_cluster_metallicity}.

\begin{figure}[h]
\vspace{-60mm}
\includegraphics[width=0.50\textwidth]{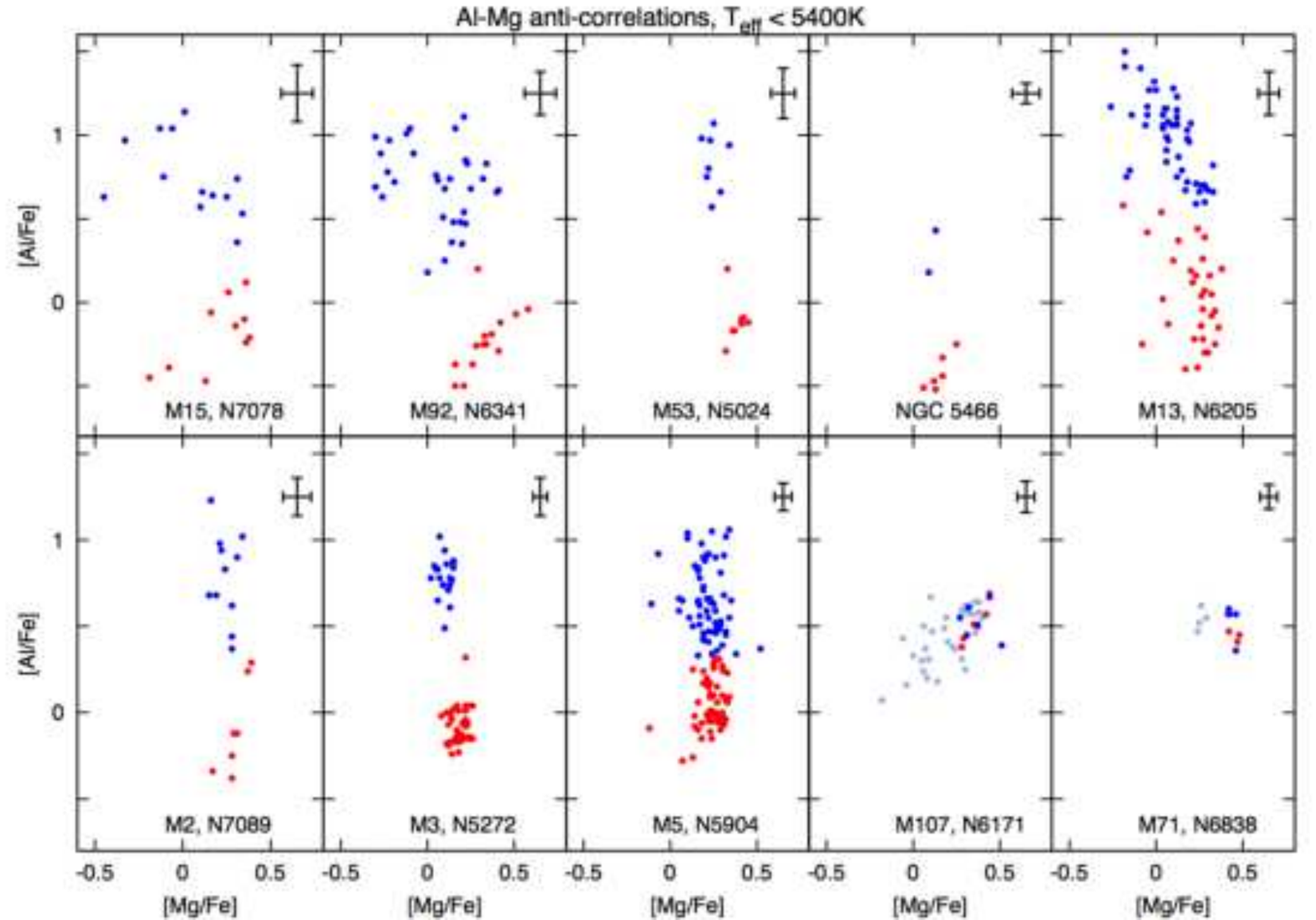}
\caption{\footnotesize Abundance anti-correlations as seen for 428 giant
star members in ten Northern Hemisphere globular clusters.
Figure from \cite{Meszaros15}.}
\label{fig:clusterabund}
\end{figure}

\subsubsection{Globular Clusters}
\label{sec:GCs}

Unlike the open clusters, where spectra of large samples of cluster members
are easily generated as part of APOGEE's systematic coverage
of the Galactic
disk, considerable investment of effort, planning and survey time was 
needed to ensure ample attention
to globular clusters.  Many globular clusters are distant, which makes even
their giant stars relatively faint and means that the stars are densely packed 
on the sky relative to the fiber-to-fiber collision radius (see \S\ref{tab:Inst_Char}).
These two facts typically necessitated many visits to specialized
fields (see Fig. \ref{fig:field_plan}), with multiple plugplate designs to accumulate
long integrations on fainter targets and to overcome the collision radius limitations
to generate reasonable sample sizes.  Nevertheless, the globular cluster visits
were deemed high priority both because the member stars are extremely 
useful calibration targets and because the chemistry of globular clusters is 
extremely interesting and relevant to understanding the evolution of Milky
Way stellar populations.

The study of globular cluster formation is undergoing a revolution
since the discovery of the commonality of multiple populations in
these systems \citep[e.g.,][]{Piotto07,Milone08,Gratton12}, which has
been established predominantly on the basis of observations collected from
the Southern Hemisphere.  APOGEE is already making important
contributions to the study of globular cluster evolution through
the determination of accurate multi-element abundances of relatively
large samples of northern cluster members.  This is illustrated
by Figure~\ref{fig:clusterabund}, which, for the majority
of the ten clusters shown (eight that were never previously studied in
this way), gives evidence for strong internal
Al variations --- in some cases anti-correlated
with Mg --- a feature commonly interpreted as the 
signature of the presence of multiple stellar populations.   
Interestingly, APOGEE has shown that some globular clusters
(M3, M53) display apparently discrete
Al-Mg distributions (i.e., Al-rich
and Al-poor groups), while others (e.g., M13, M5) may show
a more continuous distribution in the Al abundances \citep{Meszaros15}.
APOGEE data have also been used to test
previous claims of the presence of multiple populations in the well
known, unusually large, old but super-metal-rich {\it open} cluster,
NGC~6791 \citep{Cunha15}.  

A distinct advantage afforded by these APOGEE studies of globular
clusters is that the inferred chemistry of the member stars may be
directly compared to those of the numerous field stars surveyed in
precisely the same way.  An example of the power of such comparisons,
with interesting implications for our understanding of both globular
cluster and galaxy bulge formation, is the discovery of remnants
of globular cluster destruction in the Galactic bulge --- long-predicted
by theoretical arguments \citep{Tremaine75} --- through the
identification of chemical signatures typical of globular cluster
stars in an APOGEE sample of field bulge stars \citep{Schiavon15}.

\begin{figure}[h]
\vspace{-30mm}
\includegraphics[width=0.50\textwidth]{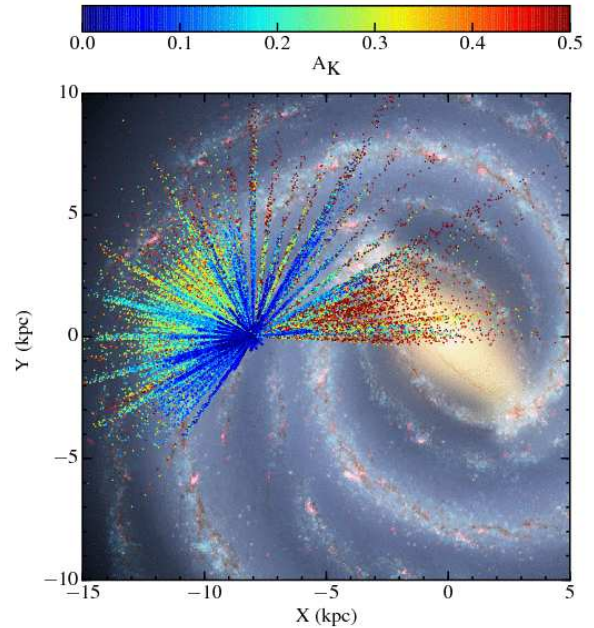}
\caption{\footnotesize 
Distribution of extinction across the Galactic plane, as measured by
comparing star-by-star photometric and spectroscopic data to derive $A_{K_s}$, 
according to the isochrone matching method discussed in \citet{Schultheis14}.  
Point positions correspond to the projected locations of the stars against
which the extinction was measured. 
Only stars from DR12 with $|Z_{GC}| < 500$ pc are shown.  Note the
substantial increase in extinction at the spiral arm towards the outer disk
and also towards the Galactic center.
(Background image credit, R. Hurt, NASA/JPL-Caltech.)
}
\label{fig:dust}
\end{figure}

\subsubsection{Mapping Interstellar Extinction}
\label{sec:dust}

The combination of APOGEE spectroscopy with available photometric 
databases, such as those provided by 2MASS or the {\it Spitzer} IRAC
surveys, increases the scope of approachable astrophysical problems.
For example, comparison of spectroscopically-derived atmospheric
properties with measured broadband colors can give key, star-by-star 
measurements of the distribution of interstellar reddening, and therefore
a powerful means by which to explore the three-dimensional distribution
of dust.  
For example, Figure \ref{fig:dust} shows the distribution of derived $K_s$
band extinction across the Galactic plane, with the extinctions
derived by comparing observed colors to the intrinsic colors expected from 
theoretical isochrones matched to the ASPCAP-derived values of
$T_{eff}$, $\log{g}$, and [M/H] (see \citealt{Schultheis14}),
and distances derived from comparing apparent and associated isochrone
absolute magnitudes.
Apart from showing global trends in dust properties, studies such as this 
can help calibrate or highlight inadequacies
in extinction maps derived by other means.

\begin{figure}[h]
\vspace{-70mm}
\includegraphics[width=0.50\textwidth]{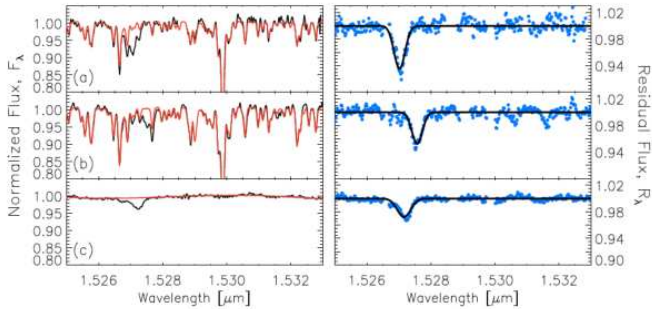}
\caption{\footnotesize ({\it Left}) Comparison of synthetic ({\it red lines}) 
to observed APOGEE
spectra ({\it black lines}) for two late type giant stars ({\it panels a and b})
and a hotter star ({\it panel c}) star.  
({\it Right}) The difference spectra ({\it blue points}) 
clearly show the DIB feature at 15,272 \AA\ , which can be fit with simple
Gaussians ({\it black lines}) to measure column density and varying 
Doppler velocities \citep{Zasowski15}.
}
\label{fig:DIBS}
\end{figure}

\subsubsection{Mapping Diffuse Interstellar Bands}
\label{sec:DIBs}

The accuracy with which the APOGEE spectral library matches the 
intrinsic spectra of typical giant stars makes possible further
explorations of interesting astrophysics that is unrelated to the
stars themselves.  For instance, disagreements between spectral
synthesis and observations may reveal the presence of absorbers/emitters
along the 
line of sight.  A compelling example of this
is shown in Fig. \ref{fig:DIBS}, where the difference between an
observed spectrum and its best-matching synthetic template reveals
the presence of a relatively strong diffuse interstellar band
\citep[DIB,][]{Herbig95}.
Using the fact that this same DIB feature is seen in the vast
majority of APOGEE spectra along disk and bulge sightlines,
\cite{Zasowski15} for the first time thoroughly mapped the spatial distribution 
and velocity field of a DIB carrier over many kpc of the Galactic disk.
Comparison of these properties to those of other
components of the ISM (dust, atomic and molecular gas) will yield
critical evidence towards the identification of the carriers of
these mysterious spectroscopic signatures --- a longstanding puzzle in
Galactic astrophysics.

\begin{figure}[h]
\vspace{-25mm}
\includegraphics[width=0.50\textwidth]{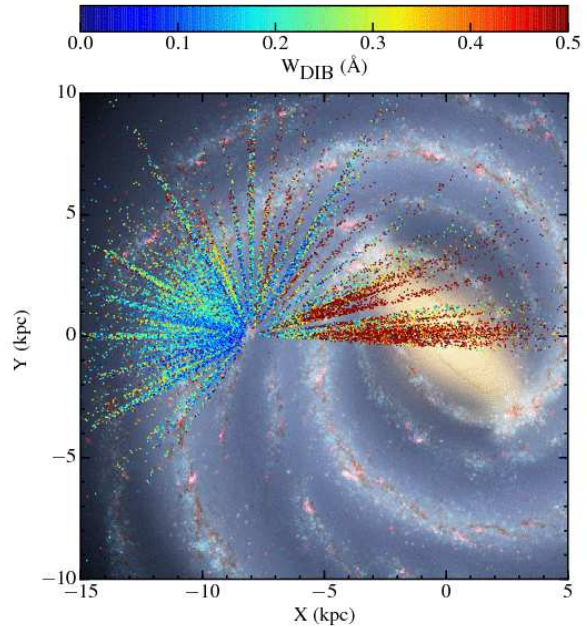}
\caption{\footnotesize Same as Fig. \ref{fig:dust}, but for the distribution of measured 
$\lambda$ 15,272 \AA\  DIB strength \citep{Zasowski15}.  As with Fig. \ref{fig:dust},
the plotted points correspond to the projected locations of the stars
against which the DIB absorption was observed.  
Only stars from DR12 having $|Z_{GC}| < 500$ pc are shown.
(Background image credit, R. Hurt, NASA/JPL-Caltech.)}
\label{fig:xyDIB}
\end{figure}

\section{Data Products, Distribution and Documentation}
\label{sec:products_docs}

\subsection{Data Products}
\label{sec:products}

There are two main sets of APOGEE data products.  The first consists
of wavelength--calibrated, normalized, 1-D restframe spectra, and
the radial velocities resulting from analysis of visit spectra.
These data products are the outputs of the data reduction and radial
velocity pipelines, described in \S\S\ref{sec:datahandle}-\ref{sec:rvs}.
The second set of data products are the stellar
parameters ($T_{\rm eff}$, $\log g$, [M/H]) and the abundances of
up to 15 elements, namely: C, N, O, Na, Mg, Al, Si, S, K, Ca, Ti, V, Fe,
Mn, and Ni.  The latter set results from application of ASPCAP
(\S\ref{sec:stelpar}) to the spectra.
To ensure
that the high level data can be reproduced by the community,
ingredients that are crucial for the stellar parameter and elemental
abundance determinations --- such as the adopted
model atmospheres  \citep{Meszaros12},
line list \citep{Shetrone15}, and grids of synthetic spectra
\citep{Zamora15} --- are also available.

\subsection{Data Releases}
\label{sec:DRs}

The data products from the three year APOGEE
survey have been included in 
two SDSS-III data releases (DRs).  
In DR10 (announced July 2013), calibrated APOGEE spectra, 
radial velocities, stellar parameters, and a limited set of elemental
abundances were made available for 57,454 stars \citep{Ahn14}.
The second public data release, DR12 (made available in January 2015)
includes the data described in \S~\ref{sec:products}
for all $\sim$$136,000$ primary science targets in the SDSS-III/APOGEE sample
\citep{Alam15,Holtzman15}.
In addition, radial velocities for the
$\sim$~17,000 hot stars\footnote{This number includes both survey-quality
and commissioning data.} used as telluric standards, as well as
spectra and derived parameters
for approximately 900 bright stars observed using the APOGEE
feed from the NMSU 1-m telescope (\S\ref{sec:standards}) are also
included in DR12.

\subsection{Data Access}
\label{sec:access}

APOGEE data can be accessed in a number of different ways
through the {\tt www.sdss.org} website.  The
SDSS-III {\it Catalog Archive Server} (CAS) contains the high level
APOGEE data products, namely radial velocities, radial velocity
dispersion, $T_{\rm eff}$, $\log g$, [M/H], and elemental abundances,
as well as information relevant to target selection,
including, e.g., coordinates, {\it 2MASS},
{\it WISE}, {\it Spitzer/IRAC}, 
and Washington+$DDO51$ photometry, proper motions, where
available, and assumed interstellar extinction.  These data can be accessed
via low level {\tt SQL} scripts, which allow users to select subsamples
according to suitable criteria in the {\tt CasJobs} link.  Lower
level data products, including the visit and combined spectra of
sources searchable via standard, basic or advanced online forms,
can be accessed with the {\it Science Archive Server} (SAS).  The SAS
also provides access to the directory tree containing the full data
set (for expert users), as well as a summary {\tt FITS} file
containing all of the catalog information listed above.  The {\tt
SkyServer} offers another way of interfacing with the data, through
a ``quick look'' tool that makes possible non-interactive visualization
of images and spectra of sources that are searchable by position
and ID, via an online form.  Finally, both the SAS and {\tt
SkyServer} link to a web tool that allows for a quick assessment
of data and model quality, through interactive viewing of the (final)
observed spectra overplotted with the best fit-fitting synthetic
spectrum.

\subsection{Documentation}
\label{sec:documentation}

The primary source of in depth information for users of APOGEE data
are the technical papers listed in Table~\ref{tab:techpapers}.
Additional, often less detailed, information can be obtained in the
webpages associated with each data release.  Users of APOGEE are
{\it strongly urged} to peruse those webpages or technical papers
for a complete 
understanding of the data quality, limitations and caveats that are
specific to each data release \citep{Ahn14,Holtzman15}.  
Users of high level data products should pay particular
attention to the documentation pertaining to ASPCAP 
\citep{Meszaros12,Meszaros13,Smith13,Zamora15,GarciaPerez15,Shetrone15},
where the 
limitations and
known systematics in the APOGEE stellar parameters and elemental
abundances are described.  On the other hand, users of APOGEE {\it spectra} will
be interested on information about possible systematics induced by
instrument features and reduction procedures \citep{Wilson15,Nidever15}.  
A good grasp of the possible biases introduced
by APOGEE's target selection procedure and field placement strategy
\citep{Zasowski13} is encouraged for those exploring global properties,
trends and correlations for stars across the Galaxy, and for those
comparing these empirical results to 
chemical and/or chemo-dynamical models of stellar populations.

\section{Future Efforts}
\label{sec:future}

\subsection{Software Development and Future Data Releases}
\label{sec:futureDR}

As part of APOGEE-2 (\S \ref{sec:APOGEE2}), we will continue to
develop our data reduction and analysis capability.  In addition,
it is expected that our understanding of the atomic and
molecular physics related to the relevant line transitions will
continue to evolve.  Therefore, to maintain consistency across the
integrated APOGEE-1 and APOGEE-2 efforts, future APOGEE data releases
will be expected to include updated analyses of the APOGEE-1 data.

\subsection{APOGEE-2 in SDSS-IV}
\label{sec:APOGEE2}

In 2012 the APOGEE team proposed to continue with an APOGEE-2 program
in Sloan Digital Sky Survey IV.  Beyond an immediate continuation of
operations on the Sloan Telescope, a key feature of this proposal is
the extension of the APOGEE-2 program into the Southern Hemisphere,
in a collaboration with the Observatories of the Carnegie Institution
for Science (OCIS) to use the du Pont 2.5-m telescope for this
purpose.  The du Pont is very much a forerunner of the Sloan
Telescope, having been designed with a wide field-of-view
\citep{Babcock77} and with a heritage in fiber plugplate spectroscopy
\citep{Shectman96} that influenced the subsequent design of the
SDSS system.  APOGEE-2 was approved as an SDSS-IV program in 2012,
and APOGEE-2 began observing at APO in September 2014.  The current
instrument at APO is 
the same as that described in \S
\ref{sec:instrumentoverview} with one important modification in
that the blue detector with``superpersistence'' has been replaced
with a cosmetically cleaner array free of this problem.

While initially it was unclear whether APOGEE-2 would operate with
two spectrographs working on the Sloan and du Pont Telescopes in
parallel or whether the existing spectrograph would be moved to the
du Pont after an initial 2 year campaign on the Sloan Telescope,
by November 2014 complete funding was secured
to build a second APOGEE spectrograph to make parallel operations
possible.  With the active participation of OCIS, Chilean universities,
and other SDSS-IV collaborators, the southern infrastructure
development, including the creation of a scaled-down version of the
Sloan fiber plugplate system (cartridges, plugging station, mapping
system, cartridge transport and loading system), is 
underway,
as is the construction of an infrastructure ``mock-up and training
facility'' in a special laboratory at the University of La Serena,
Chile.  The second APOGEE spectrograph, which is planned as a near
duplicate of the original with only minor alterations (e.g., to
include seismic mitigation and to incorporate various other ``lessons
learned"), is currently planned to begin operations on the du Pont
in mid-2015.

When it is necessary to distinguish it from the APOGEE-2 survey,
the original ``APOGEE'' survey will in the future be referred to
as ``APOGEE-1''.  Henceforth the term ``APOGEE'' will generally be
intended to refer to the combined APOGEE-1 and APOGEE-2 surveys,
which together will provide a fully comprehensive, {\it all-sky view} of
the Milky Way.

\appendix

\section{A: $H$-Band Line Maps for Individual Molecules and Atoms}
\label{sec:linemaps}

\begin{figure}[h]
\includegraphics[width=0.50\textwidth]{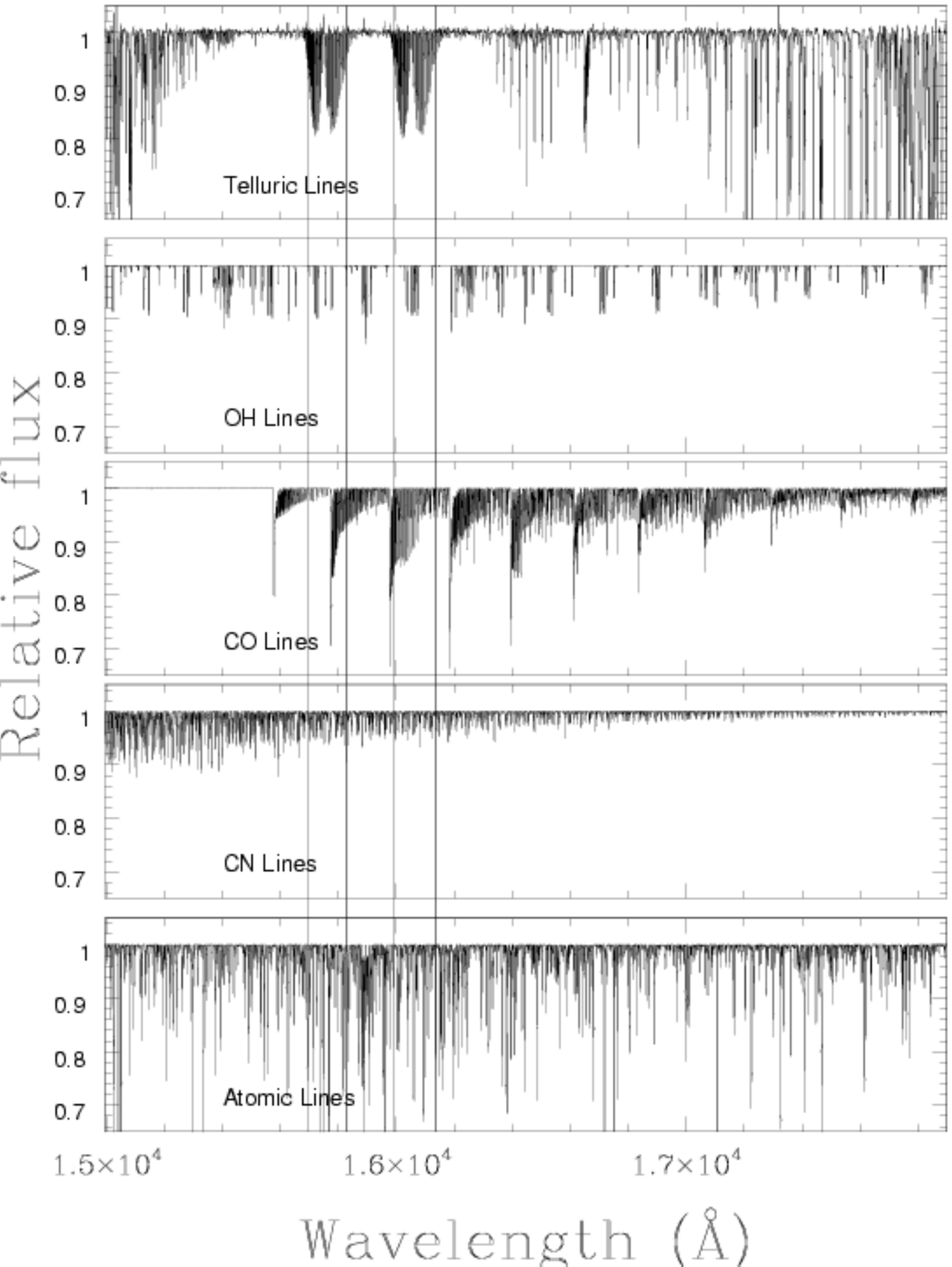}
\includegraphics[width=0.50\textwidth]{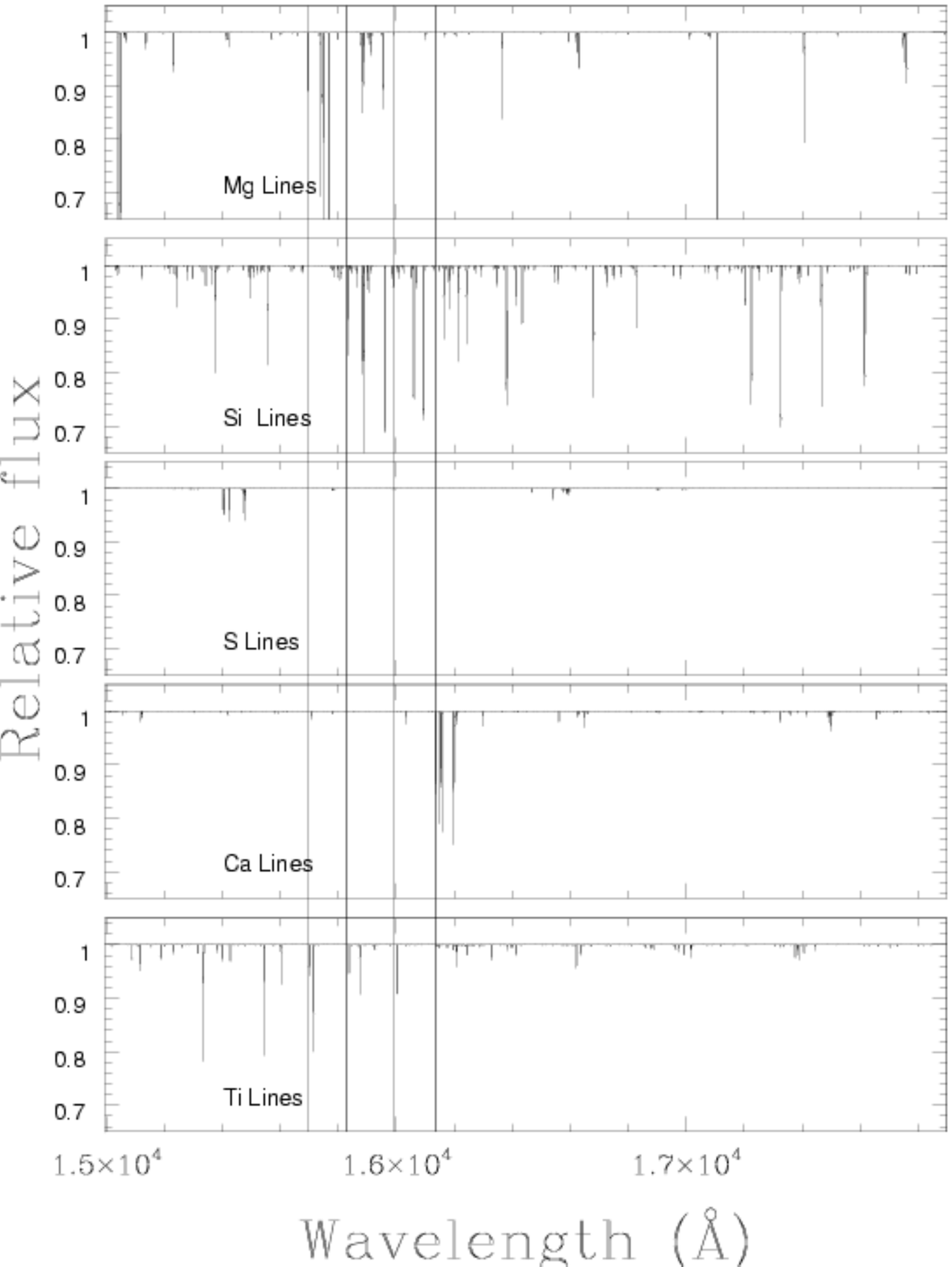}
\includegraphics[width=0.50\textwidth]{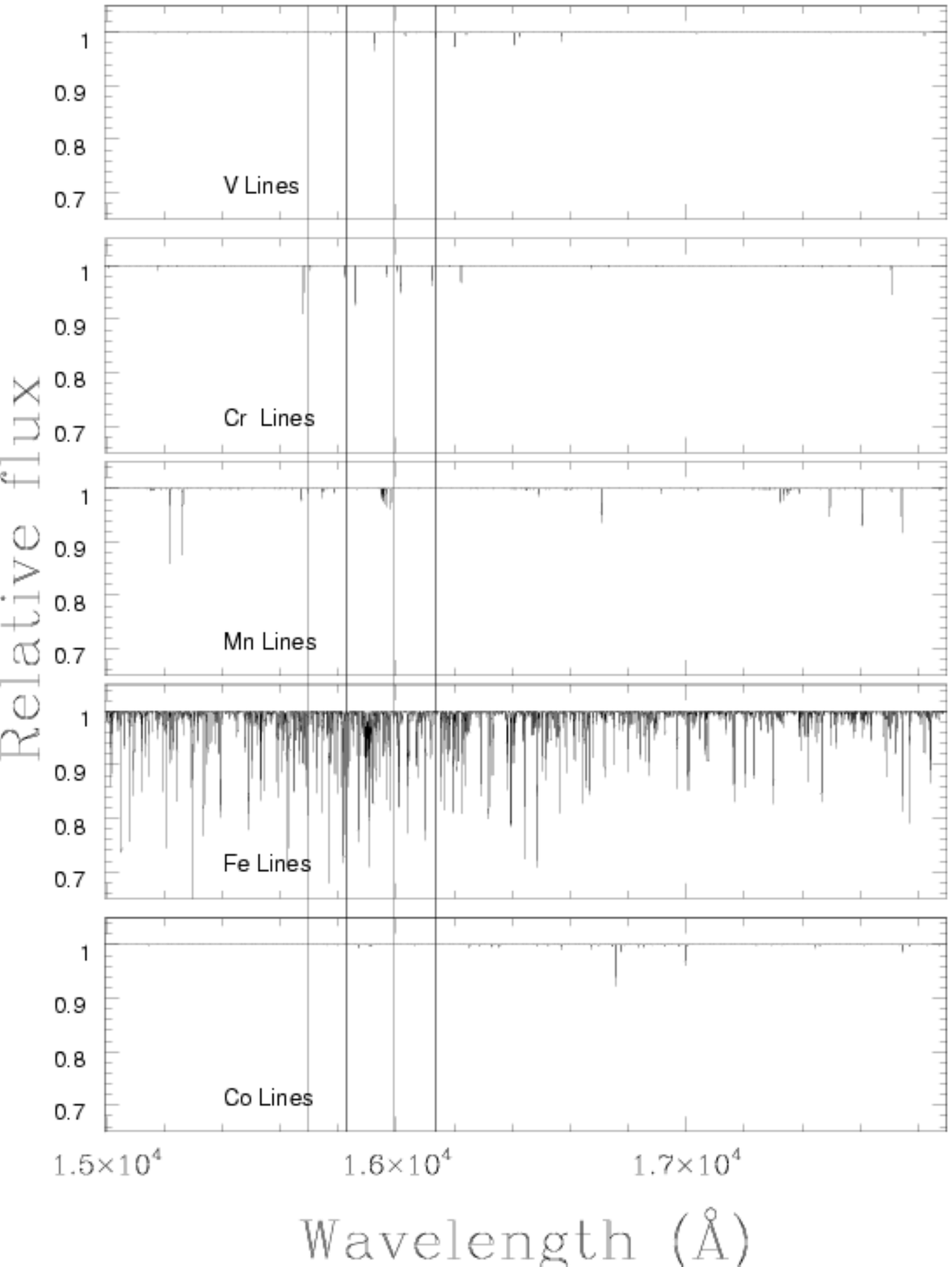}
\includegraphics[width=0.50\textwidth]{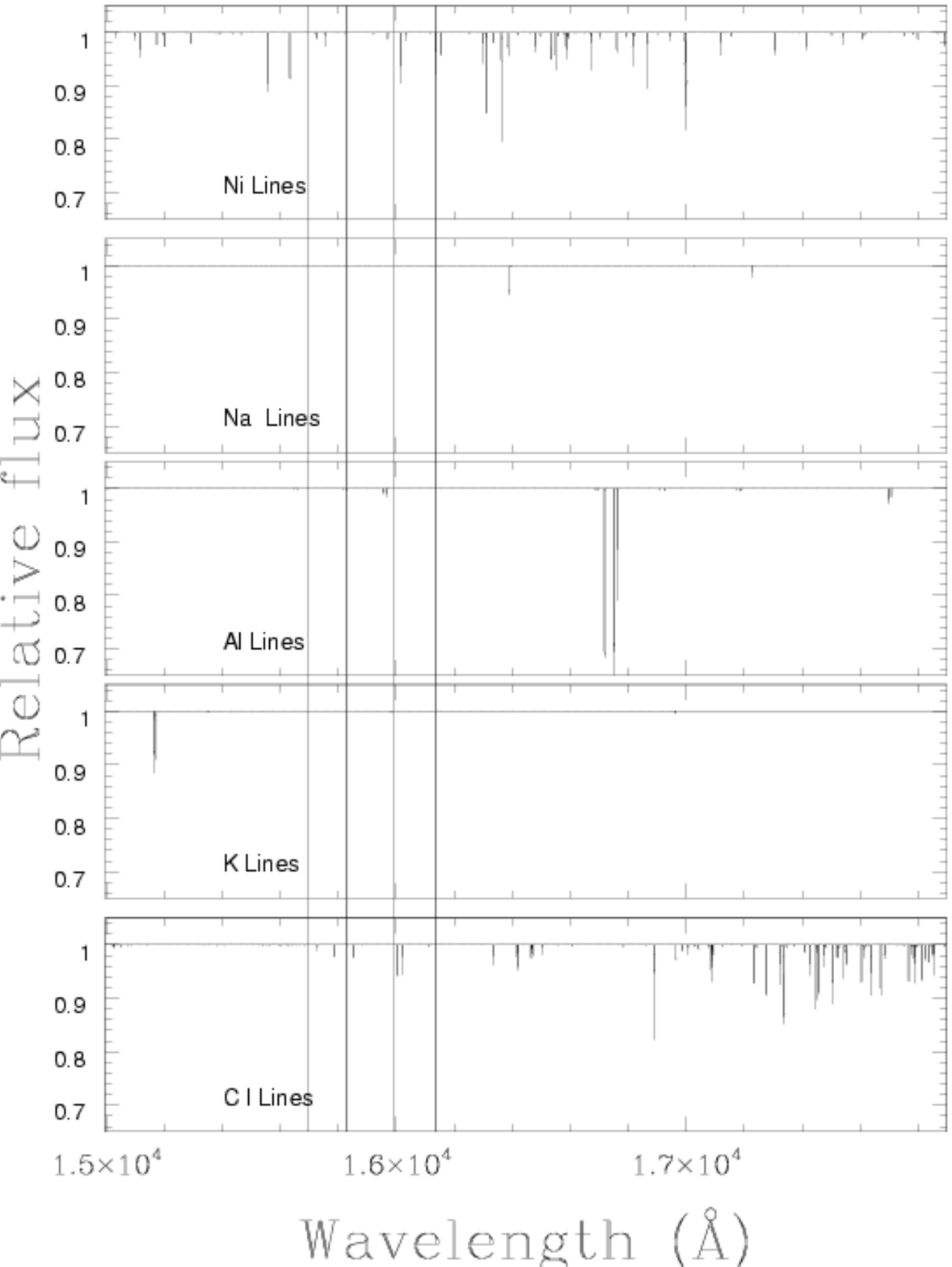}
\caption{\footnotesize Line maps of specific molecules and elements having lines/bands expressed in the $H$-band.
Generally, these are C, N, O, Na, Mg, Al, Si, S, K, Ca, Ti, V, Cr, Mn, Fe, Co and Ni. This is a useful subset of elements with which to probe most types of nucleosynthesis and chemical evolutions. 
The vertical lines in each plot indicate the ranges of the strong telluric absorption features near $1.6\mu$m.}
\label{fig:element_maps}
\end{figure}

Winnowing to the desired APOGEE wavelength range initially involved
inspection of the \cite{Hinkle95} atlas of the infrared spectrum of
Arcturus (\S \ref{sec:chemicalelements}), followed by use of synthetic
spectra, to assess the potentially useful atomic lines.  Figure
\ref{fig:element_maps} shows a telluric absorption spectrum as well
as examples of these element line maps that guided the selection of the
specific wavelengths within the $H$-band used for APOGEE.

\section{B: Calculations Used in Estimates of the APOGEE $S/N$ Requirements} 
\label{sec:SN_req}

To quantify the survey $S/N$ 
requirements we performed the following exercise:  A series of 
$H$-band spectra for RGB stars with $T_{eff}$ =
4000 K and $\log{g} = 1$, with [Fe/H] = $-2$, $-1$ were calculated.
For each case, we computed two spectra, one with scaled-solar
metal abundances ($F(0.0)$), and a second one in which the abundance of a particular
element was increased by 0.1 dex ($F(0.1)$).  From these,
we derived the minimum $S/N$ required
as the inverse of $\min |{F_i(0.1)/F_i(0.0) -1}|$, where the index $i$ runs
through all the pixels in the spectra.  The elements modeled were: C, 
N, O, Na, Mg, Al, Si, S, K, Ca, Ti, V, Mn, Fe, and Ni, and they are 
listed in Table~\ref{tab:sntab} from the most to least challenging, 
based on this analysis.

Table~\ref{tab:sntab} summarizes some of these $S/N$ per (0.15${\rm\AA}$)
pixel requirements for 0.1 dex abundance accuracies from the spectral
synthesis described for three resolutions spanning the nominal
range originally considered for APOGEE, $R = 14990$, 21414 and 29979 (as shown in
Fig.\ \ref{fig:SNR}, more resolutions than this were sampled), and
for stars with [Fe/H] $= -2.0$, $-1.0$ and 0.0.  From these numbers,
one can see that the $S/N$ requirements are higher for lower resolution
and lower metallicity, which is associated with the difficulty of
measuring variations in poorly resolved and weak lines, respectively.
According to this initial estimate, for the required
abundance accuracy to be reached for all top priority elements from
\S\ref{sec:chemicalelements}, a $S/N$ of at least $\sim$ 100-150/pixel  must be
achieved, depending on the spectral resolution of the APOGEE
spectrograph.

\begin{table}[htdp]
\caption{Required $S/N$ for detection of 0.1 dex abundance variations}
\label{tab:sntab}
\begin{center}
\begin{tabular}{rrrrrrrrrrl}
\hline
$R=$ &	15k &	21k &	30k &	15k &	21k &	30k   & 15k     &	21k &	30k \\
$[$Fe/H]= &	-2.0&	-2.0&	-2.0&	-1.0&	-1.0&	-1.0&	0.0 & 0.0&0.0  \\
Element	&&&&&&&&&& Priority \\
\hline
Na&	3648.0&	2673.7&	2050.3	& 430.6	& 309.8	& 230.0   & 78.7&	 56.0    & 41.4 & medium \\
S&	1498.4&	1067.2&	 802.0	& 232.4	& 167.2	& 125.9 & 144.0&	 104.8  & 81.6 & medium \\
V&	2089.3&	1504.7&	1124.0	& 231.0	& 164.4	& 121.4   & 61.0&	 42.4    & 30.5 & lower \\
K&	 697.5&	 505.6&	 384.4	& 105.6	& 75.3	& 56.6     & 65.0&	 44.6    & 33.1 & medium \\
Mn&	 260.0&	 184.9&	 136.2	& 73.0	& 50.9	& 36.9&	  64.8&	 46.9    & 34.9 & medium \\
Ni&	 142.0&	 101.6&	 76.7  	& 64.9	& 45.7	& 34.3&	 55.3&	 46.4    & 35.4 & top  \\
Ca&	 126.1&	 89.5&	 66.0 	& 60.2	& 42.7	& 31.5&	 57.7&	 41.0    & 30.3 & top  \\
Al&	 66.2&	 47.2&	 35.1	         & 59.6	& 41.8	& 30.4&	 60.3&	 42.1    & 31.5 & top  \\
Si&	 51.0&	 35.2&	 25.3	         & 53.4	& 38.6	& 28.6&	 43.1&	 35.7    & 29.6 & top  \\
N&	 199.7&	 147.3&	 113.4	& 52.5	& 41.7	& 32.2&	 25.4&	 21.4    & 18.1 & top  \\
Ti&	 154.1&	 110.0&	 81.8	         & 51.7	& 36.5	& 26.7&	 53.7&	 38.9    & 29.3 & medium \\
Mg&	 46.4&	 33.1&	 24.7	         & 48.9	& 36.7	& 27.7&	 34.7&	 26.4    & 20.4 & top   \\
Fe&	 57.1&	 41.6&	 31.2	         & 38.9	& 34.3	& 25.5&	 25.9&	 21.3    & 18.4 & top   \\
C&	 56.2&	 40.4&	 29.9	         & 19.5	& 14.8	& 11.3&	 10.2&	 8.3      & 6.8   & top   \\
O&	 34.5&	 24.5&	 18.1	         & 18.8	& 14.6	& 10.8&	 11.9&	 9.1      & 7.4   & top   \\
\hline
\end{tabular}
\end{center}

\end{table}

The analysis above did not account for the throughput variations 
expected along the APOGEE spectra, nor did it allow for the fact 
that for some elements, more lines can be used in the analysis than 
others. The latter effect is particularly important, as more lines 
lead to a relative decrease in the $S/N$ requirement. To 
include those effects in the estimates, we defined the following 
metric, for each element X:
\begin{equation} \label{fc}
FC(X) =\sum_{i=1}^N  \bigg[ \bigg({F_i(0.1) \over F_i(0.0)} -1\bigg) \times t_i \bigg] ^2 
\end{equation}
where the expected spectrograph throughput,as estimated 
a priori by the hardware team and 
normalized to 1.600~$\mu$m, is given by $t_i$. These computations
were performed for synthetic spectra simulated at the expected APOGEE 
sampling and resolution, including their
dependence on wavelength, as provided by the hardware team.

In most cases, one finds for the ratio spectrum that $F(0.1)/F(0.0)
< 1$ due to strengthened line absorption, but for some elements and
some wavelengths we find $F(0.1)/F(0.0) >1$ due to interactions
between species through molecular dissociation equilibrium (e.g.,
an increase in C when $C/O<1$ will cause increased CO strengths but
reduced OH aborption).

To avoid lots of pixels with very small changes from influencing the 
result, a series of cutoffs were imposed, setting the fractional 
changes to zero if they were smaller (in absolute value) than 0.005, 0.01, 
0.015, 0.02, 0.025. Obviously the larger of these numbers are 
fairly extreme; if one ignores all lines that change the flux by less than 
2-2.5\%, there are some elements that become entirely lost.
The estimated required $S/N$ is then is given by
\begin{equation} \label{sqrt}
S/N =1/\sqrt{FC}.
\end{equation}
Because the synthetic spectra are sampled at half the pixel size
(i.e., as combined from two dithered exposures), these are $S/N$
required in half the total observing time. This can be seen as the
required $S/N$ being $\sqrt{2}$ times larger.  Note this exercise
only takes into consideration the impact of elemental abundance on
line opacities.  Therefore, the results are somewhat inaccurate for
those elements that affect continuum opacity, which is dominated
by H$^{-}$ and therefore fairly sensitive to the abundances of
important electron donors.

The formal results are shown in Table~\ref{tab:snjon}.  Note that
$S/N=\infty$ means that the element is undoable, i.e., no signal
above the cutoff is detected, which means $S/N\rightarrow\infty$.
From these numbers, one can see that the overall required $S/N$ is
much lower than those from Table~\ref{tab:sntab}. This is because
the metric in Equation~\ref{fc} is roughly proportional to the
number of pixels that are sensitive to a given abundance, whereas
the metric used to generate the numbers in Table~\ref{tab:sntab}
is sensitive to the pixels contained in a few prominent features
only.  That explains why C, N, and O require such low $S/N$, due to
the many thousands of CN, CO, and OH transitions that overlap in
the $H$-band.

Interestingly, the dependence of the required $S/N$ on [Fe/H] is
much stronger than in Table~\ref{tab:sntab}.  This is probably due
to the combined effect of sensitivity per line getting lower and
lines vanishing as one goes towards lower metallicity. The latter
effect does not affect the numbers in Table~\ref{tab:sntab}, which
are only based on the single most sensitive feature.  Even though
these exercises provided a first assessment of the $S/N$ needs of the
survey, the conclusions from these tests were relatively limited,
since important effects such as line blending and limitations of
models to reproduce real spectra were not considered.
Another factor ignored
in this analysis was the
availability of continuum points either for equivalent width
measurements or to guide the comparison with synthetic spectra. At
lower temperatures and higher metallicity, continuum points are
expected to be fewer, posing a stronger requirement on the minimum
$S/N$ needed to determine the continuum accurately.  These issues
make it very difficult for one to make definitive a priori estimates of
both the overall sensitivity of the spectra to abundance variations
and the effects of line blending.

A primary conclusion obtained from this early analysis was that a precise estimate 
of the $S/N$ mandated by the abundance accuracy requirement of the 
APOGEE survey depends on variables whose effects could not be 
simulated accurately enough at the time. However, it is clear that the ideal 
$S/N$ is nicely bracketed by the two extremes resulting from the 
exercises above, being most likely closer to the numbers in 
Table~\ref{tab:sntab}, given that the metric in Equation \ref{fc} tends to 
strongly overestimate the contribution by lines that are either too 
weak or cannot be resolved. Therefore, prudence dictated a conservative 
approach in this case, and therefore we stipulated a minimum $S/N$ 
requirement of 100/pixel, which is closer to the numbers provided 
in Table~\ref{tab:sntab} and was expected to meet the abundance accuracy 
requirements for at least all of the top priority elements.

\begin{table}[htdp] \caption{Required $S/N$ for detection of 0.1 dex
abundance variations} \begin{center} \begin{tabular}{rrrrl}
\hline
cutoff at 0.005 &&& \\
\hline
    $S/N/pixel$ &&& \\
\hline
    $[Fe/H]$  &  0  & -1  & -2 & Priority \\
\hline
    Na & 28.8  & $\infty$     &   $\infty$  & medium \\
      S  & 42.8  & 120.9 &   $\infty$  & medium\\
      V  & 15.0 &  94.8 &   $\infty$  & lower\\
      K  &  28.1 &  53.7 &  $\infty$  & medium\\
    Mn &  12.1 &  25.2 & 276.9  & medium\\
      Ni & 7.3  & 13.7 &  58.4 & top \\
     Ca &   9.8  & 12.3 &  38.6 & top \\
       Al &  16.0 & 16.5  & 24.6 & top\\
    Si & 2.6  &  5.8  & 10.8 & top\\
    N  &  1.6  &  4.4  & 96.5 & top \\
       Ti & 8.0  & 13.3  & 55.0 & medium \\       
    Mg &   1.3 &   1.9 & 4.1& top \\
    Fe &   1.2 &   1.9 &   6.9 & top \\
    C  &   0.6 &   1.7 &   8.0 & top \\
    O  & 1.2  &  2.0 & 3.1 & top\\

\hline
cutoff at 0.02 &&&  \\
\hline
$S/N/pixel$ &&&  \\ 
\hline
    $[Fe/H]$  &  0  & -1  & -2 & Priority\\
\hline
Na  & 63.0 &  $\infty$ &   $\infty$   & medium\\
S   & $\infty$  &  $\infty$   & $\infty$   & medium  \\
V   & 21.5  & $\infty$  &$\infty$   & lower\\
K   & 45.5 &  $\infty$  &  $\infty$  & medium \\
Mn  & 21.8 &   $\infty$  &  $\infty$  & medium \\
Ni  & 10.7 &  25.8 &   $\infty$   & top\\
Ca  & 12.9 & 14.3  &  $\infty$   & top\\
Al  & 27.6 &  20.2 &  35.7 & top\\
Si  & 6.9 & 8.8  & 15.2  & top\\
N   & 2.1  & 18.2 &$\infty$  & top \\
Ti  & 11.2  & 16.2 &   $\infty$   & medium \\
Mg  &  4.8 &   8.3 &  11.9  & top\\
Fe  & 1.7  &  3.5  & 12.7  & top\\
C   & 0.7  &  2.4  &  $\infty$  & top  \\
O   &  1.4 &   2.4 &   3.9 & top \\

\hline 
\end{tabular} 
\end{center} 
\label{tab:snjon} 
\end{table}

\section{C: Simulations of the Sensitivity of Cadenced APOGEE
Observations to Binary Stars} \label{sec:bin_sim}

To assess the cadencing requirements to optimize the sensitivity
of APOGEE observations to the presence of binary stars a suite of
simulations of APOGEE observations of parent distributions of stellar
populations with binaries was performed.  Binary systems were
generated for three representative lines of sight in the APOGEE
survey, at Galactic coordinates of $(l, b)$ = (0, 2.5), (45, 2.5),
and (90, 0) degrees.  The distribution of primary masses were
generated using the Trilegal model \citep{Girardi05}.  For each
line of sight, we generated 300 binary systems and ran 1000 simulated
APOGEE surveys to observe these stars for each tested cadence.   The
period distribution of the binaries were adopted from \citet[][``DM91''
hereafter]{Duquennoy91}, with a cut-off at $\log{P}$ (days) $<$ 0.
Note that the \citet{Griffin85} data on red giants presented in
DM91 generally follow this period distribution, but with a deficiency
at low $\log{P}$. This difference suggests that the simulations may
have slightly overestimated the fraction of short-period binaries.
We did compensate somewhat by discarding physically impossible
binary systems with the simple constraint that the stellar radius
cannot exceed the orbital separation, but we did not consider
more sophisticated schemes such as the plausibility of systems that
would be undergoing (or have undergone) tidal interactions.  The
eccentricity distribution also follows that in DM91, which includes
circularization of orbits having $P < 100$ days.  (These
simulations pre-dated the \citealt{Duchene13} finding that the
DM91 eccentricity distribution is a much poorer fit to observed
eccentricities of binaries than a simple uniform distribution of
eccentricities.)   The adopted distribution of secondary mass ratios
is uniform between 0.1 and 1.0. The orientation of the binary orbital
axis is isotropic, and the longitude of periastron is uniformly
distributed. The time of periastron passage was randomized uniformly
between zero and the length of the period.

To simulate the observations, we defined a 3-visit observing cadence
of $O_1$, $O_2$, and $O_3$ days, where the time of the first visit
$O_1$ is always 0 and $O_2$ and $O_3$ are the times of the second
and third visit observation with respect to the first observation.
To avoid perfect integer spacing of the simulated observations, we
added random offsets to $O_1$, $O_2$, and $O_3$.  Once the observation
times and orbital parameters were defined for each binary
system, we used the IDL code ``helio\_rv'' in the IDL Astronomy
User's Library to calculate the radial velocity at each observation
date and then added measurement noise drawn from a Gaussian
distribution with $\sigma$ of 0.5~ km s$^{-1}$.  We then calculated
the average RV of each binary system. Since we adopted 0~ km s$^{-1}$
for the systemic RV of each system, this average RV corresponds to
$\Delta RV$, the difference between the measured and true velocity
of the system. Binaries with large $\Delta RV$ will have the largest
detrimental effect on kinematical measurements of the stellar
populaitons.

We explored cadences ($O_1$, $O_2$, $O_3$) over a number of possible
baselines spanning a week (0, 1, 7.1), one month (0, 7.2, 30.4),
three months (0, 15.2, 91.2), and one year (0, 30.4, 365.25).  The
detectability of each binary is defined as the maximum RV difference
between the three RV observations. The simulations showed that in
the case of three observations, the longer the baseline, the greater
number of binaries that are identified and, more importantly, the
fewer binaries with large $\Delta RV$ that are missed.  However,
the improvement in the number of large $\Delta RV$ identified between
one cadence and the next longest cadence diminished as the baseline
increased.  We also explored the effects of the timing of the second
observation in the one month cadence by varying $O_2$ between 1 and
5 days and found that there was little change in results.

\begin{figure}[h]
\vspace{-50mm}
\includegraphics[width=0.50\textwidth]{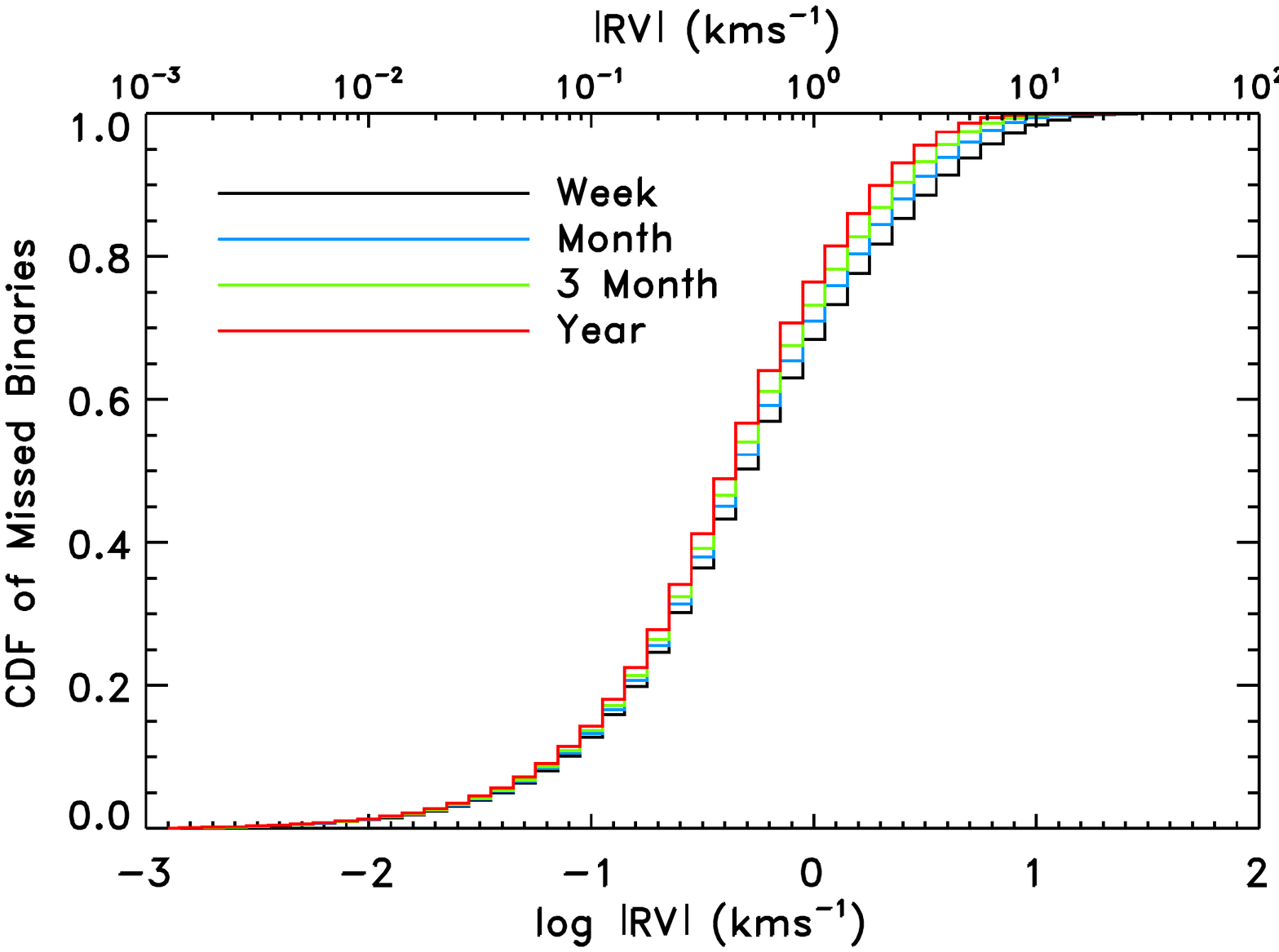}
\includegraphics[width=0.50\textwidth]{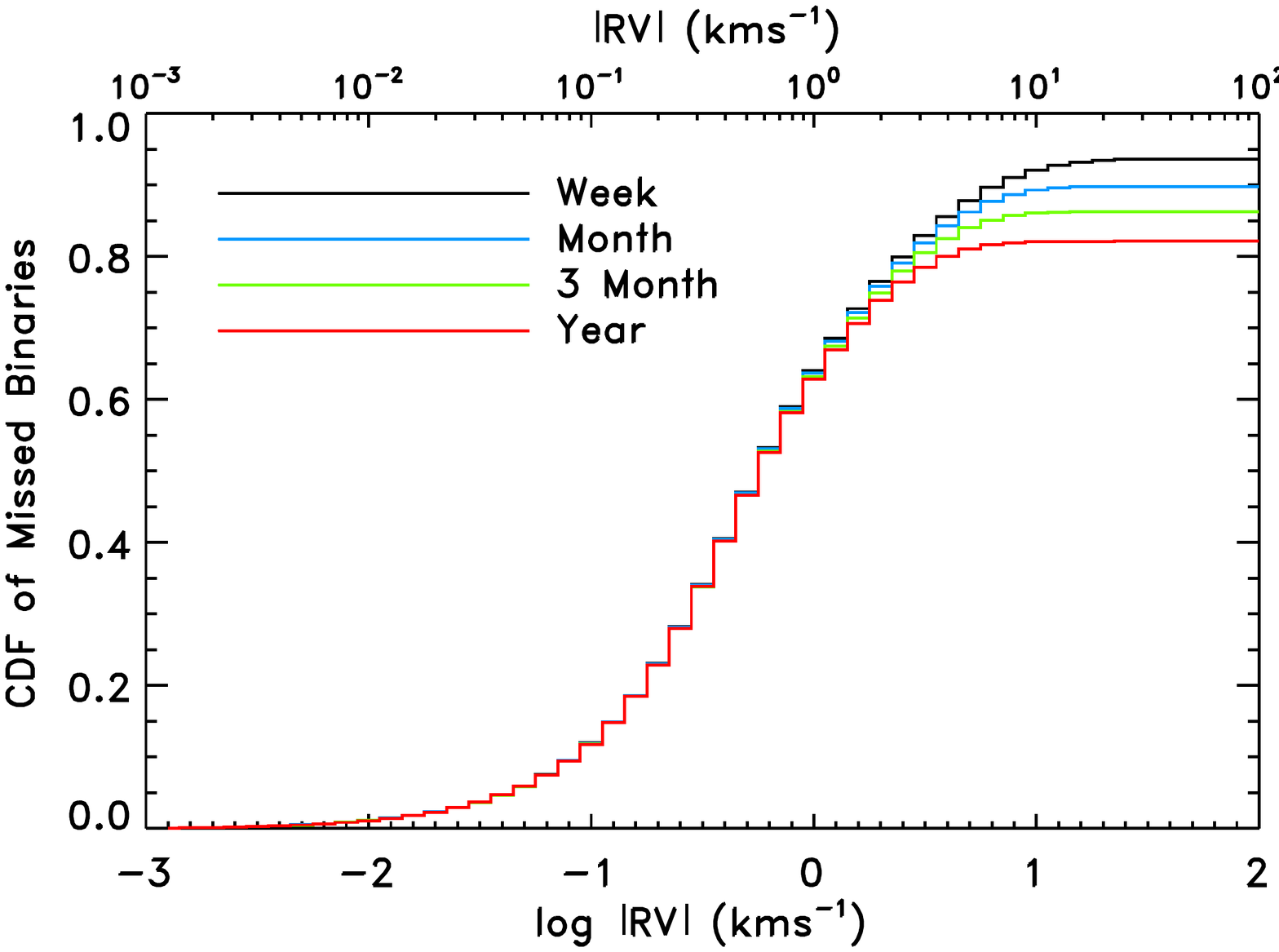}
\caption{\footnotesize Normalized cumulative distribution functions
of the fraction of stars having a given APOGEE-measured radial
velocity offset from the true systemic radial velocity for the four
different observing cadences described in the text.  Note that these
are shown only for the binaries that are not flagged as a ``likely
binary'' using the $>$4 km s$^{-1}$ pairwise-difference criterion
described in \S \ref{sec:binaries}.  In the lefthand plots the
normalizations are to the total number of binaries missed for that
cadence.  In the righthand plot, the distributions are normalized
to the total number of simulated binaries; thus, the difference in
the maximum CDF-value between each cadence corresponds to the
different fractions of ``missed binaries''.  }
\label{fig:missed_binaries}
\end{figure}

\section{D: Simulations of the Expected Galactic Distribution of APOGEE Targets} 

Section \ref{sec:fieldselectionprinciples} describes the main
philosophy that was adopted for the APOGEE field targeting plan,
while \S \ref{sec:fieldselectionevolution} describes the four major
phases through which the targeting plan evolved to the
final targeting configuration (\S \ref{sec:fieldplan}).  This
evolution was guided by application of Galactic stellar population
models --- namely the Trilegal \citep{Girardi05} and Besan\c{c}on
\citep{Robin03} Galaxy models --- to multiple strawman field placement
designs for each of the three principal field star survey regions:
disk ($|b| \sim 0^{\circ}$), bulge ($|l| < 20^{\circ}$ and $|b| <
20^{\circ}$), and halo ($|b| > 20^{\circ}$).  This modeling was
essential to the task of optimizing not only the specific locations
of fields but also the cohort distributions, and color and magnitude
limits/numbers of visits employed.  Some of the specific issues
these modeling efforts addressed were, e.g., ensuring that each
Galactic component was well sampled across the greatest possible
distance ranges, that the stellar samples were optimized to target
predominantly giant stars so as to make the largest distances most
accessible, and that each major Galactic population would be amply represented
within APOGEE.  Examples of this work and earlier targeting plans
are provided in this Appendix to demonstrate how some trade-offs
were evaluated, and how vestiges of earlier plans, modified later, 
can be found in the final APOGEE samples.

\begin{figure}
\vspace{-35mm}
\includegraphics[width=0.80\textwidth]{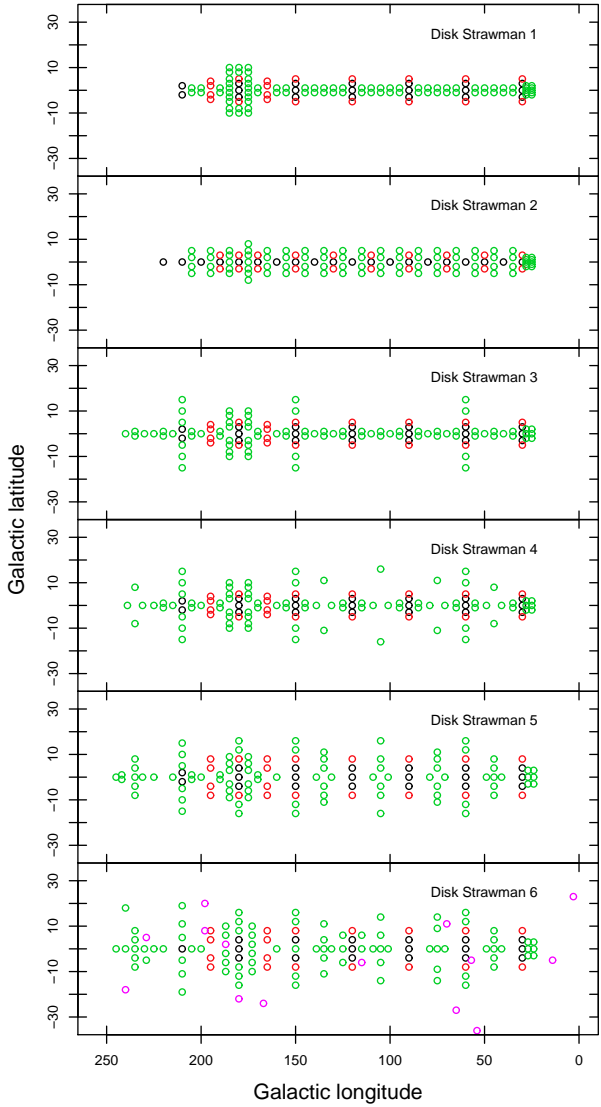}
\caption{\footnotesize 
Evolution of the design of the disk field distribution is seen from
the earliest plan, ``Disk Strawman 1'', to the nearly final disk
plan, ``Disk Strawman 6''.  Green circles represent 3-hour fields,
red circles are 10- or 12-hour fields, and black circles are 24-
or 30-hour fields.}
\label{fig:diskstrawman}
\end{figure}

Figure~\ref{fig:diskstrawman} gives some idea of the evolution of
the disk targeting strategy, and reflects several static as well
as changing considerations during the survey planning stages.  Early
field placement plans were constructed under the assumption that
75\% of APOGEE time would be conducted in conjunction with the
MARVELS survey, with only 25\% of the time with APOGEE only.  Because
MARVELS sought a total of 30 approximately one hour visits to build
its time series database, the early APOGEE targeting plans had a
greater fraction of long, many-visit fields, and was thus expected
to probe relatively fewer lines of sight, but with more depth per
direction (e.g., ``Disk Strawman 1" and ``Disk Strawman 2'' in
Figure~\ref{fig:diskstrawman}).  Because APOGEE was expected to start
in the middle of the second of three 2-year campaigns of MARVELS
fields, deep APOGEE fields were designed around 10-hour fields,
where APOGEE would ``join on'' to fields already started by MARVELS,
and 30-hour fields, which would be initiated for both APOGEE and
MARVELS simultaneously.\footnote{The 30-/10-hour plan was later
modified to a 24-/12-hour plan. In initial planning all ``visits"
would be kept to less than 60 minutes to limit the amount of MARVELS
fringe broadening from geocentric velocity changes over the exposure.}

Starting from the initial plan to very systematically and symmetrically
survey the disk while coordinating co-observing with the MARVELS
program (e.g., Disk Strawman 1 in Figure~\ref{fig:diskstrawman}), the
field plan for exploring the Galactic disk was then rearranged in
response to (a) greater expected observing time in winter due both
to the longer nights as well as how grey time was divided among
SDSS-III surveys (resulting, e.g., in the larger coverage of the
Galactic anticenter seen in models Disk Strawman 2 and later plans);
(b) the desire to increase the number of expected stars from the
Intermediate Population II ``thick disk'', which was achieved not
only by altering the latitude distribution of fields but moving the
location of the deep fields from the mid-plane to ever higher
latitudes and tweaking the cohort distributions (for example, Disk
Strawman 3 increased the number of thick disk stars with $|Z_{GC}|
> 1.5$ kpc by a factor of 20 and 4 compared to Disk Strawman 1 and
2, respectively, and these numbers were further increased by even
later models); and (c) the desire to include more calibration cluster
fields (e.g., motivating Disk Strawman 5 and 6).  The eventual
addition of many more 3-hour fields due to the MARVELS descope and
further adjustments in response to the addition of observations of
the Kepler field are not reflected in the Figure~\ref{fig:diskstrawman}
designs, but evident in the final targeting plan (Fig.~\ref{fig:field_plan}).

\begin{figure}[h]
\includegraphics[width=1.00\textwidth]{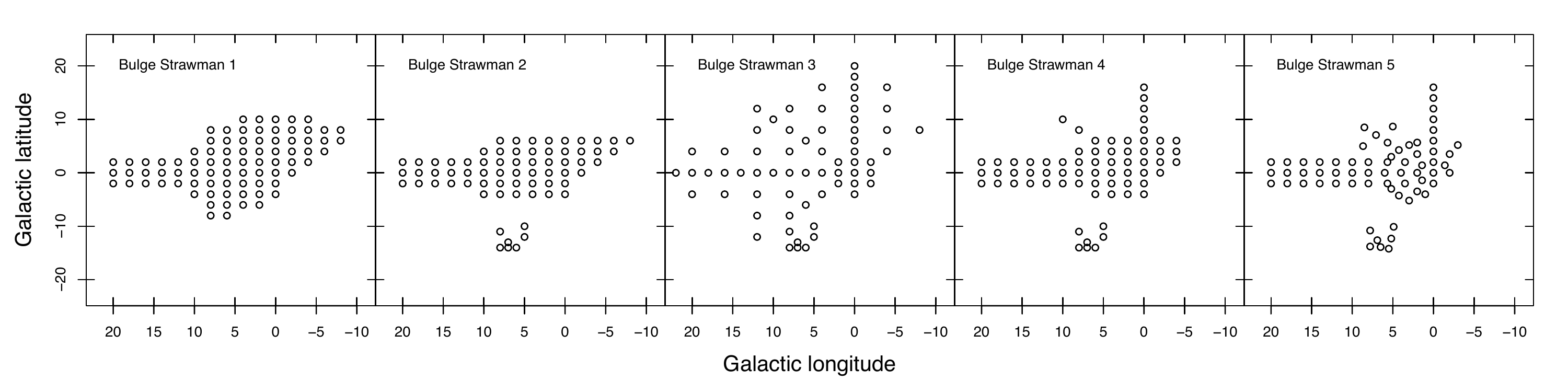}
\caption{\footnotesize 
Various strategies tested for the sampling of the bulge, 
from the earliest notions (``Bulge Strawman 1'') to that eventually adopted 
(``Bulge Strawman 6'').
}
\label{fig:bulgestrawman}
\end{figure}

Figure~\ref{fig:bulgestrawman} illustrates a similar evolution in the design
of the bulge field coverage.  Starting from simple rectilinear distributions (e.g., 
``Bulge Strawman 1'' and 2), the plans were altered to account for
(a) changes in the disk observing plan, because fields focused on observing
the inner disk compete for the same observing windows as the bulge fields
(e.g., Bulge Strawman 2 and later);
(b) the introduction of fields sampling the core of the Sgr dSph galaxy 
(Bulge Strawman 2 and later); 
(c) optimization of the limited amount of bulge accessibility from APO
to make sure that APOGEE sufficiently explored key structural axes 
(major, minor, diagonal) of the bulge (e.g., Bulge Strawman 3 and later); and
(d) exploitation of APOGEE's ability to probe highly obscured fields close to the
Galactic center that are less accessible to other spectroscopic surveys, which 
motivated a tighter concentration of the APOGEE fields around the center
(e.g., Bulge Strawman 5).

Less effort was invested in modeling halo targeting, because it was
realized early on that a large fraction of the available relevant
observing hours would already be needed to probe globular clusters
and known halo streams.  Eventually, additional simple ``picket
fence'' distributions of predominantly 3-hour fields were added to
these deep halo fields along $b =$ 30$^{\circ}$, 45$^{\circ}$,
60$^{\circ}$, and 75$^{\circ}$ (see Figure~\ref{fig:field_plan}).

\begin{figure}[h]
\includegraphics[width=1.00\textwidth]{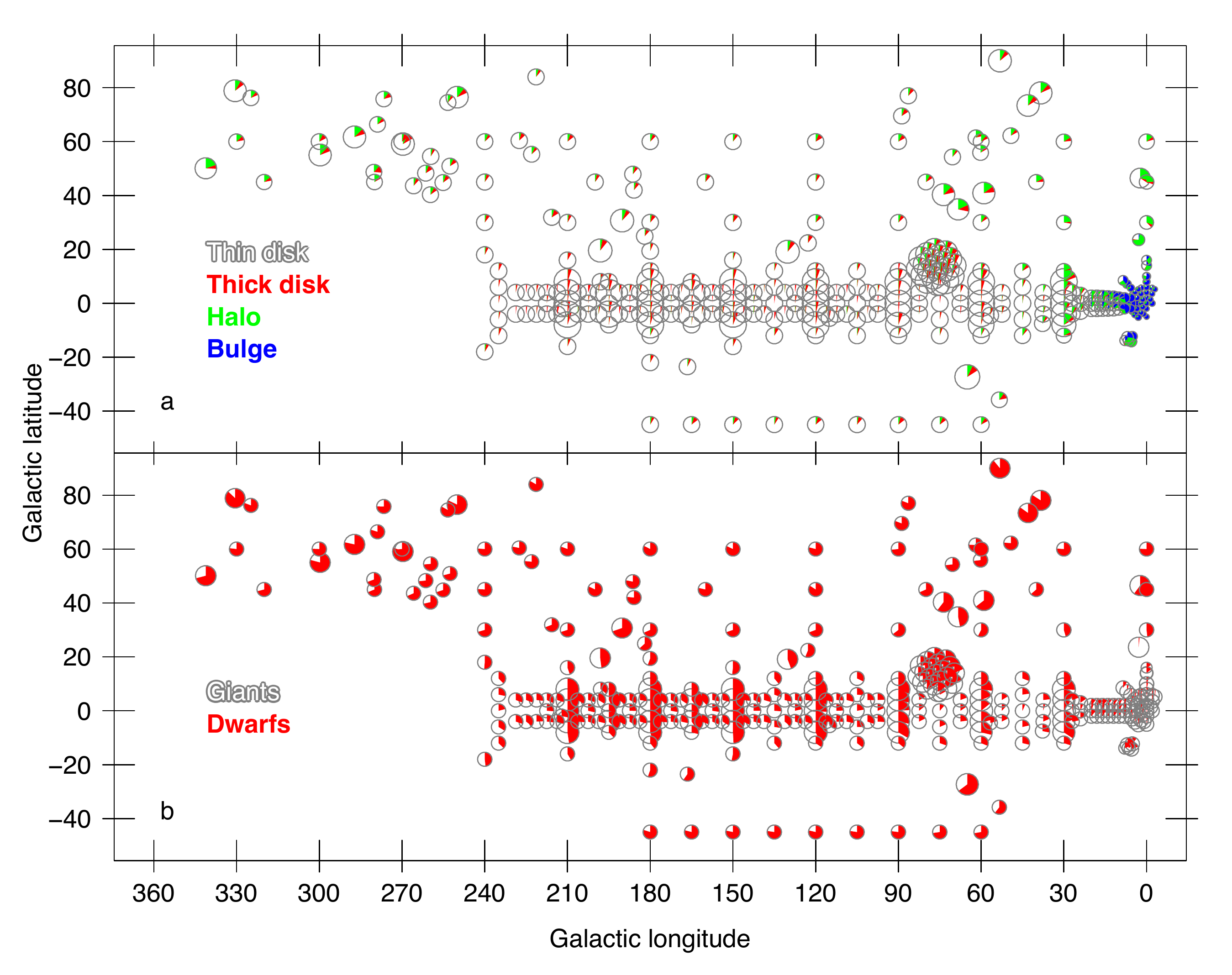}
\caption{\footnotesize Example of Trilegal modeling
of the expected distributions of  of different properties of sampled stars.
Each field position is
represented by a pie chart describing target distributions of
Galactic component membership (panel a) and evolutionary stage (panel b).
In both panels, symbol size is proportional to number of visits.
(a) Expected distributions of stars 
from each of the major Galactic stellar populations 
on a field-by-field basis.  This same distribution
is shown in an alternative, Cartesian format in Figures \ref{fig:XYApogeePlot}
and \ref{fig:RgZApogeePlot}.   According to the simulation, thin disk
(bulge) fields would be expected to be dominated by thin disk (bulge) stars.  
However, in halo fields the modeled contribution by thin disk stars is overestimated
in this particular simulation, for several reasons.
Many halo fields were placed on known 
overdensities (which increases the relative halo contribution) and globular clusters, 
where targets were selected to maximize cluster membership; 
these strategies are not taken into account in the
Trilegal simulation, which is based on a Galactic model
composed of symmetrical thin and thick disks, halo, and bulge.
Moreover, the simulation does not include the effect of the additional halo field
selection criterion
based on Washington+DDO photometry (\S\ref{sec:sources}), 
which was designed to minimize (predominantly disk) dwarf contamination,
and, indeed, was motivated by model results like that shown here.
(b) Example of Trilegal modeling of the expected giant/dwarf ratio
on a field-by-field basis, for the final, approved APOGEE plan
(after MARVELS descope).  This model shows that the 
disk fields were expected to be dominated by giants
(as borne out by the survey); on the other hand, the modeled
dwarf/giant ratio in halo fields is 
much higher than found in the actual survey, because of the 
additional implemented strategies explained above. 
}
\label{fig:PieChart}
\end{figure}

The final adopted field plan (Figure~\ref{fig:field_plan}) descends, after
several additional alterations, from the sum of the Disk Strawman 6 and
Bulge Strawman 5 plans, along with the Halo field strategy discussed above.
Many of the ensuing alterations in field positions were in response
to further reductions in the MARVELS program, which allowed additional
APOGEE fields to be included.  Further tweaking of field positions
and cohort distributions came as a direct result of further modeling
to optimize the representation of halo and thick disk as well as
the overall number of giant stars in the survey (the latter aided
by the inclusion of Washington$+DDO51$ photometry in high latitude
fields).  Figures~\ref{fig:PieChart}a and \ref{fig:PieChart}b
illustrate some of the results of such modeling.  Completing the
final field distribution shown in Figure~\ref{fig:field_plan} was not
only enabled, but further amended mid-survey, by the addition of
both the twilight observing (\S \ref{sec:twilight}) and dark time
observing (\S \ref{sec:year3}) campaigns, and modified further, of
course, by prevailing weather conditions during the APOGEE survey.

We note that throughout the modeling efforts conducted to shape the
APOGEE targeting plan large variations in expected distributions
were seen when comparing the results of the TRILEGAL and Besan\c{c}on
models.  These differences in results were, in part, due to differences
in the adopted prescriptions for the thick disk structure as well
as the model for reddening between the codes.  While tweaking of
model parameters could bring the results of the two models into
better agreement and make more consistent projections, systematic
discrepancies of one sort or another in the results of the two modeling
codes typically remained.  Such discrepancies between the models
illustrate the lack of strong observational constraints, particularly
in the Galactic mid-plane, that can be used to calibrate them ---
a problem that can now be at least partly remedied by application
of APOGEE results.

\section{E: Laboratory Astrophysics Efforts and Development of the APOGEE Line List}
\label{sec:linelists}
Physical data (line identifications, wavelengths, transition
probabilities, excitation potentials, damping constants) for line
transitions in near infrared stellar spectra are not as mature as
those for optical spectra.  Because such data are critical inputs
to the ASPCAP processing of APOGEE spectra, a significant effort
was put into canvassing the literature and collating previously
published data, whether theoretically derived or empirically measured
in the laboratory or through astrophysical observations.  In addition,
it was found necessary to collaborate with laboratory
atomic physicists (primarily at the University of Wisconsin and
Imperial College, London) to supplement  and improve the line
list database.  To ensure consistency across these multi-sourced
data and reduce their uncertainties, the catalogs of data were used
to generate synthetic spectra which were then compared to very high
resolution, Fourier Transform Spectroscopy data on well known stars:
$\mu$ Leo, $\beta$ And, $\delta$ Oph, Arcturus and the Sun
\citep{Smith13,Shetrone15} to create improved ``astrophysical''
 line lists.  The result of this enterprise has been the
creation of catalogs of data for as many as 134,000 atomic and
molecular features, meticulously checked against the cataloged and
APOGEE-observed (\S \ref{sec:standards}) spectra of the Sun and
Arcturus \citep{Shetrone15}.

\acknowledgements

We thank Whitney Richardson for help with Figure \ref{fig:plugplate}.

S.R.M. acknowledges support from National Science Foundation grant AST-1109178.

R.P.S. acknowledges support from Gemini Observatory, which is
operated by the Association of Universities for Research in Astronomy,
Inc., on behalf of the international Gemini partnership of Argentina,
Australia, Brazil, Canada, Chile, and the United States of America.

D.A.G.H. and O.Z. acknowledge support provided by the Spanish Ministry of Economy and Competitiveness under grant AYA-2011-27754.
S.Mathur acknowledges support from the NASA grant NNX12AE17G.

Sz.M. has been supported by the J{\'a}nos Bolyai Research Scholarship
of the Hungarian Academy of Sciences.

Funding for SDSS-III has been provided by the Alfred P. Sloan
Foundation, the Participating Institutions, the National Science
Foundation, and the U.S. Department of Energy Office of Science.
The SDSS-III web site is http://www.sdss3.org/.

SDSS-III is managed by the Astrophysical Research Consortium for
the Participating Institutions of the SDSS-III Collaboration including
the University of Arizona, the Brazilian Participation Group,
Brookhaven National Laboratory, University of Cambridge, Carnegie
Mellon University, University of Florida, the French Participation
Group, the German Participation Group, Harvard University, the
Instituto de Astrofisica de Canarias, the Michigan State/Notre
Dame/JINA Participation Group, Johns Hopkins University, Lawrence
Berkeley National Laboratory, Max Planck Institute for Astrophysics,
New Mexico State University, New York University, Ohio State
University, Pennsylvania State University, University of Portsmouth,
Princeton University, the Spanish Participation Group, University
of Tokyo, University of Utah, Vanderbilt University, University of
Virginia, University of Washington, and Yale University.


\begin{thebibliography}{} 

\bibitem[Abadi et al.(2003)]{Abadi03} Abadi,~M.~G., Navarro,~J.~F.,
Steinmetz,~M., \& Eke,~V.~R., 2003, \apj, 591, 4 99

\bibitem[Adibekyan et 
al.(2012)]{Adibekyan12} Adibekyan, V.~Z., Sousa, S.~G., Santos, N.~C., et al.\ 2012, \aap, 545, AA32 

\bibitem[Adibekyan et 
al.(2013)]{Adibekyan13} Adibekyan, V.~Z., Figueira, P., Santos, N.~C., et al.\ 2013, \aap, 554, AA44 

\bibitem[Ahn et al.(2014)]{Ahn14} Ahn, C.~P., Alexandroff, 
R., Allende Prieto, C., et al.\ 2014, \apjs, 211, 17 

\bibitem[Alam et al.(2015)]{Alam15} Alam, S., Albareti, F.~D., 
Allende Prieto, C., et al.\ 2015, arXiv:1501.00963 

\bibitem[Alard(2001)]{Alard01} Alard, C.\ 2001, \aap, 379, L44 

\bibitem[Allende Prieto et al.(2006)]{AllendePrieto06} Allende Prieto, C.,
Beers, T.C., Wilhelm, R. et al.\ 2006, \apj, 636, 804


\bibitem[Allende Prieto et al.(2008)]{AllendePrieto08} Allende Prieto, 
C., Majewski, S.~R., Schiavon, R., et al.\ 2008, Astronomische Nachrichten, 
329, 1018 

\bibitem[Anders et al.(2014)]{Anders14} Anders, F., Chiappini, C.,
	Santiago, B.~X. 2014, \aap, 564, 115

\bibitem[Anders et al.(2015)]{Anders15} Anders, F. et al.\ 2015 in preparation

\bibitem[Arns et al.(2010)]{Arns10a} Arns, J., Wilson, J.~C., 
Skrutskie, M., et al.\ 2010, \procspie, 7739, 


\bibitem[Asplund et al.\ (2005)]{Asplund05} Asplund, M., Grevesse, N. \&
Sauval, A.J. 2005, in ASP Conf. Ser. 336, Cosmic
Abundances as Records of Stellar Evolution and Nucleosynthesis, ed. T. G.
Barnes, III \& F. N. Bash (San Francisco, CA: ASP), 25

\bibitem[Aumer \& Sch{\"o}nrich(2015)]{Aumer15} Aumer, M., \& Sch{\"o}nrich, 
	R.\ 2015, arXiv:1507.00907 


\bibitem[Auvergne et 
al.(2009)]{Auvergne09} Auvergne, M., Bodin, P., Boisnard, L., et al.\ 2009, \aap, 506, 411 



\bibitem[Babcock(1977)]{Babcock77} Babcock, H.~W.\ 1977, \skytel, 54, 90 

\bibitem[Belokurov et al.(2006)]{Belokurov06} Belokurov, V., 
Zucker, D.~B., Evans, N.~W., et al.\ 2006, \apjl, 642, L137 

\bibitem[Belokurov et al.(2007)]{Belokurov07} Belokurov, V., 
Zucker, D.~B., Evans, N.~W., et al.\ 2007, \apj, 654, 897 

\bibitem[Bensby et al.(2003)]{Bensby03} Bensby, T., Feltzing, S.,
\& Lundstr{\"o}m, I.\ 2003, \aap, 410, 527

\bibitem[Bensby et al.(2014)]{Bensby14} Bensby, T., Feltzing, S.,
\& Oey, M.~S. 2014, \aap, 562, 71

\bibitem[Bizyaev et al.(2006)]{Bizyaev06} Bizyaev, D., Smith, 
V.~V., Arenas, J., et al.\ 2006, \aj, 131, 1784 

\bibitem[Bird et al.(2013)]{Bird13} Bird, J.~C., Kazantzidis, 
S., Weinberg, D.~H., et al.\ 2013, \apj, 773, 43

\bibitem[Borucki et al.(2010)]{Borucki10} Borucki, W.~J., Koch, 
D., Basri, G., et al.\ 2010, Science, 327, 977 

\bibitem[Bournaud et al.(2009)]{Bournaud09}
  Bournaud,~F., Elmegreen,~B.~G., Martig,~M. 2009, \apj, 707, 1

\bibitem[Bovy et al.(2012)]{Bovy12} Bovy, J., Allende Prieto, 
C., Beers, T.~C., et al.\ 2012, \apj, 759, 131 

\bibitem[Bovy et al.(2014)]{Bovy14}
  Bovy,~J., Nidever,~D.~L., Rix,~H.-W., et al.\ 2014, \apj, 790, 127

\bibitem[Bovy et al.(2015)]{Bovy15} Bovy, J., Bird, J.~C., 
Garc{\'{\i}}a P{\'e}rez, A.~E., et al.\ 2015, \apj, 800, 83 
  
\bibitem[Bressan et al.(2012)]{Bressan12a}
Bressan,~A., Marigo,~P., Girardi,~L., et al. 2012, \mnras, 427, 127

\bibitem[Brunner et al.(2010)]{Brunner10} Brunner, S., Burton, 
A., Crane, J., et al.\ 2010, \procspie, 7735,  

\bibitem[Burgh et al.(2007)]{Burgh07} Burgh, E.~B., Bershady, 
M.~A., Westfall, K.~B., \& Nordsieck, K.~H.\ 2007, \pasp, 119, 1069 

\bibitem[Cabrera-Lavers et 
al.(2007)]{Cabrera-Lavers07} Cabrera-Lavers, A., Hammersley, P.~L., Gonz{\'a}lez-Fern{\'a}ndez, C., et al.\ 2007, \aap, 465, 825 

\bibitem[Cardelli et al.(1989)]{Cardelli89} Cardelli, J.~A., 
Clayton, G.~C., \& Mathis, J.~S.\ 1989, \apj, 345, 245 

\bibitem[Carlberg et al.(2015)]{Carlberg15} Carlberg, J.~K., 
Smith, V.~V., Cunha, K., et al.\ 2015, arXiv:1501.05625 

\bibitem[Castelli 
\& Kurucz(2004)]{Castelli04} Castelli, F., \& Kurucz, R.~L.\ 2004, arXiv:astro-ph/0405087 

\bibitem[Cenarro et al.(2004)]{Cenarro04} Cenarro, A.~J., 
S{\'a}nchez-Bl{\'a}zquez, P., Cardiel, N., 
\& Gorgas, J.\ 2004, \apjl, 614, L101 

\bibitem[Chaplin et al.(2010)]{Chaplin10} Chaplin, W.~J., 
Appourchaux, T., Elsworth, Y., et al.\ 2010, \apjl, 713, L169 

\bibitem[Chaplin 
\& Miglio(2013)]{Chaplin13} Chaplin, W.~J., \& Miglio, A.\ 2013, \araa, 51, 353 

\bibitem[Chiappini(2013)]{Chiappini13} Chiappini, C.\ 2013, 
Astronomische Nachrichten, 334, 595 

\bibitem[Chiappini et al.(2015)]{Chiappini15} Chiappini, C., Anders,
F., Rodrigues, T.~S., et al.\ 2015, \aap, 576, L12

\bibitem[Chiappini et al.(2001)]{Chiappini01} Chiappini, C., 
Matteucci, F., \& Romano, D.\ 2001, \apj, 554, 1044 

\bibitem[Chiappini et al.(2003)]{Chiappini03} Chiappini, C., 
Romano, D., \& Matteucci, F.\ 2003, \mnras, 339, 63 

\bibitem[Chojnowski et al.(2015)]{Chojnowski15} Chojnowski, S.~D., 
Whelan, D.~G., Wisniewski, J.~P., et al.\ 2015, \aj, 149, 7 


\bibitem[Churchwell et al.(2009)]{Churchwell09} Churchwell, E., 
Babler, B.~L., Meade, M.~R., et al.\ 2009, \pasp, 121, 213 

\bibitem[Ciddor (1996)]{Ciddor96} Ciddor, P.E. 1996, Applied
Optics, 35, 1566

\bibitem[Cirasuolo et al.\ (2014)]{Cirasuolo14} Cirasuolo, M.,
Afonso, J., Carollo, M., et al. 2014, in Society of Photo-Optical
Instrumentation Engineers (SPIE) Conference Series, Vol. 9147,
Society of Photo-Optical Instrumentation Engineers (SPIE) Conference
Series

\bibitem[Clough et al.\ (2005)]{Clough05} Clough, S.~A.,
Shephard, M.~W., Mlawer, E.~J., Delamere, J.~S., Iacono, M.~J.,
Cady-Pereira, K., Boukabara, S. \& Brown, P.~D. 2005, J. Quant. Spectrosc.
Radiat. Transfer, 91, 233

\bibitem[Conroy et al.(2014)]{Conroy14} Conroy, C., Graves, 
G.~J., \& van Dokkum, P.~G.\ 2014, \apj, 780, 33 

\bibitem[Cottaar et al.(2014)]{Cottaar14} Cottaar, M., Covey, 
K.~R., Meyer, M.~R., et al.\ 2014, \apj, 794, 125 

\bibitem[Cui et al.(2012)]{Cui12} Cui, X.-Q., Zhao, Y.-H. Chi, Y.-Q., et
al.\ 2012, RAA, 12, 1197

\bibitem[Cunha 
\& Smith(2006)]{Cunha06} Cunha, K., \& Smith, V.~V.\ 2006, \apj, 651, 491 

\bibitem[Cunha et al.(2007)]{Cunha07} Cunha, K., Sellgren, K., 
Smith, V.~V., et al.\ 2007, \apj, 669, 1011 

\bibitem[Cunha et al.(2015)]{Cunha15} Cunha, K., Smith, V.~V., Johnson,
	J.~A. 2015, \apj, 798, 41

\bibitem[Dalton et al.(2014)]{Dalton14} Dalton, G., Trager, S.,
Abrams, D. C., et al. 2014, in Society of Photo-Optical Instrumentation
Engineers (SPIE) Conference Series, Vol. 9147, Society of Photo-Optical
Instrumentation Engineers (SPIE) Conference Series

\bibitem[Dehnen(1998)]{Dehnen98} Dehnen, W.\ 1998, \aj, 115, 2384 

\bibitem[de Jong et al.\ (2014)]{deJong14} de Jong, R.~S., Barden,
S., Bellido-Tirado, O. et al. 2014,  in Society of Photo-Optical
Instrumentation Engineers (SPIE) Conference Series, Vol. 9147,
Society of Photo-Optical Instrumentation Engineers (SPIE) Conference
Series

\bibitem[Deshpande et al.\ (2013)]{Deshpande13} Deshpande, R., Bender,
C.~F., Mahadevan, S. et al.\ 2013, \aj, 146, 156

\bibitem[Duch{\^e}ne 
\& Kraus(2013)]{Duchene13} Duch{\^e}ne, G., \& Kraus, A.\ 2013, \araa, 51, 269 

\bibitem[Duquennoy 
\& Mayor(1991)]{Duquennoy91} Duquennoy, A., \& Mayor, M.\ 1991, \aap, 248, 485 

\bibitem[Edvardsson et 
al.(1993)]{Edvardsson93} Edvardsson, B., Andersen, J., Gustafsson, B., et al.\ 1993, \aap, 275, 101 

\bibitem[Eggen(1958)]{Eggen58} Eggen, O.~J.\ 1958, \mnras, 118, 65 

\bibitem[Eggen(1998)]{Eggen98} Eggen, O.~J.\ 1998, \aj, 116, 782 

\bibitem[Eikenberry et al.(2014)]{Eikenberry14} Eikenberry, S.~S., 
Chojnowski, S.~D., Wisniewski, J., et al.\ 2014, \apjl, 784, LL30 

\bibitem[Eisenstein et al.(2011)]{Eisenstein11} Eisenstein, D.~J., 
Weinberg, D.~H., Agol, E., et al.\ 2011, \aj, 142, 72 

\bibitem[Epstein et al.(2014)]{Epstein14} Epstein, C.~R.
Elsworth, Y.~P., Johnson, J.~A. 2014, \apj, 785, L28

\bibitem[Famaey et 
al.(2005)]{Famaey05} Famaey, B., Jorissen, A., Luri, X., et al.\ 2005, \aap, 430, 165 

\bibitem[Fleming et al.(2015)]{Fleming15} Fleming, S.~W., 
Mahadevan, S., Deshpande, R., et al.\ 2015, \aj, 149, 143 

\bibitem[Foster et al.\ (2015)]{Foster15} Foster, J.~B., Cottaar, M.,
Covey, K.~R. et al.\ 2015, \apj, 799 136

\bibitem[Fran{\c c}ois et al.(2004)]{Francois04} Fran{\c c}ois, P.,
Matteucci, F., Cayrel, R., et al.\ 2004, \aap, 421, 613

\bibitem[Freeman \& Bland-Hawthorn(2002)]{Freeman02} Freeman, K.,
\& Bland-Hawthorn, J.\ 2002, \araa, 40, 487

\bibitem[Freeman et al.(2013)]{Freeman13} Freeman, K., Ness, M., 
Wylie-de-Boer, E., et al.\ 2013, \mnras, 428, 3660 

\bibitem[Frinchaboy et al.(2013)]{Frinchaboy13} Frinchaboy, P.M., Thompson,
	B., Jackson, K.M. et al.\ 2013, \apj, 777, L1
	
\bibitem[Frinchaboy et al.(2015)]{Frinchaboy15} Frinchaboy, P.M., Thompson,
O'Connell, J. 2015, in preparation
	
\bibitem[Fuhrmann(2004)]{Fuhrmann04} Fuhrmann, K.\ 2004, 
Astronomische Nachrichten, 325, 3 

\bibitem[Fulbright et al.(2007)]{Fulbright07} Fulbright, J.~P., 
McWilliam, A., \& Rich, R.~M.\ 2007, \apj, 661, 1152 

\bibitem[Gai et al.(2011)]{Gai11} Gai, N., Basu, S., Chaplin, 
W.~J., \& Elsworth, Y.\ 2011, \apj, 730, 63 

\bibitem[Garc{\'{\i}}a P{\'e}rez et al.(2015)]{GarciaPerez15} Garc{\'{\i}}a P{\'e}rez, A.~E., Allende Prieto, C., 
Holtzman, J.~A., et al.\ 2015, in preparation

\bibitem[Ge et al.(2008)]{Ge08} Ge, J., Mahadevan, S., Lee, 
B., et al.\ 2008, Extreme Solar Systems, 398, 449 

\bibitem[Geisler(1984)]{Geisler84} Geisler, D.\ 1984, \pasp, 96, 723 

\bibitem[Gilliland et al.(2010)]{Gilliland10} Gilliland, R.~L., 
Brown, T.~M., Christensen-Dalsgaard, J., et al.\ 2010, \pasp, 122, 131 

\bibitem[Gilmore et al.\ (2012)]{Gilmore12} Gilmore, G., Randich, S.,
	Asplund, M. et al.\ 2012, The Messenger, 147, 25
	
\bibitem[Gilmore et 
al.(1989)]{Gilmore89} Gilmore, G., Wyse, R.~F.~G., \& Kuijken, K.\ 1989, \araa, 27, 555 

\bibitem[Girardi et al.(2005)]{Girardi05} Girardi, L., Groenewegen,
M.~A.~T., Hatziminaoglou, E., \& da Costa, L.\ 2005, \aap, 436, 895

\bibitem[Gonzalez et al.(2011)]{Gonzalez11} Gonzalez, O.~A., Rejkuba, M.
Zoccali, M. et al. 2011, \aap, 530, 54

\bibitem[Gratton, Carretta \& Bragaglia (2012)]{Gratton12}
   Gratton, R.G., Carretta, E. \& Bragaglia, A. 2012, A\&ARv, 20, 50

\bibitem[Griffin(1985)]{Griffin85} Griffin, R.~F.\ 1985, NATO 
Advanced Science Institutes (ASI) Series C, 150, 1 

\bibitem[Grillmair(2009)]{Grillmair09} Grillmair, C.~J.\ 2009, 
\apj, 693, 1118 

\bibitem[Gunn et al.(2006)]{Gunn06} Gunn, J.~E., Siegmund, 
W.~A., Mannery, E.~J., et al.\ 2006, \aj, 131, 2332 

\bibitem[Hammersley et al.(2000)]{Hammersley00} Hammersley, P.~L., 
Garz{\'o}n, F., Mahoney, T.~J., L{\'o}pez-Corredoira, M., 
\& Torres, M.~A.~P.\ 2000, \mnras, 317, L45

\bibitem[Hayden et al.(2014)]{Hayden14} Hayden, M.~R., Holtzman, J.~A.,
	Bovy, J. 2014, \aj, 147, 116
	
\bibitem[Hayden et al.(2015)]{Hayden15} Hayden, M.~R., Bovy, J., Holtzman, J.~A.,
	 2015, \aj, submitted

\bibitem[Hekker et al.(2011)]{Hekker11} Hekker, S., Gilliland, 
R.~L., Elsworth, Y., et al.\ 2011, \mnras, 414, 2594 

\bibitem[Herbig(1995)]{Herbig95} Herbig, G.~H.\ 1995, \araa, 33, 19 

\bibitem[Hinkle et al.(1995)]{Hinkle95} Hinkle, K., Wallace, L.,
\& Livingston, W.~C.\ 1995, Infrared atlas of the Arcturus spectrum,
0.9-5.3 microns, San Francisco, Calif.~: Astronomical Society of
the Pacific, 1995.

\bibitem[Holtzman et al.(2015)]{Holtzman15} Holtzman, J.~.A., Shetrone, M., Allende Prieto, C., 
Y., et al.\ 2015, in preparation

\bibitem[Ivezi{\'c} et 
al.(2012)]{Ivezic12} Ivezi{\'c}, {\v Z}., Beers, T.~C., \& Juri{\'c}, M.\ 2012, \araa, 50, 251

\bibitem[Jordi et al.(2010)]{Jordi2010} Jordi, C., Gebran, M., Carrasco, J.~M., 
et al.\ 2010, \aap, 523, A48 

\bibitem[Junqueira et al.(2015)]{Junqueira15} Junqueira, T.~C., 
Chiappini, C., L{\'e}pine, J.~R.~D., Minchev, I., 
\& Santiago, B.~X.\ 2015, arXiv:1503.00926 

\bibitem[Juri{\'c} et al.(2008)]{Juric08} Juri{\'c}, M., 
Ivezi{\'c}, {\v Z}., Brooks, A., et al.\ 2008, \apj, 673, 864 

\bibitem[Karakas(2010)]{Karakas10} Karakas, A.~I.\ 2010, \mnras, 
403, 1413 

\bibitem[Kewley \& Ellison(2008)]{Kewley08} Kewley, L.~J., \& Ellison, S.~L.\ 2008, 
\apj, 681, 1183 


\bibitem[Koesterke et al.(2008)]{2008ApJ...680..764K} Koesterke, L., 
Allende Prieto, C., \& Lambert, D.~L.\ 2008, \apj, 680, 764 

\bibitem[Koesterke(2009)]{Koesterke09} Koesterke, L.\ 2009, 
American Institute of Physics Conference Series, 1171, 73 

\bibitem[Kubryk et al.(2014)]{Kubryk14} Kubryk, M., Prantzos, 
N., \& Athanassoula, E.\ 2014, arXiv:1412.4859 

\bibitem[Kurucz (1993)]{Kurucz93} Kurucz, R.~L. 1993, ATLAS9 Stellar
Atmosphere Programs and 2 km s$^{-1}$ grid, Kurucz CD-ROM No. 13
(Cambridge, MA: Smithsonian Astrophysical Observatory)

\bibitem[Laird (1986)]{Laird86} Laird, J.~B.\ 1986, \apj, 303, 
718 

\bibitem[Latham et al.(1989)]{Latham89} Latham D.~W., Stefanik, R.~P.,
	Mazeh, T., Mayor, M. \& Burki, G., Nature, 339, L38

\bibitem[L{\'o}pez-Corredoira et al.(2002)]{LopezCorredoira02}
L{\'o}pez-Corredoira, M., Cabrera-Lavers, A., Garz{\'o}n, F., \&
Hammersley, P.~L.\ 2002, \aap, 394, 883

\bibitem[Majewski(1993)]{Majewski93} Majewski, S.~R.\ 1993, \araa, 31, 575 

\bibitem[Majewski et al.(2012)]{Majewski12} Majewski, S.~R., 
Nidever, D.~L., Smith, V.~V., et al.\ 2012, \apjl, 747, L37 

\bibitem[Majewski et al.(2013)]{Majewski13} Majewski, S.~R., 
Hasselquist, S., {\L}okas, E.~L., et al.\ 2013, \apjl, 777, L13 

\bibitem[Majewski et al.(2015)]{Majewski15} Majewski, S.~R., Law, D.~R.,
Hasselquist, S. \& Damke, G. 2015, in Lessons from the Local Group, 
eds. Freeman, K., Elmegreen, B., Block, D. \& Woolway, M., Springer:Heidelberg, pp. 231-242

\bibitem[Majewski et al.(2000)]{Majewski00} Majewski, S.~R., 
Ostheimer, J.~C., Kunkel, W.~E., \& Patterson, R.~J.\ 2000, \aj, 120, 2550 


\bibitem[Majewski et al.(2003)]{Majewski03} Majewski, S.R., Skrutskie,
M.F., Weinberg, M.D. \& Ostheimer, J.C.  2003, \apj, 599, 1082

\bibitem[Majewski et al.(2010)]{Majewski10} Majewski, S.~R., 
Wilson, J.~C., Hearty, F., Schiavon, R.~R., 
\& Skrutskie, M.~F.\ 2010, IAU Symposium, 265, 480

\bibitem[Majewski et al.(2011)]{Majewski11} Majewski, S.~R., 
Zasowski, G., \& Nidever, D.~L.\ 2011, \apj, 739, 25 


\bibitem[Malo et al.(2013)]{Malo13} Malo, L., Doyon, R., 
Lafreni{\`e}re, D., et al.\ 2013, \apj, 762, 88 

\bibitem[Martig et al.(2015)]{Martig15} Martig, M., Rix, H.-W., Silva
Aguirre, V. et al. 2015, \mnras, submitted, arXiv:1412.3453

\bibitem[Masseron \& Gilmore (2015)]{Masseron15} Masseron, T. \& Gilmore,
G. 2015, arXiv:1503.00537

\bibitem[Matteucci(2001)]{Matteucci01} Matteucci, F.\ 2001, The Chemical 
Evolution of the Galaxy, 
Astrophysics and Space Science Library, Vol. 253, Dordrecht: Kluwer Academic Publishers

\bibitem[McWilliam(1997)]{McWilliam97} McWilliam, A.\ 1997, \araa, 35, 503 

\bibitem[McWilliam 
\& Zoccali(2010)]{McWilliam10} McWilliam, A., \& Zoccali, M.\ 2010, \apj, 724, 1491 


\bibitem[Merrill 
\& Ridgway(1979)]{Merrill79} Merrill, K.~M., \& Ridgway, S.~T.\ 1979, \araa, 17, 9 


\bibitem[M\'esz\'aros et al.(2012)]{Meszaros12} M\'esz\'aros, Sz., Allende
Prieto, C., Edvardsson, B. et al.\ 2012, \aj, 144, 120

\bibitem[M{\'e}sz{\'a}ros et al.(2013)]{Meszaros13} 
M{\'e}sz{\'a}ros, S., Holtzman, J., Garc{\'{\i}}a P{\'e}rez, A.~E., et al.\ 
2013, \aj, 146, 133 

\bibitem[M\'esz\'aros et al.(2015)]{Meszaros15} M\'esz\'aros, Sz., Martell,
	S.~L., Shetrone, M. 2015, \aj, in press, arXiv:1501.05127

\bibitem[Michel et al.(2008)]{Michel08} Michel, E., Baglin, A., 
Weiss, W.~W., et al.\ 2008, Communications in Asteroseismology, 156, 73 

\bibitem[Miller \& Brown (2004)]{Miller04} Miller, C.~E. \& Brown, L.~R.
2004, J. Mol. Spectrosc., 228, 329

\bibitem[Milone et al.(2000)]{Milone00} Milone, A., Barbuy, B., 
\& Schiavon, R.~P.\ 2000, \aj, 120, 131 


\bibitem[Milone et al.(2008)]{Milone08} Milone, A. P., Bedin, L.
R., Piotto, G., et al. 2008, \apj, 673, 241

\bibitem[Minchev \& Famaey(2010)]{Minchev10}
  Minchev,~I. \& Famaey,~B. 2010, \apj, 722, 112

\bibitem[Minchev et 
al.(2013)]{Minchev13} Minchev, I., Chiappini, C., \& Martig, M.\ 
2013, \aap, 558, A9 

\bibitem[Minchev et 
al.(2014)]{Minchev14} Minchev, I., Chiappini, C., \& Martig, M.\
2014, \aap, 572, A92

\bibitem[Molloy et al.(2015)]{Molloy15} Molloy, M., Smith, 
M.~C., Evans, N.~W., \& Shen, J.\ 2015, arXiv:1505.04245 

\bibitem[Montes et al.(2001)]{Montes01} Montes, D., 
L{\'o}pez-Santiago, J., G{\'a}lvez, M.~C., et al.\ 2001, \mnras, 328, 45 

\bibitem[Morrison et al.(2000)]{Morrison00} Morrison, H.~L., 
Mateo, M., Olszewski, E.~W., et al.\ 2000, \aj, 119, 2254 

\bibitem[Mosser et 
al.(2010)]{Mosser10} Mosser, B., Belkacem, K., Goupil, M.-J., et al.\ 2010, \aap, 517, AA22 

\bibitem[Nelder \& Meade (1965)]{Nelder65} Nelder, J.A. \& Meade, R.
1965, Comput.~J., 7, 308

\bibitem[Ness et al.(2015)]{Ness15} Ness, M. et al.\ 2015, in preparation

\bibitem[Newberg et al.(2007)]{Newberg07} Newberg, H.~J., Yanny, 
B., Cole, N., et al.\ 2007, \apj, 668, 221 

\bibitem[Nidever et al.(2012)]{Nidever12} Nidever, D.~L., 
Zasowski, G., Majewski, S.~R., et al.\ 2012, \apjl, 755, L25 

\bibitem[Nidever et al.(2014)]{Nidever14} Nidever, D.~L., Bovy, J., Bird,
	J.~C. 2014, \apj, 796, 38

\bibitem[Nidever et al.(2015)]{Nidever15} Nidever, D.~L., Holtzman, J.~A., Allende Prieto, C.~A., et
al.\ 2015, \aj, submitted, arXiv:1501.03742

\bibitem[Nissen 
\& Schuster(2010)]{Nissen10} Nissen, P.~E., \& Schuster, W.~J.\ 2010, \aap, 511, LL10 

\bibitem[Nomoto et 
al.(2013)]{Nomoto13} Nomoto, K., Kobayashi, C., \& Tominaga, N.\ 2013, \araa, 51, 457 


\bibitem[Owen et al.(1998)]{Owen98} Owen, R.~E., Buffaloe, 
M.~J., Leger, R.~F., et al.\ 1998, Fiber Optics in Astronomy III, 152, 98 

\bibitem[Perryman et al.(2001)]{Perryman01} Perryman, M. A. C., de Boer,
	K. S., Gilmore, G., et al. 2001, \aap, 369, 339 

\bibitem[Pinsonneault et al.(2014)]{Pinsonneault14} Pinsonneault, 
M.~H., Elsworth, Y., Epstein, C., et al.\ 2014, \apjs, 215, 19 

\bibitem[Piotto et al.(2007)]{Piotto07} Piotto, G., Bedin, L. R.,
Anderson, J., et al. 2007, \apj, 661, L53

\bibitem[Proctor 
\& Sansom(2002)]{Proctor02} Proctor, R.~N., \& Sansom, A.~E.\ 2002, \mnras, 333, 517 

\bibitem[Rayner et al.(2009)]{Rayner09} Rayner, J.~T., Cushing, 
M.~C., \& Vacca, W.~D.\ 2009, \apjs, 185, 289 

\bibitem[Reddy et al.(2006)]{Reddy06} Reddy, B.~E., Lambert, 
D.~L., \& Allende Prieto, C.\ 2006, \mnras, 367, 1329 

\bibitem[Reyl{\'e} et al.(2009)]{Reyle09} Reyl{\'e},
C., Marshall, D.~J., Robin, A.~C., \& Schultheis, M.\ 2009, \aap,
495, 819

\bibitem[Rich 
\& Origlia(2005)]{Rich05} Rich, R.~M., \& Origlia, L.\ 2005, \apj, 634, 1293 

\bibitem[Rich et al.(2007)]{Rich07} Rich, R.~M., Reitzel, 
D.~B., Howard, C.~D., \& Zhao, H.\ 2007, \apjl, 658, L29 

\bibitem[Rix \& Bovy(2013)]{Rix13} Rix, H.-W., \&
Bovy, J.\ 2013, \aapr, 21, 61

\bibitem[Robin et 
al.(2012)]{Robin12} Robin, A.~C., Marshall, D.~J., Schultheis, M., \& Reyl{\'e}, C.\ 2012, \aap, 538, 106 

\bibitem[Robin et 
al.(2003)]{Robin03} Robin, A.~C., Reyl{\'e}, C., Derri{\`e}re, S., \& Picaud, S.\ 2003, \aap, 409, 523 

\bibitem[Rocha-Pinto et al.(2004)]{Rocha-Pinto04} Rocha-Pinto, H.~J., 
Majewski, S.~R., Skrutskie, M.~F., Crane, J.~D., 
\& Patterson, R.~J.\ 2004, \apj, 615, 732 

\bibitem[Rockosi et al.(2009)]{Rockosi09} Rockosi, C., Beers, T.~C.,
	Majewski, S., Schiavon, R. \& Eisenstein, D. 2009, in Astronomy,
	Vol. 2010, astro2010, The Astronomy and Astrophysics Decadal
	Survey, 14
	
\bibitem[Rothman et al.(2013)]{Rothman13} Rothman, L.~S., Gordon, I.~E.,
	Babikov, Y. et al.\ 2013, J. Mol. Spectrosc., 130, 4	

\bibitem[Ryde et 
al.(2010)]{Ryde10} Ryde, N., Gustafsson, B., Edvardsson, B., et al.\ 2010, \aap, 509, AA20 

\bibitem[Saglia et al.(2002)]{Saglia02} Saglia, R.~P., Maraston, 
C., Thomas, D., Bender, R., \& Colless, M.\ 2002, \apjl, 579, L13 

\bibitem[Santiago et al.(2015)]{Santiago15a}
Santiago, B.~X., Brauer, D.~E., Anders, F., et al.\ 2015, \aap, submitted (arXiv:1501.05500)
 
\bibitem[Schiavon(2007)]{Schiavon07} Schiavon, R.~P.\ 2007, \apjs, 
171, 146 

\bibitem[Schiavon(2010)]{Schiavon10} Schiavon, R.~P.\ 2010, 
Publication of Korean Astronomical Society, 25, 83 

\bibitem[Schiavon et al.(2013)]{Schiavon13} Schiavon, R.~P., 
Caldwell, N., Conroy, C., et al.\ 2013, \apjl, 776, L7 

\bibitem[Schiavon et al.(2015)]{Schiavon15} Schiavon, R.~P., et al. \ 2015, \mnras, submitted.

\bibitem[Sch{\"o}nrich 
\& Binney(2009)]{Schonrich09} Sch{\"o}nrich, R., \& Binney, J.\ 2009, \mnras, 396, 203

\bibitem[Schlegel et al.(1998)]{Schlegel98} Schlegel, D.~J., 
Finkbeiner, D.~P., \& Davis, M.\ 1998, \apj, 500, 525 

\bibitem[Schultheis et al.\ (2014)]{Schultheis14} Schultheis, M., Zasowski,
	G., Allende Prieto, C. et al.\ 2014, \aj, 148, 24

\bibitem[Sellwood \& Binney(2002)]{Sellwood02} 
  Sellwood, J.~A., \& Binney, J.~J.\ 2002, \mnras, 336, 785

\bibitem[Shectman et al.(1996)]{Shectman96} Shectman, S.~A., 
Landy, S.~D., Oemler, A., et al.\ 1996, \apj, 470, 172 

\bibitem[Shetrone et al.(2015)]{Shetrone15} Shetrone, M., Bizyaev, D.~M.,
Lawler, J., et al.\ 2015, in preparation

\bibitem[Shimansky et al.(2003)]{Shimansky03} Shimansky, V.~V., 
Bikmaev, I.~F., Galeev, A.~I., et al.\ 2003, Astronomy Reports, 47, 750 

\bibitem[Siegmund et al.(1998)]{Siegmund98} Siegmund, W.~A., Owen, 
R.~E., Granderson, J., et al.\ 1998, Fiber Optics in Astronomy III, 152, 92 

\bibitem[Skrutskie et al.(2006)]{Skrutskie06} Skrutskie, M.~F., 
Cutri, R.~M., Stiening, R., et al.\ 2006, \aj, 131, 1163 

\bibitem[Skrutskie \& Wilson (2015)]{Skrutskie15} Skrutskie, M.~F. \&
Wilson, J.~C. 2015, arXiv:1503.08918

\bibitem[Smee et al.(2013)]{Smee13} Smee, S.~A., Gunn, J.~E., 
Uomoto, A., et al.\ 2013, \aj, 146, 32 

\bibitem[Smith et al.(2013)]{Smith13} Smith, V.~V., Cunha, K., 
Shetrone, M.~D., et al.\ 2013, \apj, 765, 16 '

\bibitem[Smith et al.(2002)]{Smith02} Smith, V.~V., Hinkle, 
K.~H., Cunha, K., et al.\ 2002, \aj, 124, 3241 

\bibitem[Smith 
\& Lambert(1985)]{Smith85} Smith, V.~V., \& Lambert, D.~L.\ 1985, \apj, 294, 326 

\bibitem[Smith 
\& Lambert(1986)]{Smith86} Smith, V.~V., \& Lambert, D.~L.\ 1986, \apj, 311, 843 

\bibitem[Smith 
\& Lambert(1990)]{Smith90} Smith, V.~V., \& Lambert, D.~L.\ 1990, \apjs, 72, 387 

\bibitem[Soubiran et 
al.(2010)]{Soubiran10} Soubiran, C., Le Campion, J.-F., Cayrel de Strobel, G., \& Caillo, A.\ 2010, \aap, 515, AA111 

\bibitem[Stasi{\'n}ska et al.(2012)]{Stasinska12} Stasi{\'n}ska, 
G., Prantzos, N., Meynet, G., et al.\ 2012, EAS Publications Series, 54, 
255 

\bibitem[Steinmetz et al.\ (2006)]{Steinmetz06} Steinmetz, M., Zwitter,
	T., Siebert, A., et al.\ 2006, \aj, 132, 1645

\bibitem[Tinsley(1979)]{Tinsley79} Tinsley, B.~M.\ 1979, \apj, 229, 1046 

\bibitem[Tinsley(1980)]{Tinsley80} Tinsley, B.~M.\ 1980, \fcp, 5, 
287 
	
\bibitem[Terrien et al.(2014)]{Terrien14} Terrien, R.~C., 
Mahadevan, S., Deshpande, R., et al.\ 2014, \apj, 782, 61 

\bibitem[Tomkin et al.\ (1985)]{Tomkin85} Tomkin, J., Lambert, 
D.~L., \& Balachandran, S.\ 1985, \apj, 290, 289 

\bibitem[Tremaine et al.\ (1975)]{Tremaine75} Tremaine, S.~D., 
Ostriker, J.~P., \& Spitzer, L., Jr.\ 1975, \apj, 196, 407 

\bibitem[Tsuji \& Nakajima(2014)]{Tsuji14} Tsuji, T. \& Nakajima, T. 2014,
\pasj, 66, 98

\bibitem[Tumlinson(2010)]{Tumlinson10} Tumlinson, J.\ 2010, \apj, 
708, 1398

\bibitem[Udry et al.(1997)]{Udry97} Udry, S., Mayor, M., 
Andersen, J., et al.\ 1997, Hipparcos - Venice '97, 402, 693 

\bibitem[van Saders 
\& Pinsonneault(2013)]{vanSaders13} van Saders, J.~L., \& Pinsonneault, M.~H.\ 2013, \apj, 776, 67 

\bibitem[Venn et al.(2004)]{Venn04} Venn, K.~A., Irwin, M., 
Shetrone, M.~D., et al.\ 2004, \aj, 128, 1177 

\bibitem[Ventura et al.(2013)]{Ventura13} Ventura, P., Di 
Criscienzo, M., Carini, R., \& D'Antona, F.\ 2013, \mnras, 431, 3642 

\bibitem[Vivas et al.(2001)]{Vivas01} Vivas, A.~K., Zinn, R., 
Andrews, P., et al.\ 2001, \apjl, 554, L33 

\bibitem[Wallerstein (1962)]{Wallerstein62} Wallerstein, G.\ 1962, 
\apjs, 6, 407 

\bibitem[Wheeler et al.(1989)]{Wheeler89} Wheeler, J.~C., Sneden, C., 
\& Truran, J.~W., Jr.\ 1989, \araa, 27, 279 

\bibitem[Willman et al.(2005)]{Willman05} Willman, B., Dalcanton, 
J.~J., Martinez-Delgado, D., et al.\ 2005, \apjl, 626, L85

\bibitem[Wilson et al.(2010)]{Wilson10a} Wilson, J.~C., Hearty, 
F., Skrutskie, M.~F., et al.\ 2010, \procspie, 7735, 

\bibitem[Wilson et al.(2015)]{Wilson15} Wilson, J.~C., Hearty, 
F., Skrutskie, M.~F., et al.\ 2015, in preparation 

\bibitem[Wood et al.\ (2014)]{Wood14} Wood, M.P., Lawler, J.E. \& Shetrone,
M.D. 2014, \apj, 787, L16

\bibitem[Woosley 
\& Weaver(1995)]{Woosley95} Woosley, S.~E., \& Weaver, T.~A.\ 1995, \apjs, 101, 181 

\bibitem[Worthey et al.(1992)]{Worthey92} Worthey, G., Faber, 
S.~M., \& Gonzalez, J.~J.\ 1992, \apj, 398, 69 

\bibitem[Wright et al.(2010)]{Wright10} Wright, E.~L., 
Eisenhardt, P.~R.~M., Mainzer, A.~K., et al.\ 2010, \aj, 140, 1868 

\bibitem[Yanny et al.(2009)]{Yanny09} Yanny, B., Rockosi, C., Newberg, H.J.
et al.\ 2009, \aj, 137, 4377

\bibitem[Yong et al.(2012)]{Yong12} Yong, D., Carney, B.~W., 
\& Friel, E.~D.\ 2012, \aj, 144, 95 

\bibitem[York et al.(2000)]{York00} York, D.~G., Adelman, J., 
Anderson, J.~E., Jr., et al.\ 2000, \aj, 120, 1579 

\bibitem[Zamora et al.(2015)]{Zamora15} Zamora, O., Garc{\'{\i}}a-Hern{\'a}ndez, D.~A., 
Allende Prieto, C., et al.\ 2015, in preparation

\bibitem[Zasowski et al.(2013)]{Zasowski13} Zasowski, G., Johnson, 
J.~A., Frinchaboy, P.~M., et al.\ 2013, \aj, 146, 81 

\bibitem[Zasowski et al.(2015)]{Zasowski15} Zasowski, G., M\'enard, B.,
	Bizyaev, D. et al.\ 2015, \apj, 798, 35

\bibitem[Zhao et al.(2009)]{Zhao09} Zhao, B., Ge, J., 
\& Groot, J.\ 2009, \procspie, 7440, 74401E 

\bibitem[Zucker et al.(2012)]{Zucker12} Zucker, D. B., de Silva,
G., Freeman, K. et al.\ 2012, in ASP
Conf. Ser. 458, Galactic Archaeology: Near-Field Cosmology and the
Formation of the Milky Way, ed. W. Aoki, M. Ishigaki, T. Suda, T.
Tsujimoto, \& N. Arimoto (San Francisco, CA: ASP), 421



\end{thebibliography}
\end{document}